\documentclass[12pt]{article}
\usepackage[breaklinks=true,colorlinks=true,linkcolor=blue,urlcolor=blue,citecolor=blue]{hyperref}

\usepackage{amsmath,amssymb,amsfonts,amsthm,epsfig,bm}
\usepackage{graphicx,float,enumerate,natbib}
\usepackage[titletoc,title]{appendix}
\usepackage[]{color}
\usepackage{bm}
\usepackage{hyperref}
\usepackage{setspace}
\usepackage{epstopdf}
\usepackage{ragged2e}
\usepackage{arydshln}

\usepackage[font=small]{caption}
\usepackage{apalike}

\newcommand{\citeh}[1]{\citeauthor{#1} \citeyear{#1}}
\renewcommand{\citep}[1]{(\citeauthor{#1} \citeyear{#1})}

\begin{document}

\title{Erlang mixture modeling for Poisson process intensities}
\author{Hyotae Kim and Athanasios Kottas
\thanks{
Hyotae Kim (hkim153@ucsc.edu) is Ph.D. student and Athanasios Kottas 
(thanos@soe.ucsc.edu) is Professor, Department of Statistics, 
University of California, Santa Cruz.}
}

\date{}
\maketitle

\begin{abstract}
\noindent
We develop a prior probability model for temporal Poisson process intensities
through structured mixtures of Erlang densities with common scale parameter, 
mixing on the integer shape parameters. The mixture weights are constructed 
through increments of a cumulative intensity function which is modeled 
nonparametrically with a gamma process prior. Such model specification 
provides a novel extension of Erlang mixtures for density estimation
to the intensity estimation setting. The prior model structure supports general
shapes for the point process intensity function, and it also enables effective 
handling of the Poisson process likelihood normalizing term resulting in 
efficient posterior simulation. The Erlang mixture modeling approach is further 
elaborated to develop an inference method for spatial Poisson processes. 
The methodology is examined relative to existing Bayesian
nonparametric modeling approaches, including empirical comparison with Gaussian process 
prior based models, and is illustrated with synthetic and real data examples. 
\end{abstract}

\noindent
KEY WORDS: Bayesian nonparametrics; Erlang mixtures; Gamma process; 
Markov chain Monte Carlo; Non-homogeneous Poisson process.

\section{Introduction}
\label{sec:intro}

Poisson processes play a key role in both theory and applications of point processes.
They form a widely used class of stochastic models for point patterns that arise in 
biology, ecology, engineering and finance among many other disciplines. 
%In economics, for example, interest may lie in the number of customers who visit a shop in a 
%bounded time interval or more broadly, in the pattern of the number of customers visiting the 
%shop over time. We can come up with a Poisson process for the function accounting for the 
%number of customers over time. Likewise, we can explore the change in population of a species 
%in an area through the Poisson process. 
%The Poisson process is characterized by its intensity function where our modeling framework 
%aims at the function. One reason for the common usage of the Poisson process is the simplicity 
%in the form of the likelihood. 
The relatively tractable form of the non-homogeneous Poisson process (NHPP)
likelihood is one of the reasons for the popularity of NHPPs in
applications involving point process data.

Theoretical background for the Poisson process can be found, for example, in 
\cite{Kingman:1993} and \cite{Daley_Vere-Jones:2003}. 
Regarding Bayesian nonparametric modeling and inference, prior probability 
models have been developed for the NHPP mean measure 
%NHPP cumulative intensity function
%$\int_{0}^{t} \lambda(u) \, \text{d}u$, for $t \in \mathbb{R}^{+}$, in the temporal case 
\cite[e.g.,][]{Lo1982,Lo:1992}, and mainly for the intensity function
of NHPPs over time and/or space.
Modeling methods for NHPP intensities include: mixtures of non-negative
kernels with weighted gamma process priors for the mixing 
measure \cite[e.g.,][]{Lo_Weng:1989,WI1998,Ishwaran_James:2004,KNWJ2014}; 
piecewise constant functions driven by Voronoi tessellations with 
Markov random field priors \cite[][]{Heikkinen_Arjas:1998,Heikkinen_Arjas:1999};
Gaussian process priors for logarithmic or logit transformations of the intensity 
\cite[e.g.,][]{Moller_et_al.:1998,Brix_Diggle:2001,AMM2009,Rodrigues_Diggle:2012};
and Dirichlet process mixtures for the NHPP density, i.e., the intensity function normalized 
in the observation window \cite[e.g.,][]{AK2006,KS2007,Taddy_Kottas:2012}. 
%
%Prior models for the NHPP intensity typically require complex
%computational methods for inference. The modeling approach 
%that builds from the NHPP density utilizes well established posterior 
%simulation methods for Dirichlet process mixtures, and it thus facilitates 
%extensions to different types of hierarchical settings 
%\cite[e.g.,][]{Taddy2010,KBMPO2012,XKS2015,RWK2017}. 
%However, this approach relies on a prior structure that models separately 
%the NHPP density and the total intensity over the observation window.
%

Here, we seek to develop a flexible and computationally efficient model for 
NHPP intensity functions over time or space. We focus on temporal
intensities to motivate the modeling approach and to detail the
methodological development, and then extend the model for spatial NHPPs.
The NHPP intensity over time  
is represented as a weighted combination of Erlang densities indexed by their integer
shape parameters and with a common scale parameter. Thus, different from existing mixture 
representations, the proposed mixture model is more structured with each Erlang density identified 
by the corresponding mixture weight. The non-negative mixture weights are defined through 
increments of a cumulative intensity on $\mathbb{R}^{+}$. Under certain conditions, the Erlang 
mixture intensity model can approximate in a pointwise sense general intensities on $\mathbb{R}^{+}$
(see Section \ref{subsec:model}). A gamma process prior is assigned to the 
primary model component, that is, the cumulative intensity that defines the mixture weights. 
Mixture weights driven by a gamma process prior result in flexible intensity function shapes, 
and, at the same time, ready prior-to-posterior updating given the observed point pattern. 
Indeed, a key feature of the model is that it can be implemented with an efficient Markov chain
Monte Carlo (MCMC) algorithm that does not require approximations, complex computational 
methods, or restrictive prior modeling assumptions in order to handle the NHPP likelihood 
normalizing term. 
%$\exp( - \int_{0}^{T} \lambda(u) \, \text{d}u )$. 
%The proposed prior model for NHPP intensities can be viewed as an extension of Erlang 
%mixtures for density estimation, for which the mixture weights are defined through increments
%of a distribution function on $\mathbb{R}^{+}$.
%
The intensity model is extended to the two-dimensional setting through products 
of Erlang densities for the mixture components, with the weights built
from a measure modeled again with a gamma process prior. The extension
to spatial NHPPs retains the appealing aspect of computationally
efficient MCMC posterior simulation.

The paper is organized as follows. Section \ref{sec:method} presents
the modeling and inference methodology for NHPP intensities over time.  
%including discussion relative to existing Bayesian nonparametric
%methods. 
The modeling approach for temporal NHPPs is illustrated through synthetic and real data 
in Section \ref{sec:egs}. Section \ref{spatial_NHPP} develops the model for spatial NHPP 
intensities, including two data examples. Finally, Section \ref{sec:disc} concludes with a 
discussion of the modeling approach relative to existing Bayesian 
nonparametric models, as well as of possible extensions of the methodology.

\section{Methodology for temporal Poisson processes}
\label{sec:method}

The mixture model for NHPP intensities is developed in Section \ref{subsec:model},
including discussion of model properties and theoretical justification. 
Sections \ref{subsec:priors} and \ref{subsec:MCMC} present a prior specification approach
and the posterior simulation method, respectively. 
%and in Section \ref{subsec:literature}, 
%we discuss the proposed modeling approach relative to existing methods. 

\subsection{The mixture modeling approach}
\label{subsec:model}

A NHPP on $\mathbb{R}^{+}$ can be defined through its intensity function, $\lambda(t)$, 
for $t \in \mathbb{R}^{+}$, a non-negative and locally integrable function such that:  
(a) for any bounded $B \subset \mathbb{R}^+$, the number of events in $B$, $N(B)$, is Poisson 
distributed with mean $\Lambda(B)=$ $\int_{B} \lambda(u) \, \text{d}u$; and (b) given $N(B)=n$, 
the times $t_{i}$, for $i=1,...,n$, that form the point pattern in $B$ arise independently and 
identically distributed (i.i.d.) according to density $\lambda(t)/\Lambda(B)$. 
Consequently, the likelihood for the NHPP intensity function, based on the point pattern 
$\{ 0 < t_{1} < ... < t_{n} < T \}$ observed in time window $(0,T)$, is proportional to 
$\exp( - \int_{0}^{T} \lambda(u) \, \text{d}u ) \prod_{i=1}^{n} \lambda(t_{i})$.

Our modeling target is the intensity function, $\lambda(t)$. 
We denote by $\text{ga}(\cdot \mid \alpha,\beta)$ the gamma density 
(or distribution, depending on the context) with mean $\alpha/\beta$.
%Denote generically by $\text{gamma}(\alpha,\beta)$ the gamma distribution with mean 
%$\alpha/\beta$, and by $\text{ga}(t \mid \alpha,\beta)$, for $t \in \mathbb{R}^{+}$, the 
%corresponding density. 
The proposed intensity model involves a structured mixture of 
Erlang densities, $\text{ga}(t \mid j,\theta^{-1})$, mixing on the integer shape parameters, 
$j$, with a common scale parameter $\theta$. The non-negative mixture weights are 
defined through increments of a cumulative intensity function, $H$, on $\mathbb{R}^{+}$, 
which is assigned a gamma process prior. More specifically,
\begin{equation}
\label{mixture_model}
\begin{array}{c}
\lambda(t) \equiv \lambda(t \mid H,\theta) \, =  \,
\sum\limits_{j=1}^{J} \omega_{j} \, \text{ga}(t \mid j,\theta^{-1}), \enspace t \in \mathbb{R}^{+} \\
\omega_{j} =  H(j\theta) - H((j-1)\theta), \enspace \enspace H \sim \mathcal{G}(H_0,c_0) ,
\end{array}
\end{equation}
where $\mathcal{G}(H_0,c_0)$ is a gamma process specified through $H_{0}$,
a (parametric) cumulative intensity function, and $c_{0}$, a positive scalar parameter
\citep{Kalbfleisch:1978}. For any $t \in \mathbb{R}^{+}$, $\text{E}(H(t))=$
$H_{0}(t)$ and $\text{Var}(H(t))=$ $H_{0}(t)/c_{0}$, and thus $H_{0}$ plays the role of 
the centering cumulative intensity, whereas $c_{0}$ is a precision parameter.
As an independent increments process, the $\mathcal{G}(H_0,c_0)$ prior for $H$ 
implies that, given $\theta$, the mixture weights are independent 
$\text{ga}(\omega_{j} \mid c_{0} \, \omega_{0j}(\theta),c_{0} )$ distributed, where 
$\omega_{0j}(\theta) =$ $H_{0}(j\theta) - H_{0}((j-1)\theta)$. As shown in Section 
\ref{subsec:MCMC}, this is a key property of the prior model with respect to 
implementation of posterior inference.

The model in (\ref{mixture_model}) is motivated by Erlang mixtures for density estimation,
under which a density $g$ on $\mathbb{R}^{+}$ is represented as $g(t) \equiv$ $g_{J,\theta}(t)=$ 
$\sum\nolimits_{j=1}^{J} p_{j} \, \text{ga}(t \mid j,\theta^{-1})$, for $t \in \mathbb{R}^{+}$.
Here, $p_{j}=$ $G(j\theta) - G((j-1)\theta)$, where $G$ is a distribution function on 
$\mathbb{R}^{+}$; the last weight can be defined as $p_{J}=$ $1 - G((J-1)\theta)$ to ensure
that $(p_{1},...,p_{J})$ is a probability vector. Erlang mixtures can approximate general densities 
on the positive real line, in particular, as $\theta \rightarrow 0$ and $J \rightarrow \infty$,
$g_{J,\theta}$ converges pointwise to the density of distribution function $G$ that defines 
the mixture weights. This convergence property can be obtained from
more general results from the probability 
literature that studies Erlang mixtures as extensions of Bernstein polynomials to the positive 
real line \cite[e.g.,][]{Butzer1954}; a convergence proof specifically for the distribution function 
of $g_{J,\theta}$ can be found in \cite{Lee_Lin:2010}.
Density estimation on compact sets via Bernstein polynomials has been explored in the 
Bayesian nonparametrics literature following the work of \cite{Petrone1999a,Petrone1999b}.
Regarding Bayesian nonparametric modeling with Erlang mixtures, we are only aware of 
\cite{XKSK2021} where renewal process inter-arrival distributions are modeled with 
mixtures of Erlang distributions, using a Dirichlet process prior \citep{Ferguson:1973}
for distribution function $G$.
\cite{VDP2008} study a parametric Erlang mixture model for density estimation on 
$\mathbb{R}^{+}$, working with a Dirichlet prior distribution for the mixture weights.

Therefore, the modeling approach in (\ref{mixture_model}) exploits the structure of the 
Erlang mixture density model to develop a prior for NHPP intensities, 
using the density/distribution function and intensity/cumulative intensity function connection 
to define the prior model for the mixture weights. In this context, the gamma process prior 
for cumulative intensity $H$ is the natural analogue to the Dirichlet process prior for 
distribution function $G$; recall that the Dirichlet process can be defined through 
normalization of a gamma process \cite[e.g.,][]{ghosal-vandervaart}.
To our knowledge, this is a novel construction for NHPP intensities that has not been 
explored for intensity estimation in either the classical or Bayesian nonparametrics literature.
The following lemma, which can be obtained applying Theorem 2 from \cite{Butzer1954},
provides theoretical motivation and support for the mixture model.

\vspace{0.15cm}
\noindent 
{\bf Lemma.} Let $h$ be the intensity function of a NHPP on $\mathbb{R}^{+}$, with 
cumulative intensity function 
$H(t)=$ $\int_{0}^{t} h(u) \, \text{d}u$, such that $H(t)=$ $O(t^{m})$, as $t \rightarrow \infty$, 
for some $m>0$. Consider the mixture intensity model $\lambda_{J,\theta}(t)=$ 
$\sum\nolimits_{j=1}^{J} \{ H(j\theta) - H((j-1)\theta) \} \, \text{ga}(t \mid j,\theta^{-1})$, 
for $t \in \mathbb{R}^{+}$. Then, as $\theta \rightarrow 0$ and $J \rightarrow \infty$,
$\lambda_{J,\theta}(t)$ converges to $h(t)$ at every point $t$ where $h(t)=$ $\text{d} H(t)/\text{d}t$.
 
\vspace{0.25cm}

The form of the prior model for the intensity in (\ref{mixture_model}) allows ready expressions 
for other NHPP functionals. For instance, the total intensity over the observation time 
window $(0,T)$ is given by $\int_{0}^{T} \lambda(u) \, \text{d}u =$
$\sum\nolimits_{j=1}^{J} \omega_{j} K_{j,\theta}(T)$, where 
$K_{j,\theta}(T)=$ $\int_{0}^{T} \text{ga}(u \mid j,\theta^{-1}) \, \text{d}u$ is the $j$-th Erlang 
distribution function at $T$. In the context of the MCMC posterior simulation method, 
this form enables efficient handling of the NHPP likelihood normalizing constant.
Moreover, the NHPP density on interval $(0,T)$ can be expressed as a mixture of truncated
Erlang densities. More specifically, 
\begin{equation}
\label{formula_NHPP_density}
f(t) \, = \, \frac{\lambda(t)}{\int_{0}^{T} \lambda(u) \, \text{d}u} \, = \,
\sum_{j=1}^{J} \omega^{*}_{j} \, k(t \mid j,\theta), \enspace t \in (0,T), 
\end{equation}
where $\omega^{*}_{j}=$ 
$\omega_{j} K_{j,\theta}(T) / \{ \sum\nolimits_{r=1}^{J} \omega_{r} K_{r,\theta}(T) \}$, 
and $k(t \mid j,\theta)$ is the $j$-th Erlang density truncated on $(0,T)$.

\begin{figure}[t!]
\centering	
\includegraphics[width=0.32\textwidth]{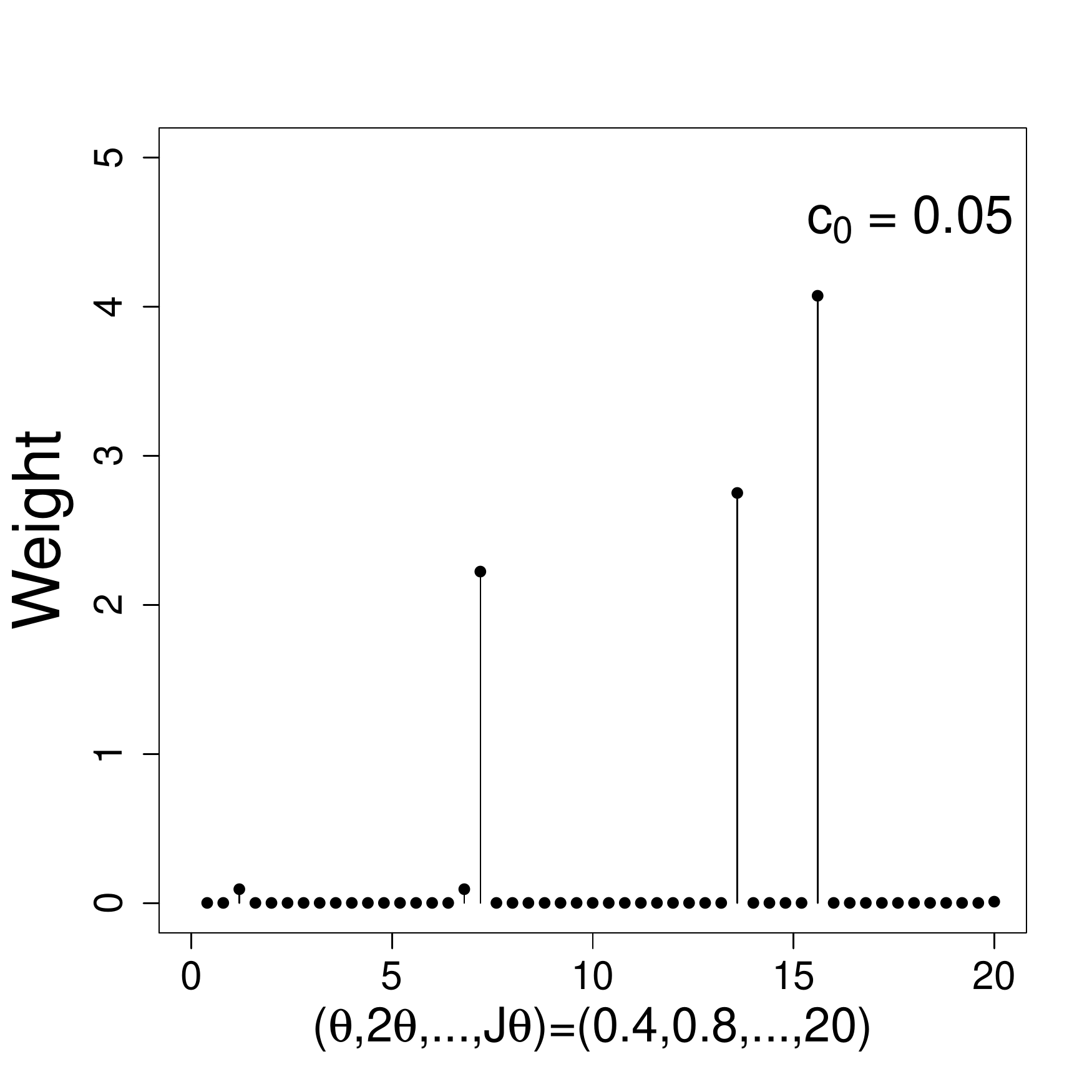}
\includegraphics[width=0.32\textwidth]{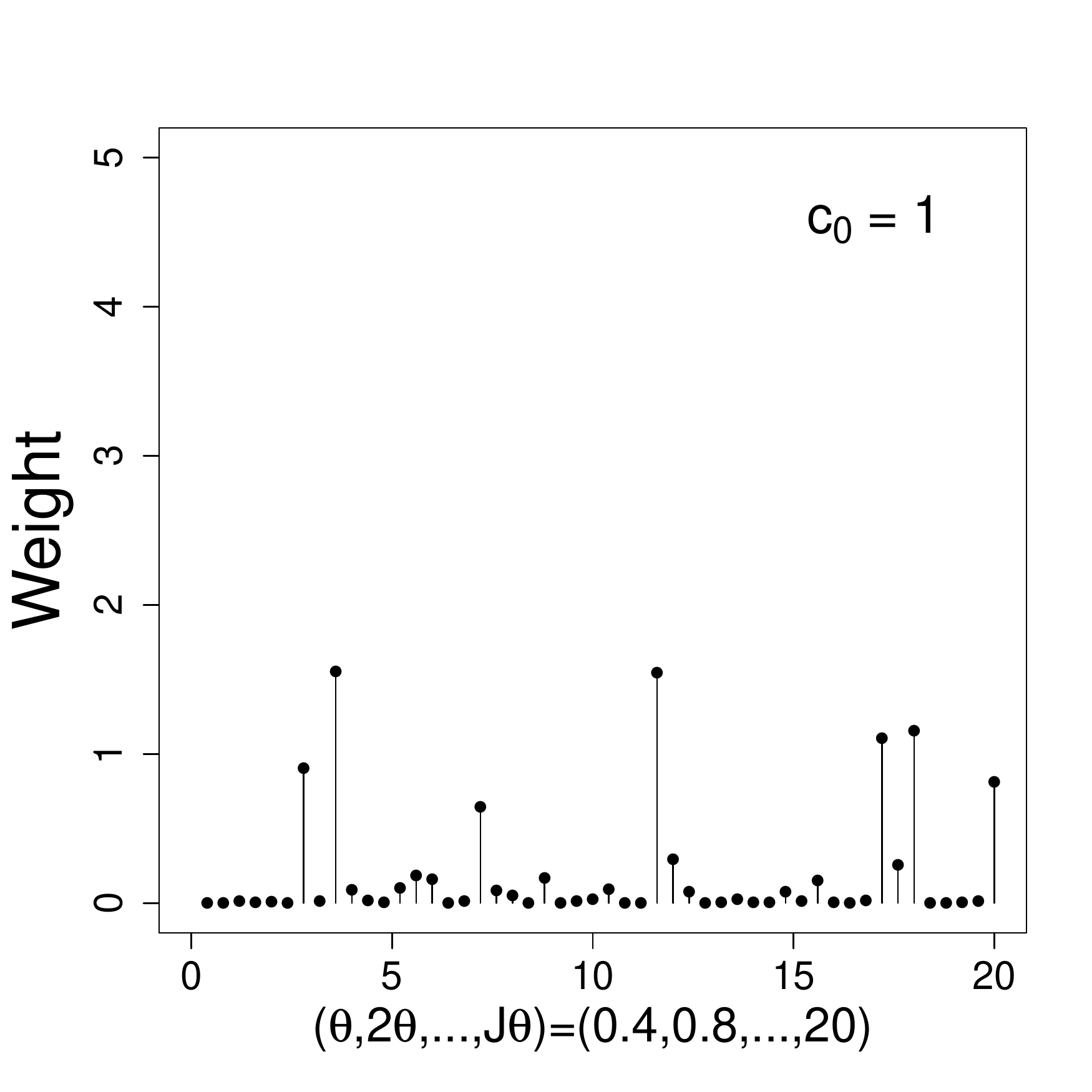}
\includegraphics[width=0.32\textwidth]{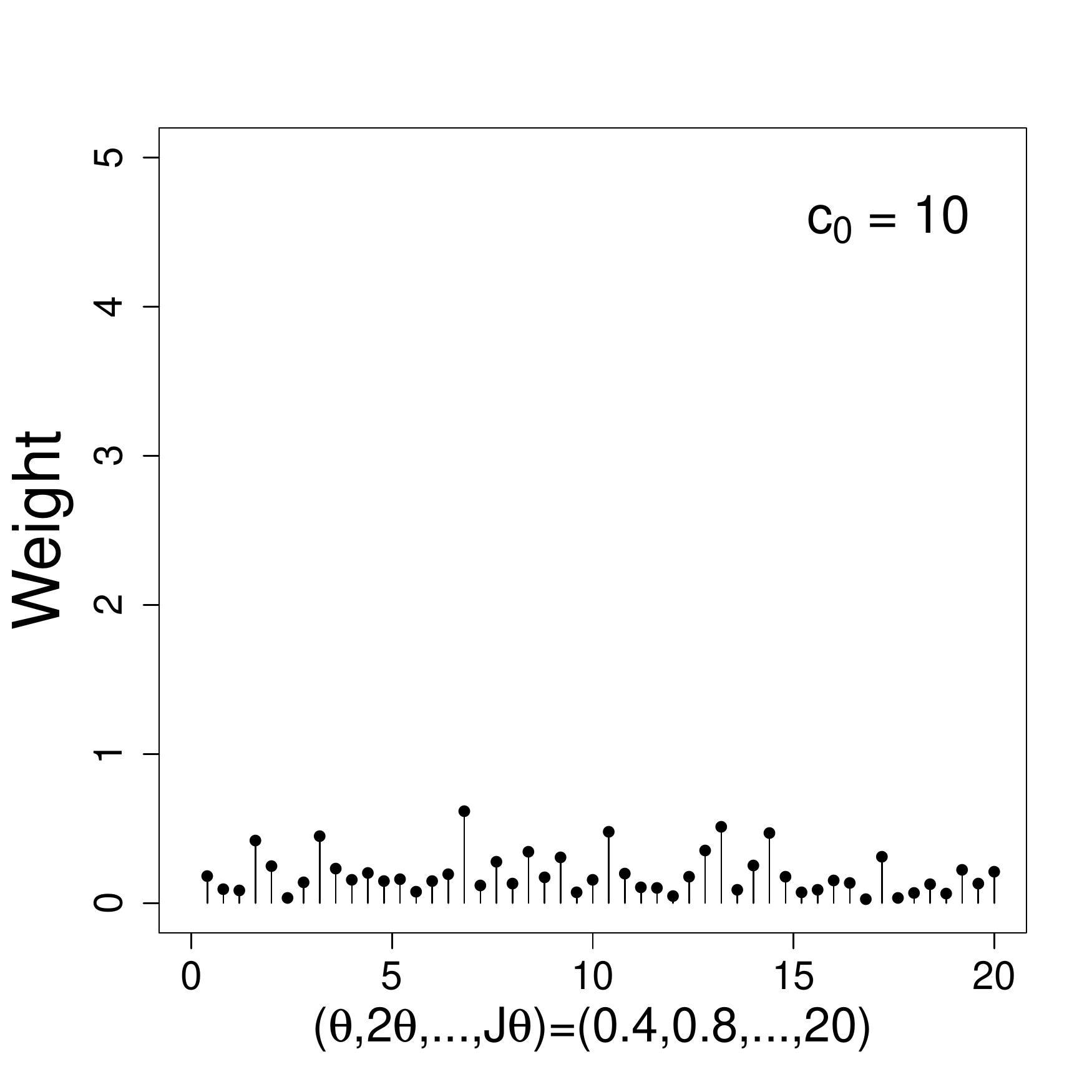}\\
\includegraphics[width=0.32\textwidth]{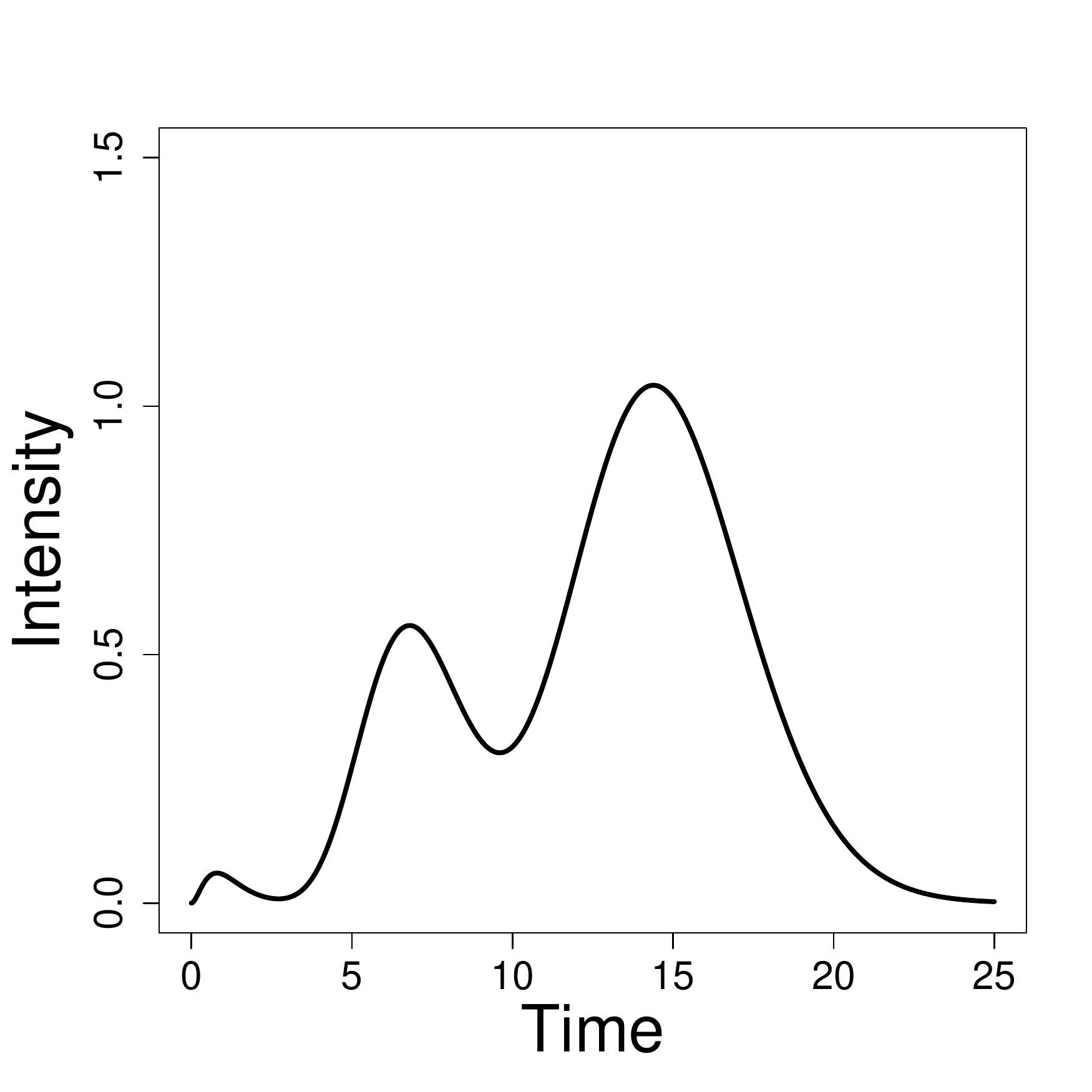}        
\includegraphics[width=0.32\textwidth]{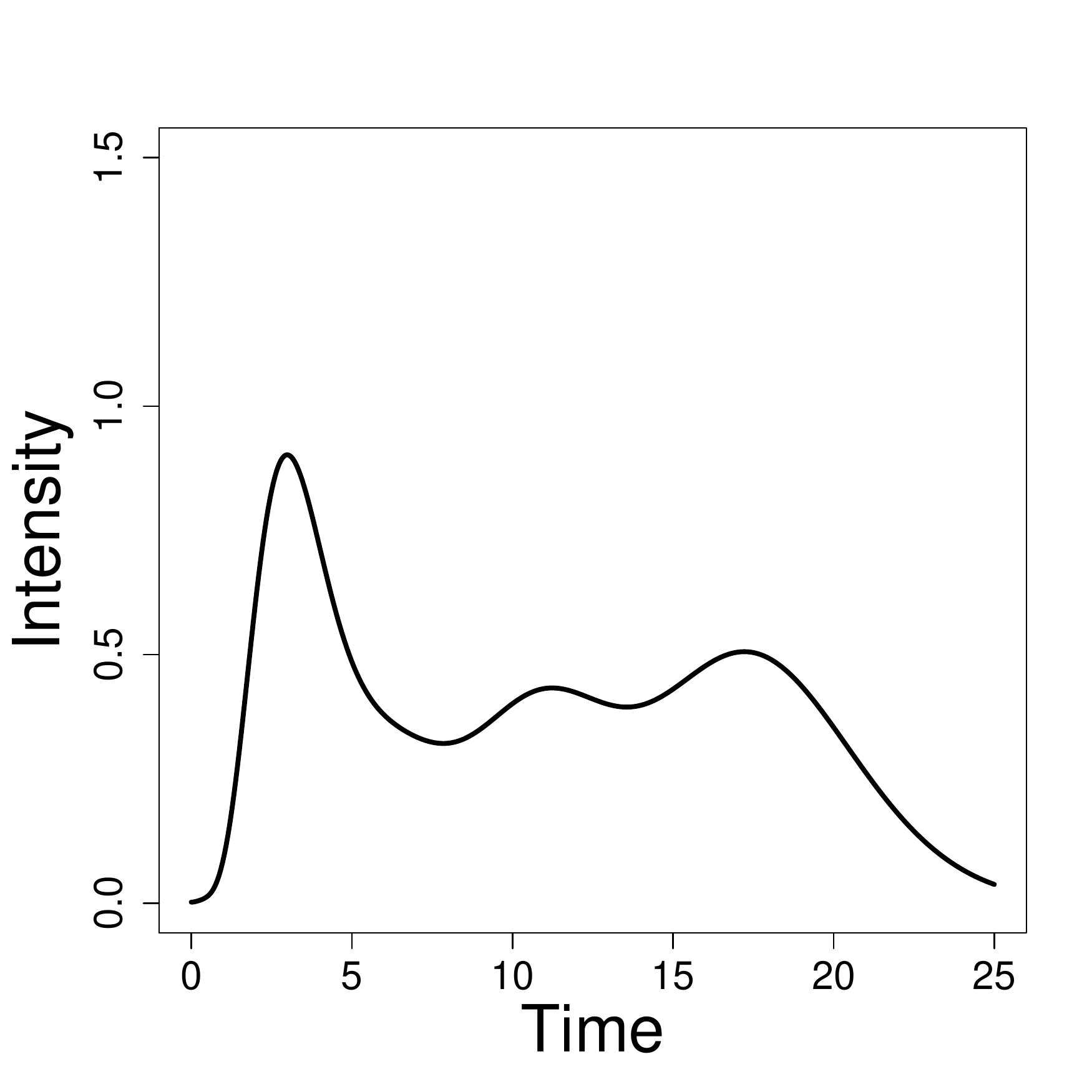}
\includegraphics[width=0.32\textwidth]{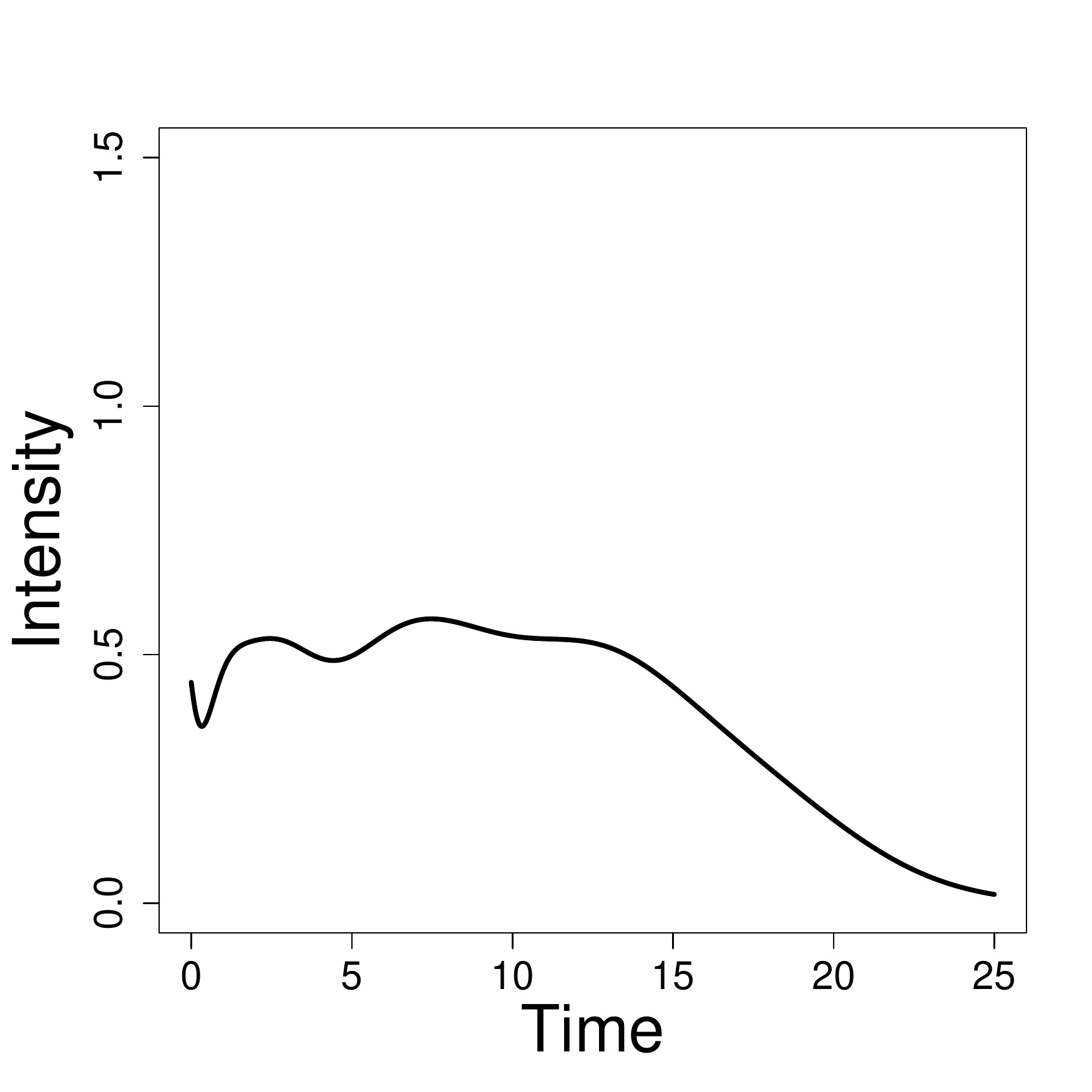}\\	  	  
\caption{Prior realizations for the mixture weights (top panels) and the
corresponding intensity function (bottom panels) for three different values of the
gamma process precision parameter, $c_0 = 0.05, 1, 10$. 
In all cases, $J = 50$, $\theta = 0.4$, and $H_0(t) = t/2$.}
\label{fig:prior_weights_c0}
\end{figure}

Regarding the role of the different model parameters, we reiterate that (\ref{mixture_model}) 
corresponds to a structured mixture. The Erlang densities,
$\text{ga}(t \mid j,\theta^{-1})$, play the role of basis functions in the representation for 
the intensity. In this respect, of primary importance is the flexibility of the nonparametric
prior for the cumulative intensity function $H$ that defines the mixture weights.
In particular, the gamma process prior provides realizations 
for $H$ with general shapes that can concentrate on different time intervals, thus favoring 
different subsets of the Erlang basis densities through the corresponding $\omega_{j}$.
Here, the key parameter is the precision parameter $c_{0}$, which controls the
variability of the gamma process prior around $H_{0}$, and thus the
effective mixture weights. As an illustration, Figure \ref{fig:prior_weights_c0}
shows prior realizations for the weights $\omega_{j}$ (and the resulting intensity function) 
for different values of $c_{0}$, keeping all other model parameters the same. 
Note that as $c_{0}$ decreases, so does the number of practically non-zero weights.

The prior mean for $H$ is taken to be $H_{0}(t)=$ $t/b$, 
i.e., the cumulative intensity (hazard) of an exponential distribution
with scale parameter $b > 0$. Although it is 
possible to use more general centering functions, such as the Weibull $H_{0}(t)=$ $(t/b)^{a}$, 
the exponential form is sufficiently flexible in practice, as demonstrated with the 
synthetic data examples of Section \ref{sec:egs}. Based on the role of
$H$ in the intensity mixture model, we typically anticipate
realizations for $H$ that are different from the centering function $H_{0}$, 
and thus, as discussed above, the more important gamma process parameter is $c_{0}$.
Moreover, the exponential form for $H_{0}$ allows for an analytical
result for the prior expectation of the Erlang mixture intensity model.
Under $H_{0}(t)=$ $t/b$, the prior expectation for the weights is
given by $\text{E}(\omega_{j} \mid \theta,b)=$ $\theta/b$. Therefore,
conditional on all model hyperparameters, the expectation of
$\lambda(t)$ over the gamma process prior can be written as
\[
\text{E}(\lambda(t) \mid b,\theta) = 
\frac{\theta}{b} \sum\limits_{j=1}^{J} \text{ga}(t \mid j,\theta^{-1})
= \frac{ \exp( - (t/\theta) )}{b} 
\sum\limits_{m=0}^{J-1} \frac{ (t/\theta)^{m} }{m!}, \enspace t \in \mathbb{R}^{+} ,
\]
which converges to $b^{-1}$, as $J \rightarrow \infty$, for any $t \in \mathbb{R}^{+}$
(and regardless of the value of $\theta$ and $c_{0}$). In practice,
the prior mean for the intensity function 
is essentially constant at $b^{-1}$ for $t \in (0,J \theta)$, which, as
discussed below, is roughly the effective support of the NHPP intensity. 
This result is useful for prior specification as it distinguishes the role
of $b$ from that of parameters $\theta$ and $c_{0}$.

\begin{figure}[t!]
\centering
\includegraphics[width=\textwidth]{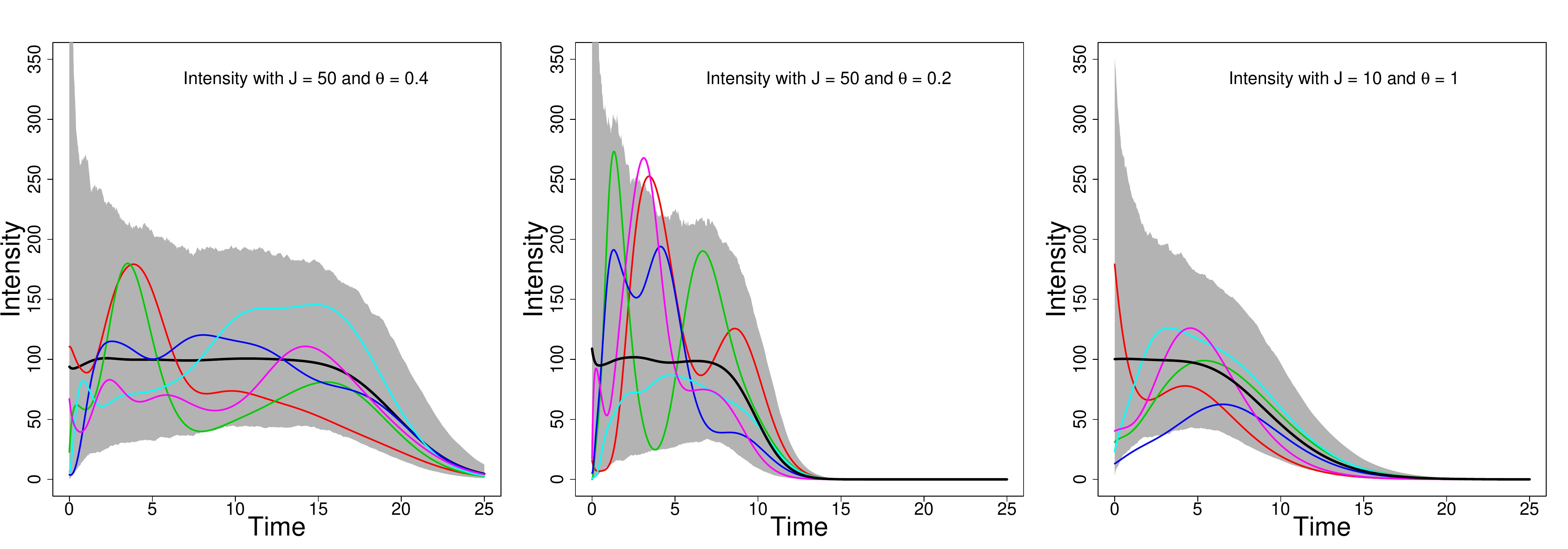}
\caption{Prior mean (black line), prior $95\%$ interval bands (shaded area),
and five individual prior realizations for the intensity under the Erlang mixture model in 
(\ref{mixture_model}) with $(\theta,J)=$ $(0.4,50)$ (left panel), $(\theta,J)=$ $(0.2,50)$ 
(middle panel), and $(\theta,J)=$ $(1,10)$ (right panel). In all cases, the gamma process 
prior is specified with $c_{0}=0.01$ and $H_{0}(t)=t/0.01$.}
\label{fig:inten_Jtheta}
\end{figure}

Also key are the two remaining model parameters, the number of Erlang basis densities $J$, 
and their common scale parameter $\theta$. Parameters $\theta$ and $J$ interact to control 
both the effective support and shape of NHPP intensities arising under (\ref{mixture_model}). 
Regarding intensity shapes, as the lemma suggests, smaller values of $\theta$ and larger 
values of $J$ generally result in more variable, typically multimodal intensities. Moreover, 
the representation for $\lambda(t)$ in (\ref{mixture_model}) utilizes Erlang basis densities 
with increasing means $j \theta$, and thus $(0, J \theta)$ can be used as a proxy for the 
effective support of the NHPP intensity. Of course, the mean underestimates the effective 
support, a more accurate guess can be obtained using, say, the 95\% percentile of the last 
Erlang density component. For an illustration, Figure \ref{fig:inten_Jtheta} plots five 
prior intensity realizations under three combinations of $(\theta,J)$ values, with 
$c_{0}=0.01$ and $b=0.01$ in all cases. 
Also plotted are the prior mean and $95\%$ interval bands for the intensity, based on 1000 realizations from the prior model.
The left panel corresponds to the largest value for $J \theta$ and, consequently, to the widest
effective support interval. The value of $J \theta$ is the same for the middle and right panels, 
resulting in similar effective support. However, the intensities in the middle panel show 
larger variability in their shapes, as expected since the value of $J$ is increased and the 
value of $\theta$ decreased relative to the ones in the right panel.

\subsection{Prior specification}
\label{subsec:priors}

To complete the full Bayesian model, we place prior distributions on the 
parameters $c_{0}$ and $b$ of the gamma process prior for $H$, and on the 
scale parameter $\theta$ of the Erlang basis densities. 
%Moreover, $\theta$ is assigned an inverse gamma prior, $\text{IG}(a_{\theta},b_{\theta})$, 
%with mean $b_{\theta}/(a_{\theta}-1)$ (provided $a_{\theta} > 1$). 
%In practice, the parametric form of the prior distributions for $c_{0}$, $b$ and 
%$\theta$ is less important than the choice of their hyperparameters. 
%
A generic approach to specify these hyperpriors can be obtained using the 
observation time window $(0,T)$ as the effective support of the NHPP intensity.

We work with exponential prior distributions for parameters 
$c_{0}$ and $b$.
Using the prior mean for the intensity function, which as discussed in 
Section \ref{subsec:model} is roughly constant at $b^{-1}$ within the
time interval of interest, the total intensity in $(0,T)$ can be
approximated by $T/b$. Therefore,  taking the size $n$ of the observed 
point pattern, as a proxy for the total intensity in $(0,T)$, we
can use $T/n$ to specify the mean of the exponential prior
distribution for $b$. 
Given its role in the gamma process prior, we anticipate that small
values of $c_{0}$ will be important to allow prior variability around 
$H_{0}$, as well as sparsity in the mixture weights. Experience from prior 
simulations, such as the ones shown in Figure \ref{fig:prior_weights_c0}, 
is useful to guide the range of ``small'' values. 
Note that the pattern observed in Figure \ref{fig:prior_weights_c0}
is not affected by the length of the observation window. In general, a 
value around $10$ can be viewed as a conservative guess at a high
percentile for $c_{0}$. For the data examples of Section \ref{sec:egs}, 
we assigned an exponential prior with mean $10$ to $c_{0}$, observing 
substantial learning for this key model hyperparameter with its posterior 
distribution supported by values (much) smaller than $1$.

Also given the key role of parameter $\theta$ in controlling the intensity shapes, we 
recommend favoring sufficiently small values in the prior for $\theta$, especially if prior 
information suggests a non-standard intensity shape. Recall that
$\theta$, along with $J$, control the effective support of the
intensity, and thus ``small'' values for $\theta$ should be assessed
relative to the length of the observation window. Again, prior simulation, 
as in Figure \ref{fig:inten_Jtheta}, is a useful tool.
A practical approach to specify the prior range of $\theta$ values 
involves reducing the Erlang mixture model to
the first component. The corresponding (exponential) density has mean
$\theta$, and we thus use $(0,T)$ as the effective prior range for $\theta$. 
Because $T$ is a fairly large upper bound, and since we wish to favor smaller
$\theta$ values, rather than an exponential prior, we use a Lomax
prior, $p(\theta) \propto$ $(1 + d_{\theta}^{-1} \theta)^{-3}$, with
shape parameter equal to $2$ (thus implying infinite variance), and 
median $d_{\theta} (\sqrt{2} - 1)$. The value of the scale parameter, $d_{\theta}$, 
is specified such that $\text{Pr}(0 < \theta < T) \approx 0.999$. This simple 
strategy is effective in practice in identifying a plausible range of $\theta$ values. 
For the synthetic data examples of Section \ref{sec:egs}, for which $T=20$,
we assigned a Lomax prior with scale parameter $d_{\theta}=1$ to $\theta$, 
obtaining overall moderate prior-to-posterior learning for $\theta$.

Finally, we work with fixed $J$, the value of which can be specified exploiting the 
role of $\theta$ and $J$ in controlling the support of the NHPP intensity. In particular, 
$J$ can be set equal to the integer part of $T/\theta^{*}$, where
$\theta^{*}$ is the prior median for $\theta$. More conservatively, 
this value can be used as a lower bound for values of $J$ to be
studied in a sensitivity analysis, especially for applications where
one expects non-standard shapes for the intensity function. 
In practice, we recommend conducting prior sensitivity analysis for all model 
parameters, as well as plotting prior realizations and prior uncertainty bands for the 
intensity function to graphically explore the implications of different prior choices. 
%Results from both of these strategies are provided for the data examples 
%of Section \ref{sec:egs}.
%

The number of Erlang basis densities is the only model parameter which is not assigned 
a hyperprior. Placing a prior on $J$ complicates significantly the posterior simulation 
method, as it necessitates use of variable-dimension MCMC techniques, while offering
relatively little from a practical point of view. The key observation is again that the 
Erlang densities play the role of basis functions rather than of kernel densities in 
traditional (less structured) finite mixture models. 
Also key is the nonparametric nature of the prior for function $H$ that defines the mixture 
weights which {\it select} the Erlang densities to be used in the representation of the intensity. 
This model feature effectively guards against over-fitting if one conservatively
chooses a larger value for $J$ than may be necessary. In this respect, the flexibility afforded 
by random parameters $c_{0}$ and $\theta$ is particularly useful. Overall, we have found that 
fixing $J$ strikes a good balance between computational tractability and model flexibility in
terms of the resulting inferences.

\subsection{Posterior simulation}
\label{subsec:MCMC}

Denote as before by $\{ 0 < t_{1} < ... < t_{n} < T \}$ the point pattern observed in time 
window $(0,T)$. Under the Erlang mixture model of Section \ref{subsec:model}, 
the NHPP likelihood is proportional to
{\small 
\begin{eqnarray}
\exp\left( - \int_{0}^{T} \lambda(u) \, \text{d}u \right) \prod_{i=1}^{n} \lambda(t_{i}) & = &
\exp\left( - \sum\nolimits_{j=1}^{J} \omega_{j} K_{j,\theta}(T) \right) 
\prod_{i=1}^{n} \left\{ \sum\limits_{j=1}^{J} \omega_{j} \, \text{ga}(t_{i}
\mid j,\theta^{-1}) \right\} \notag \\
 & = & \prod_{j=1}^{J} \exp( - \omega_{j} K_{j,\theta}(T) ) 
\prod_{i=1}^{n} \left\{ \left( \sum\nolimits_{r=1}^{J} \omega_{r} \right) 
\sum\limits_{j=1}^{J} 
\left( \frac{\omega_{j}}{ \sum\nolimits_{r=1}^{J} \omega_{r} } \right)
\, \text{ga}(t_{i} \mid j,\theta^{-1}) \right\} , \notag
\end{eqnarray}
}
where $K_{j,\theta}(T)=$ $\int_{0}^{T} \text{ga}(u \mid j,\theta^{-1})
\, \text{d}u$ is the $j$-th Erlang distribution function at $T$.

For the posterior simulation approach, we augment the likelihood with
auxiliary variables $\bm{\gamma}=$ $\{\gamma_{i} : i=1,\ldots,n\}$, 
where $\gamma_{i}$ identifies the Erlang basis density to which time 
event $t_{i}$ is assigned. Then, the augmented, hierarchical model for the data
can be expressed as follows:
\begin{eqnarray}
\{ t_{1},...,t_{n}\} \mid \bm{\gamma},\bm{\omega},\theta & \sim &
\prod_{j=1}^{J} \exp( - \omega_{j} K_{j,\theta}(T) ) 
\prod_{i=1}^{n} \left\{ \left( \sum\nolimits_{r=1}^{J} \omega_{r} \right) 
\text{ga}(t_{i} \mid \gamma_{i},\theta^{-1}) \right\}  \notag \\
\gamma_{i} \mid \bm{\omega} & \stackrel{i.i.d.}{\sim} & 
\sum_{j=1}^{J}  \left( \frac{\omega_{j}}{ \sum\nolimits_{r=1}^{J} \omega_{r} } \right)
\delta_{j}(\gamma_{i}),  \quad i=1,...,n   \notag \\
\theta, c_{0}, b, \bm{\omega} & \sim & p(\theta) \, p(c_{0}) \, p(b) \,
\prod_{j=1}^{J} \text{ga}(\omega_{j} \mid c_{0} \, \omega_{0j}(\theta),c_{0} ) ,
\end{eqnarray}
where $\bm{\omega}=$ $\{ \omega_j: j=1,...,J \}$, and 
$p(\theta)$, $p(c_{0})$, and $p(b)$ denote the priors for $\theta$, $c_{0}$, and $b$.
Recall that, under the exponential distribution form for $H_{0}=t/b$,
we have $\omega_{0j}(\theta)=$ $\theta/b$.

We utilize Gibbs sampling to explore the posterior distribution. The sampler involves 
ready updates for the auxiliary variables $\gamma_{i}$, and,
importantly, also for the mixture weights $\omega_{j}$. More specifically, the posterior full 
conditional for each $\gamma_{i}$ is a discrete distribution on $\{ 1,..., J \}$ such that 
$\text{Pr}(\gamma_{i} = j \mid \theta,\bm{\omega},\text{data}) \propto$
$\omega_{j} \, \text{ga}(t_{i} \mid j,\theta^{-1})$, for $j=1,...,J$.

Denote by $N_{j}=$ $| \{ t_{i}: \gamma_{i} = j \} |$, for $j=1,...,J$,
that is, $N_{j}$ is the number of time points assigned to the $j$-th
Erlang basis density. The posterior full conditional distribution for $\omega$
is derived as follows:
\begin{eqnarray}
p(\bm{\omega} \mid \theta,c_{0},b,\bm{\gamma},\text{data}) & \propto &
\left\{ \prod_{j=1}^{J} \exp( - \omega_{j} K_{j,\theta}(T) ) \right\}
\left( \sum\nolimits_{r=1}^{J} \omega_{r} \right)^{n}  \notag \\
 & \times & \left\{ \prod_{j=1}^{J} \omega_{j}^{N_{j}} 
\left( \sum\nolimits_{r=1}^{J} \omega_{r} \right)^{-N_{j}} \right\}
\left\{ \prod_{j=1}^{J} \text{ga}(\omega_{j} \mid c_{0} \,
\omega_{0j}(\theta),c_{0} ) \right\}  \notag \\
 & \propto & \prod_{j=1}^{J} \exp( - \omega_{j} K_{j,\theta}(T) ) \, \omega_{j}^{N_{j}}
\, \text{ga}(\omega_{j} \mid c_{0} \, \omega_{0j}(\theta),c_{0} ) \notag \\
 & = & \prod_{j=1}^{J} \text{ga}(\omega_{j} \mid N_{j} + c_{0} \, \omega_{0j}(\theta),
K_{j,\theta}(T) + c_{0} ) , \notag
\end{eqnarray}
where we have used the fact that $\sum_{j=1}^{J} N_{j} = n$. 
Therefore, given the other parameters and the data, the mixture weights are 
independent, and each $\omega_{j}$ follows a gamma posterior full conditional 
distribution. This is a practically important feature of the model
in terms of convenient updates for the mixture weights, and with respect to 
efficiency of the posterior simulation algorithm as it pertains to this key 
component of the model parameter vector.

Finally, each of the remaining parameters, $c_{0}$, $b$, and $\theta$, 
is updated with a Metropolis-Hastings (M-H) step, using a log-normal proposal 
distribution in each case.
%parameter $\theta$ and the hyperparameters, $c_{0}$ and $b$, of the
%gamma process prior for $H$ are updated with Metropolis-Hastings (M-H) steps. 
%A log-normal proposal distribution is employed for the M-H step
%used to update each of these parameters.

%
%---------------------------------------------------------------------
%

\subsection{Model extensions to incorporate marks}

Here, we discuss how the Erlang mixture prior
for NHPP intensities can be embedded in semiparametric models for 
point patterns that include additional information on marks.
%or subject-specific covariates.}

Consider the setting where, associated with each
observed time event $t_{i}$, marks $\bm{y}_{i} \equiv$ $\bm{y}_{t_{i}}$
are recorded (marks are only observed when an event is observed). Without 
loss of generality, we assume that marks are continuous variables taking 
values in mark space ${\cal M} \subseteq \mathbb{R}^{d}$, for $d \geq 1$.
%(Marks are only observed when an event is observed, but we use here the 
%simpler notation $\bm{y}$ and $\bm{y}_{i}$ instead of the more accurate 
%$\bm{y}_{t}$ and $\bm{y}_{t_{i}}$.) 
As discussed in \cite{Taddy_Kottas:2012}, a nonparametric prior for the 
intensity of the temporal process, ${\cal T}$, can be combined with a mark 
distribution to construct a semiparametric model for marked NHPPs. In particular, 
consider a generic marked NHPP 
$\{ (t, \bm{y}_{t}): t \in {\cal T}, \bm{y}_{t} \in {\cal M} \}$, that is: 
the temporal process ${\cal T}$ is a NHPP on $\mathbb{R}^{+}$ with intensity
function $\lambda$; and, conditional on ${\cal T}$, the marks 
$\{ \bm{y}_{t}: t \in {\cal T} \}$ are mutually independent. Now, assume 
that, conditional on ${\cal T}$, the marks have density $m_{t}$ that depends
only on $t$ (i.e., it does not depend on any earlier time $t^{\prime} < t$).
Then, by the ``marking" theorem (e.g., \citeh{Kingman:1993}), we have that 
the marked NHPP is a NHPP on the extended space $\mathbb{R}^{+} \times {\cal M}$ 
with intensity $\lambda^{*}(t,\bm{y}_{t}) =$ $\lambda(t) \, m_{t}(\bm{y}_{t})$.
Therefore, the likelihood for the observed marked point pattern 
$\{ (t_{i},\bm{y}_{i}): i = 1,...,n \}$ can be written as 
$\exp\left( - \int_{0}^{T} \lambda(u) \, \text{d}u \right) \prod_{i=1}^{n} \lambda(t_{i})
\prod_{i=1}^{n} m_{t_{i}}(\bm{y}_{i})$ (the integral 
$\int_{0}^{T} \int_{{\cal M}} \lambda^{*}(u,\bm{z}) \, \text{d}u \text{d}\bm{z}$
in the normalizing term reduces to 
$\int_{0}^{T} \lambda(u) \, \text{d}u$, since $m_{t}$ is a density).
Hence, the MCMC method of Section \ref{subsec:MCMC} can be extended for marked 
NHPP models built from the Erlang mixture prior for intensity $\lambda$, and 
any time-dependent model for the mark density $m_{t}$.

\section{Data examples}
\label{sec:egs}

To empirically investigate inference under the proposed model,
we present three synthetic data examples corresponding 
to decreasing, increasing, and bimodal intensities. We also 
consider the coal-mining disasters data set, which is commonly used 
to illustrate NHPP intensity estimation. 

%
%Convergence and mixing of the MCMC algorithm 
%was assessed graphically through trace plots of the intensity function
%evaluated at specific time points within the observation window. 
%An example is given in Figure \ref{fig:dec_tra} for the synthetic data
%of Section \ref{subsec:dec}. The trace plots in Figure \ref{fig:dec_tra}
%are representative of the mixing observed in all other data examples. 
%Regarding model parameters, the highest autocorrelation was 
%observed in posterior samples for parameter $\theta$. 
%
%\begin{figure}[t!]
%\centering
%\includegraphics[width=12.5cm,height=9.5cm]{./figs/dec_traces.pdf}
%\caption{Synthetic data from temporal NHPP with decreasing intensity.
%Trace plots of posterior samples for the intensity function 
%evaluated at time points $t=5$, $10$, $15$, $20$.}
%	  \label{fig:dec_tra}
%\end{figure}
%

We used the approach of Section \ref{subsec:priors} to specify the priors 
for $c_{0}$, $b$ and $\theta$, and the value for $J$. In particular, we used 
the exponential prior for $c_{0}$ with mean $10$ for all data examples. 
For the three synthetic data sets (for which $T=20$), we used the Lomax prior 
for $\theta$ with shape parameter equal to $2$ and scale parameter equal to $1$.
Prior sensitivity analysis results for the synthetic data example 
of Section \ref{subsec:bim} are provided in the Supplementary Material. Overall, 
results from prior sensitivity analysis (also conducted for all other data examples) 
suggest that the prior specification approach of Section \ref{subsec:priors} is effective 
as a general strategy. Moreover, more dispersed priors for parameters $c_{0}$, $b$ 
and $\theta$ have little to no effect on the posterior distribution for these 
parameters and essentially no effect on posterior estimates for the NHPP 
intensity function, even for point patterns with relatively small size, such 
as the one ($n = 112$) for the data example of Section \ref{subsec:bim}.

The Supplement provides also computational details about the MCMC 
posterior simulation algorithm, including study of the effect of the number of basis 
densities ($J$) and the size of the point pattern ($n$) on effective sample size and 
computing time.

\subsection{Decreasing intensity synthetic point pattern}
\label{subsec:dec}

The first synthetic data set involves $491$ time points generated in
time window $(0,20)$ from a NHPP with intensity function 
$\beta^{-1} \alpha (\beta^{-1} t)^{\alpha - 1}$, where $(\alpha,\beta)=$ 
$(0.5, 8\times 10^{-5})$. This form corresponds to the hazard function 
of a Weibull distribution with shape parameter less than $1$, thus 
resulting in a decreasing intensity function.

\begin{figure}[t!]
\centering
\includegraphics[width=0.85\textwidth]{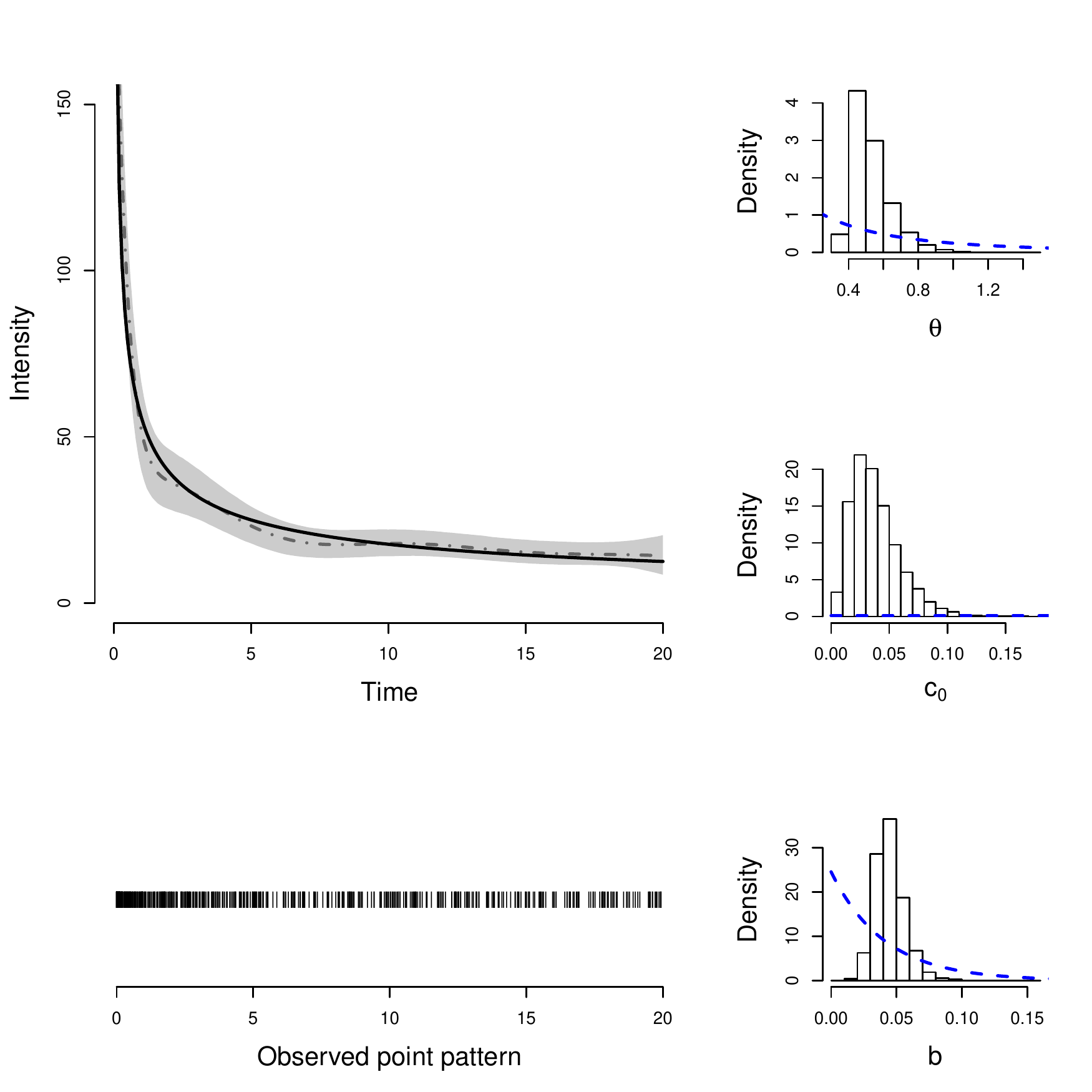}\\
\caption{Synthetic data from temporal NHPP with decreasing intensity.
The top left panel shows the posterior mean estimate (dashed-dotted line) 
and posterior 95\% interval bands (shaded area) for the intensity function. 
The true intensity is denoted by the solid line. The point
pattern is plotted in the bottom left panel. The three plots on the
right panels display histograms of the posterior samples for the model
hyperparameters, along with the corresponding prior densities (dashed lines).}
	  \label{fig:dec_est}
\end{figure}

The Erlang mixture model was applied with $J=50$, and an exponential prior for 
$b$ with mean $0.04$. The model captures the decreasing pattern of the data 
generating intensity function; see Figure \ref{fig:dec_est}. We note that there 
is significant prior-to-posterior learning in the intensity function estimation; 
the prior intensity mean is roughly constant at value about $25$ with prior uncertainty 
bands that cover almost the entire top left panel in Figure \ref{fig:dec_est}. 
Prior uncertainty bands were similarly wide for all other data examples.

\subsection{Increasing intensity synthetic point pattern}
\label{subsec:inc}

\begin{figure}[t!]
\centering
\includegraphics[width=0.85\textwidth]{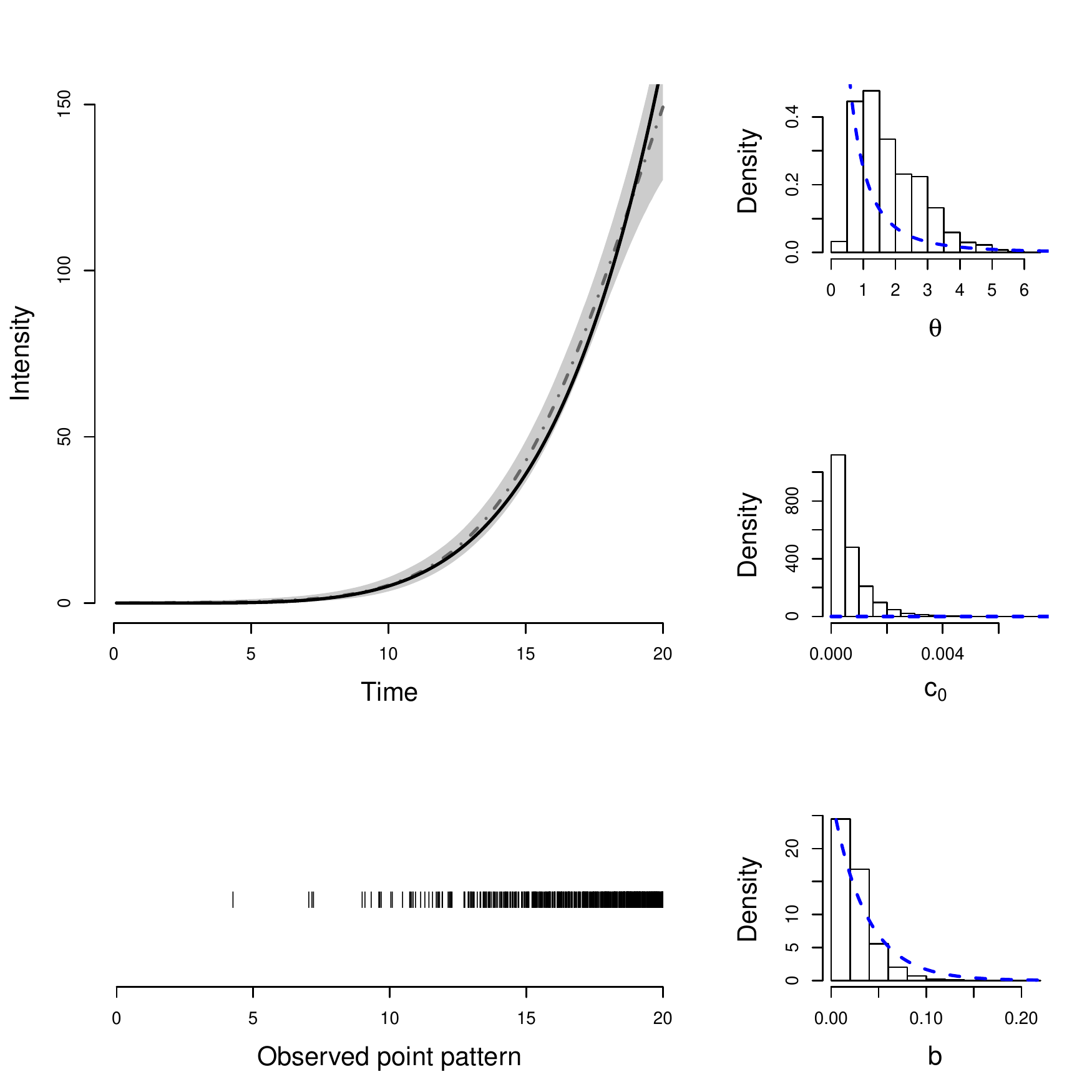}\\
\caption{Synthetic data from temporal NHPP with increasing intensity.
The top left panel shows the posterior mean estimate (dashed-dotted line) and 
posterior 95\% interval bands (shaded area) for the intensity function. The
true intensity is denoted by the solid line. The point
pattern is plotted in the bottom left panel. The three plots on the
right panels display histograms of the posterior samples for the model
hyperparameters, along with the corresponding prior densities (dashed lines).}
	  \label{fig:inc_est}
\end{figure}

We consider again the form $\beta^{-1} \alpha (\beta^{-1} t)^{\alpha - 1}$
for the NHPP intensity function, but here with $(\alpha,\beta)=$ $(6,7)$
such that the intensity is increasing. A point pattern comprising
$565$ points was generated in time window $(0,20)$. The Erlang mixture model 
was applied with $J=50$, and an exponential prior for $b$ with mean $0.035$. 
Figure \ref{fig:inc_est} reports inference results.
This example demonstrates the model's capacity to effectively
recover increasing intensity shapes over the bounded observation window, 
even though the Erlang basis densities are ultimately decreasing.

\subsection{Bimodal intensity synthetic point pattern}
\label{subsec:bim}

\begin{figure}[t!]
\centering
\includegraphics[width=0.33\textwidth]{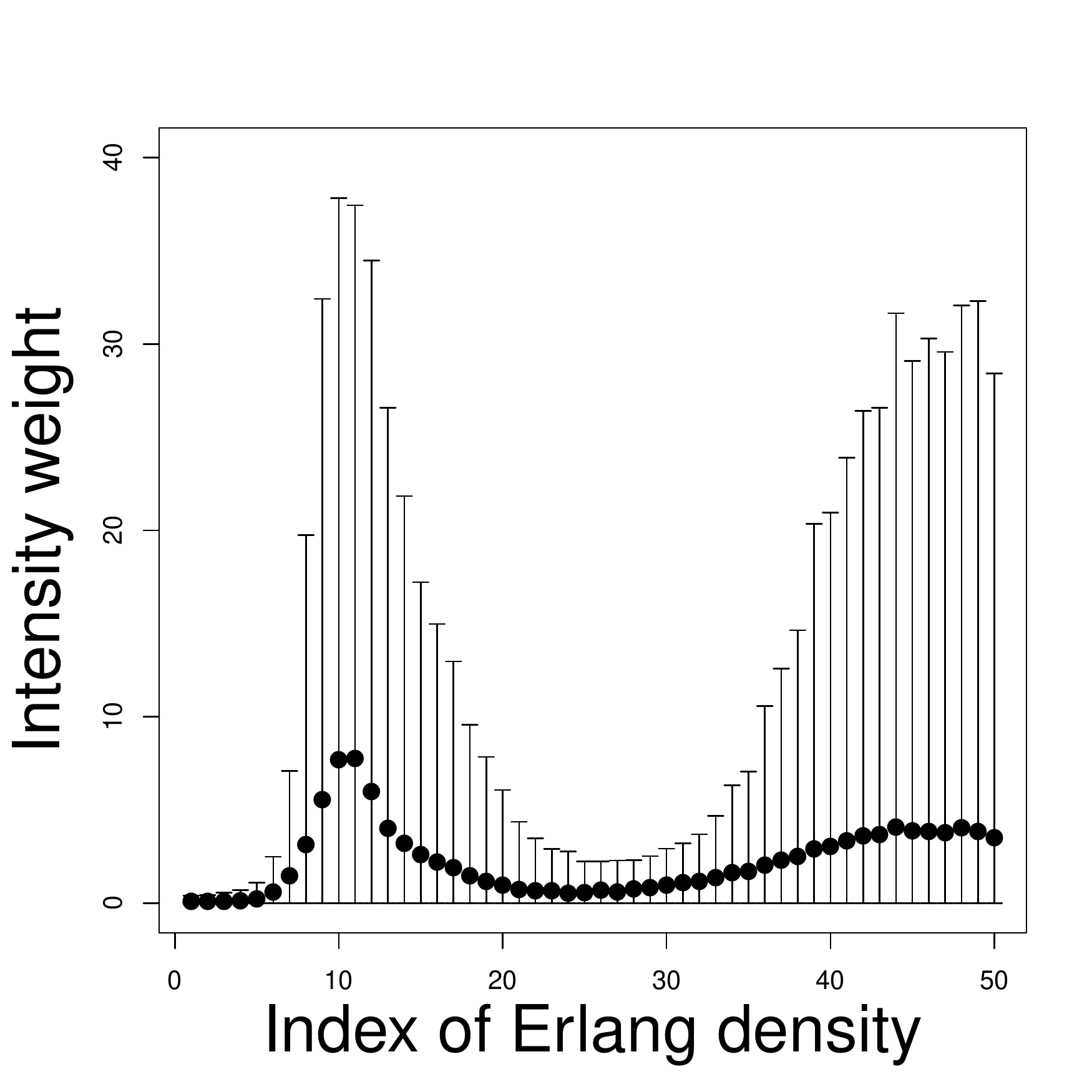}\includegraphics[width=0.33\textwidth]{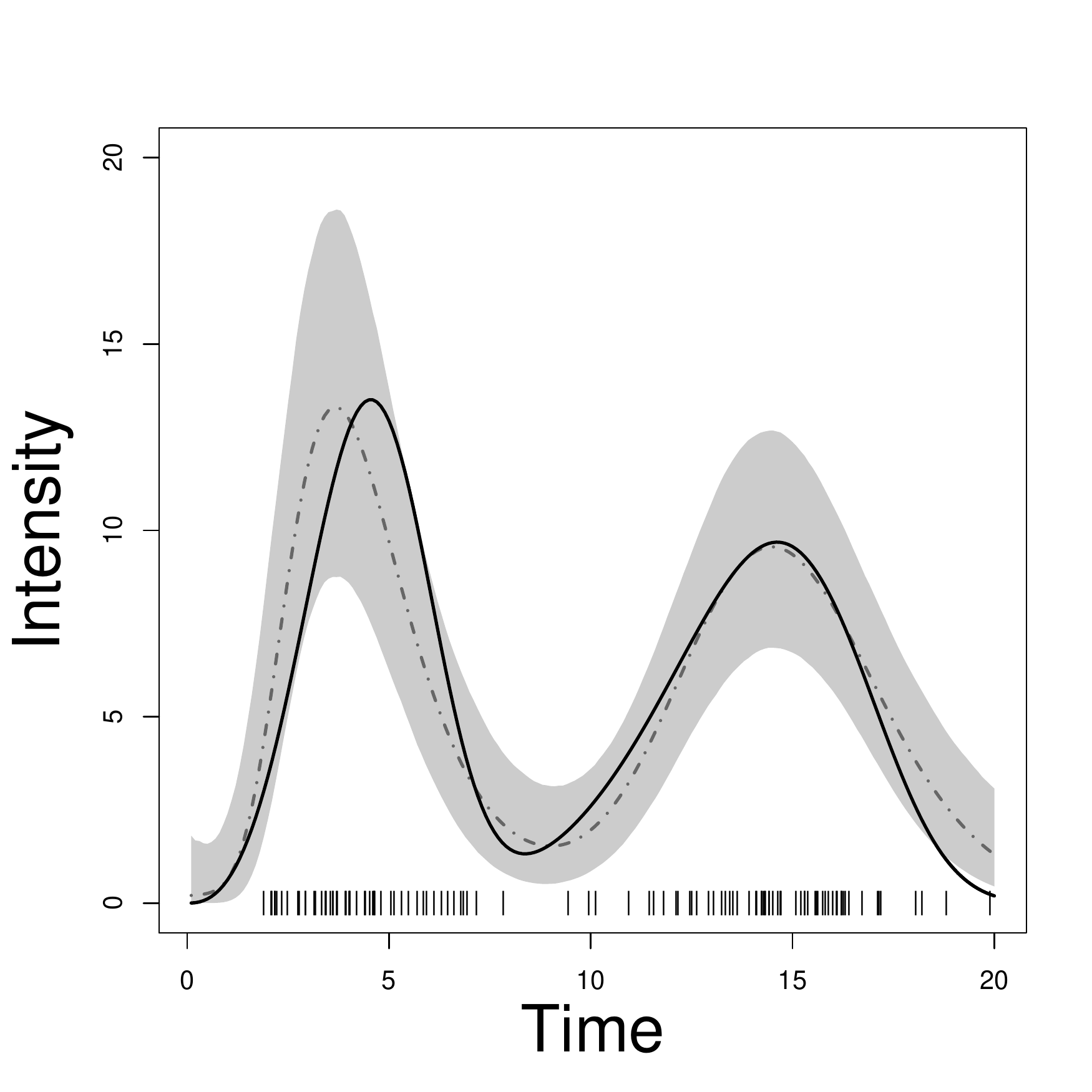}\includegraphics[width=0.33\textwidth]{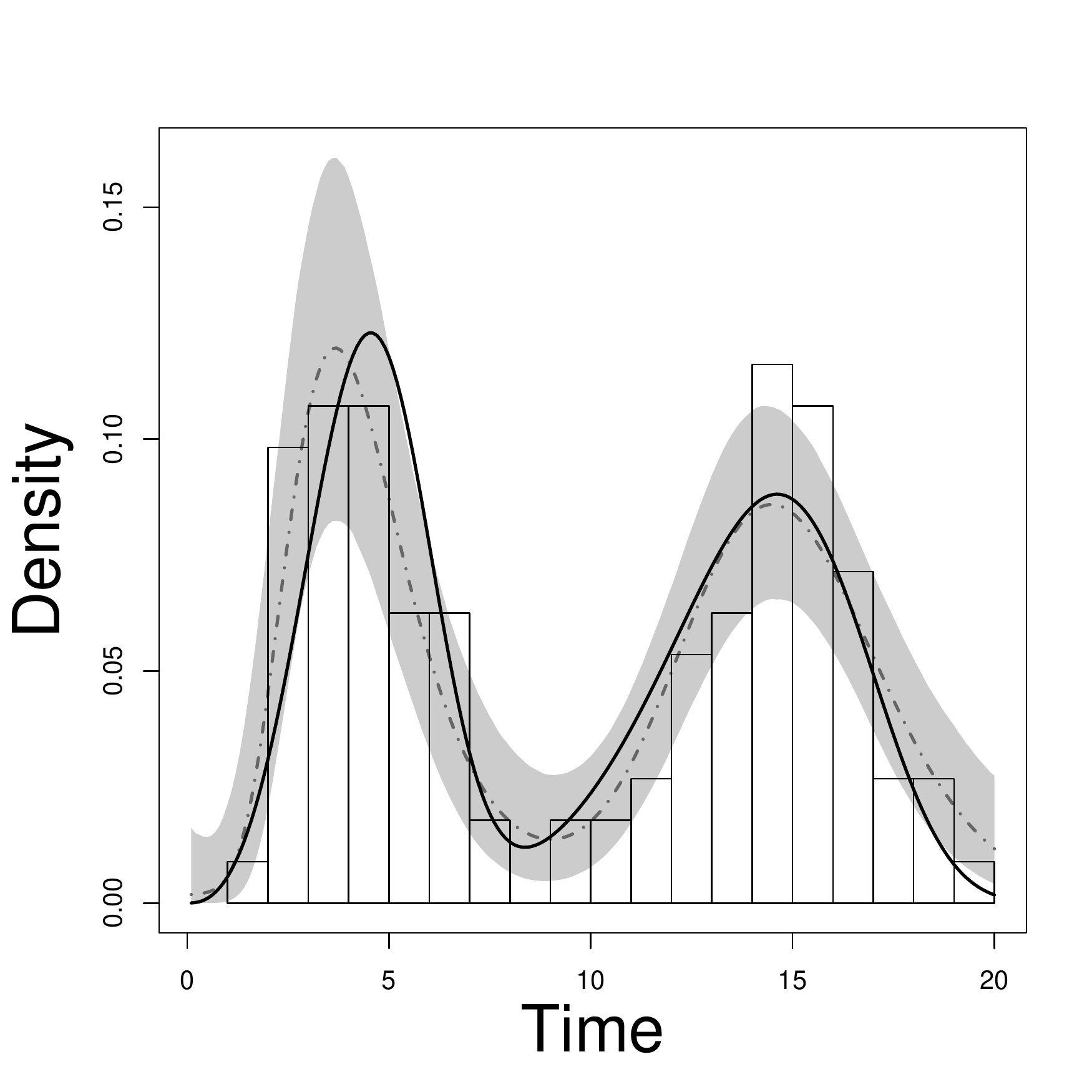}\\
\includegraphics[width=0.33\textwidth]{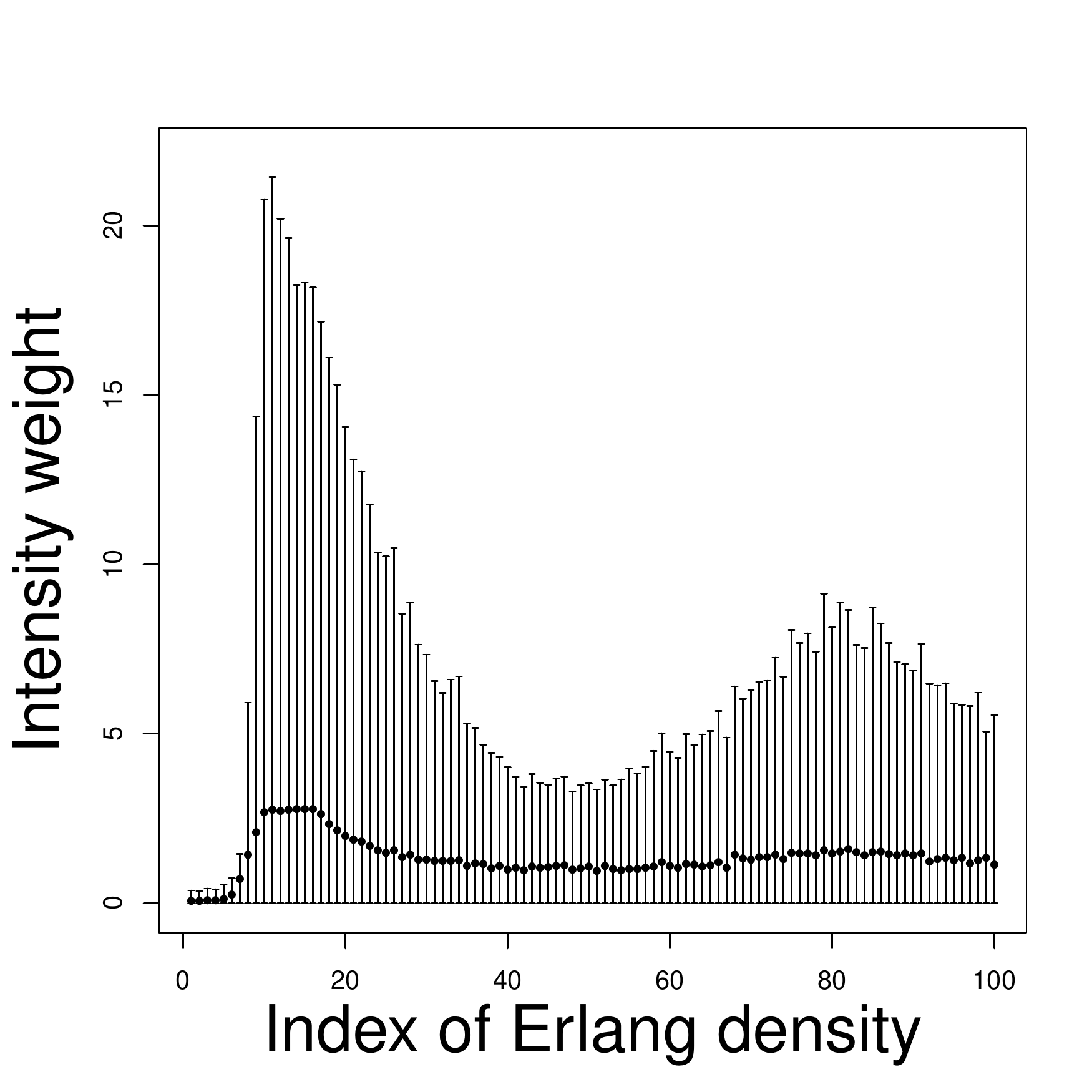}\includegraphics[width=0.33\textwidth]{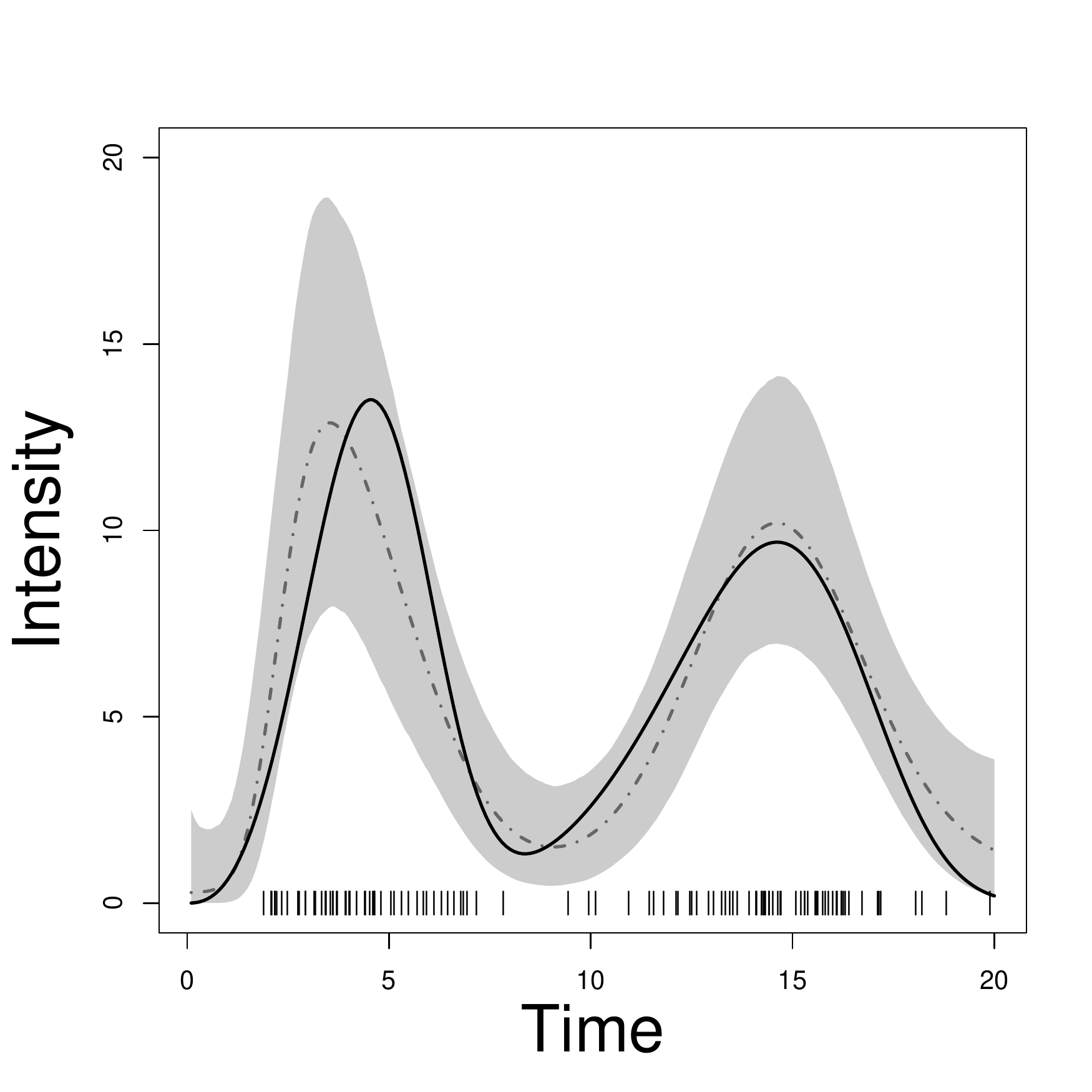}\includegraphics[width=0.33\textwidth]{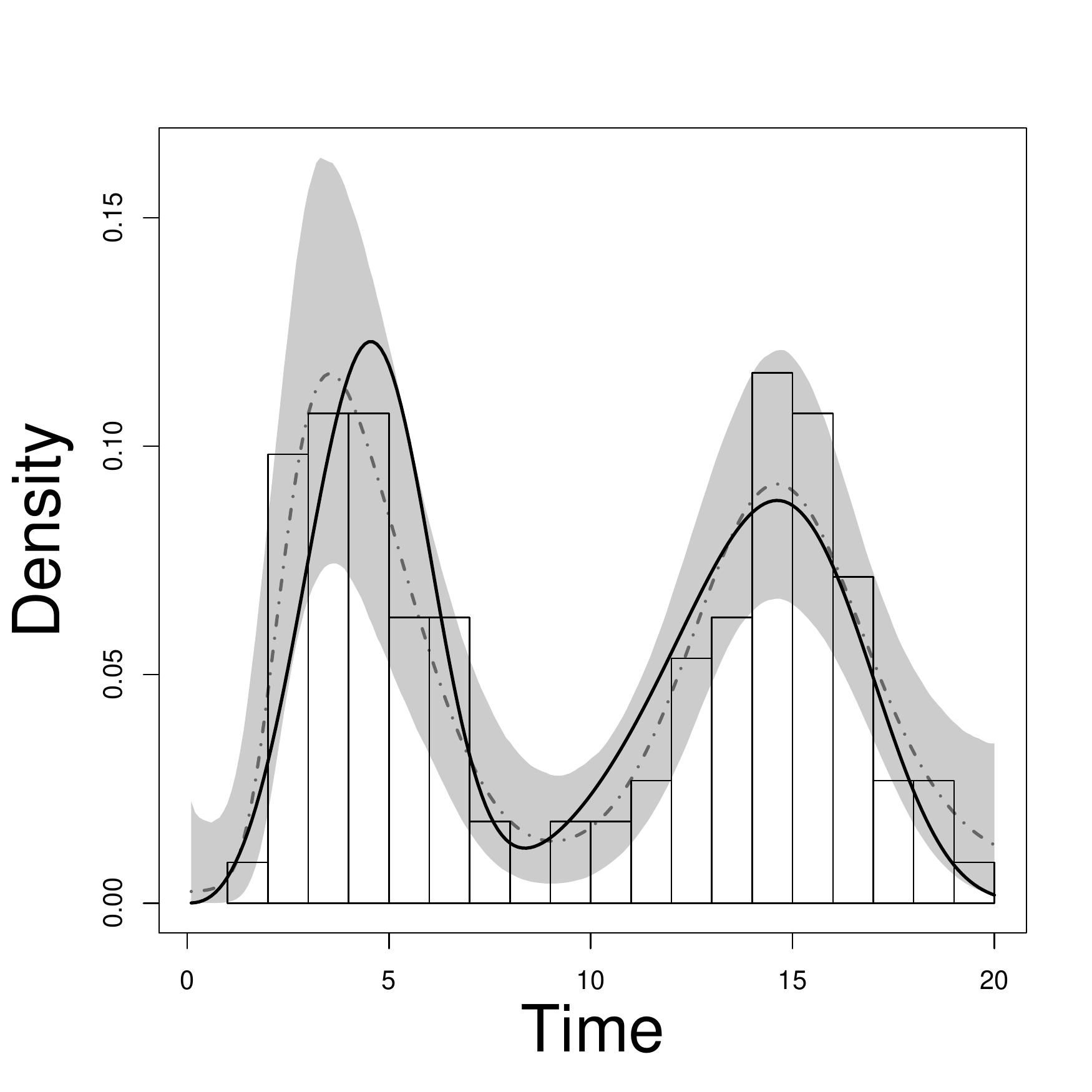}\\
\caption{Synthetic data from temporal NHPP with bimodal intensity. 
Inference results are reported under $J=50$ (top row) and $J=100$ (bottom row).
The left column plots the posterior means (circles) and 90\% interval estimates (bars) of 
the weights for the Erlang basis densities. The middle column displays the posterior mean 
estimate (dashed-dotted line) and posterior 95\% interval bands (shaded area) for the 
NHPP intensity function. The true intensity is denoted by the solid line. The bars on the 
horizontal axis indicate the point pattern. The right column plots the posterior mean 
estimate (dashed-dotted line) and posterior 95\% interval bands (shaded area) for the 
NHPP density function on the observation window. The histogram corresponds to the 
simulated times that comprise the point pattern.}
\label{fig:bim_est}
\end{figure}

The data examples in Sections \ref{subsec:dec} and \ref{subsec:inc}
illustrate the model's capacity to uncover monotonic intensity shapes,  
associated with a parametric distribution different from the Erlang
distribution that forms the basis of the mixture intensity model. 
Here, we consider a point pattern generated from a NHPP with 
a more complex intensity function, $\lambda(t)=$ $50 \, \text{We}(t \mid 3.5,5)$ +
$60 \, \text{We}(t \mid 6.5,15)$, where $\text{We}(t \mid \alpha,\beta)$ 
denotes the Weibull density with shape parameter $\alpha$ and mean 
$\beta \, \Gamma(1 + 1/\alpha)$. This specification results in a bimodal
intensity within the observation window $(0,20)$ where a synthetic point 
pattern of $112$ time points is generated; see Figure \ref{fig:bim_est}.

%\begin{figure}[t!]
%\centering
%\includegraphics[width=0.335\textwidth]{./figs/bim_smn_weight_snsj50.pdf}
%\includegraphics[width=0.335\textwidth]{./figs/bim_smn_weight_snsj100.pdf}\\
%\includegraphics[width=0.335\textwidth]{./figs/bim_smn_insty_snsj50.pdf}
%\includegraphics[width=0.335\textwidth]{./figs/bim_smn_insty_snsj100.pdf}\\
%\includegraphics[width=0.335\textwidth]{./figs/bim_smn_dnsty_snsj50.pdf}
%\includegraphics[width=0.335\textwidth]{./figs/bim_smn_dnsty_snsj100.pdf}\\
%\caption{Synthetic data from temporal NHPP with bimodal intensity. 
%Inference results are reported under $J=50$ (left column) and $J=100$ (right column).
%The top row plots the posterior means (circles) and 90\% interval estimates (bars) of 
%the weights for the Erlang basis densities. The middle row displays the posterior mean 
%estimate (dashed-dotted line) and posterior 95\% interval bands (shaded area) for the 
%NHPP intensity function. The true intensity is denoted by the solid line. The bars on the 
%horizontal axis indicate the point pattern. The bottom row plots the posterior mean 
%estimate (dashed-dotted line) and posterior 95\% interval bands (shaded area) for the 
%NHPP density function on the observation window. The histogram corresponds to the 
%simulated times that comprise the point pattern.}
%\label{fig:bim_est}
%\end{figure}

We used an exponential prior for $b$ with mean $0.179$.
Anticipating an underlying intensity with less standard shape than in 
the earlier examples, we compare inference results under $J=50$ and 
$J=100$; see Figure \ref{fig:bim_est}.
The posterior point and interval estimates capture effectively the bimodal 
intensity shape, especially if one takes into account the relatively small
size of the point pattern. (In particular, the histogram of the simulated random 
time points indicates that they do not provide an entirely accurate depiction 
of the underlying NHPP density shape.) The estimates are somewhat more accurate under 
$J=100$. The estimates for the mixture weights (left column of Figure \ref{fig:bim_est})
indicate the subsets of the Erlang basis densities that are utilized under
the two different values for $J$. The posterior mean of $\theta$ was $0.366$ under
$J=50$, and $0.258$ under $J=100$, that is, as expected, inference for $\theta$
adjusts to different values of $J$ such that $(0,J\theta)$ provides roughly the 
effective support of the intensity.

\subsection{Coal-mining disasters data}
\label{subsec:real}

Our real data example involves the ``coal-mining disasters'' data
(e.g., \citeh{Andrews_Herzberg:1985}, p. 53-56), a standard dataset
used in the literature to test NHPP intenstiy estimation methods. The
point pattern comprises the times (in days) of $n=191$ explosions of 
fire-damp or coal-dust in mines resulting in 10 or more casualties 
from the accident. The observation window consists of 40,550 days,  
from March 15, 1851 to March 22, 1962.

%To choose an appropriate $J$, we looked at the data distribution in
%which two or three modes were detected (see, the middle panel of
%Figure \ref{fig:real_est}); thus, 100 components model was selected
%with the corresponding inverse gamma prior with shape 21, scale 9000,
%and mean 450 for $\theta$. The effective support $(0,45000]$ is large
%enough to cover the interval $(0,T]=(0,40550]$ on which the intensity
%function of NHPP is defined. We also tried the models with $J=50$ and
%$J=150$; however, 50 components model had a limitation in representing
%the fluctuated data pattern. 150 components model was
%indistinguishable from the 100 components model in the standard, ``the
%time-rescaling theorem'' (e.g., \citeh{Daley_Vere-Jones:2003}), that
%we adopted for the model assessment. The sample size $n=191$ with the
%$T=40550$ suggested a gamma(2000,10) for $b$ as its prior
%distribution. we placed gamma(2,100) prior on $c_0$ as in the bimodal 
%intensity example.

\begin{figure}[t!]
\centering
\includegraphics[width=0.475\textwidth]{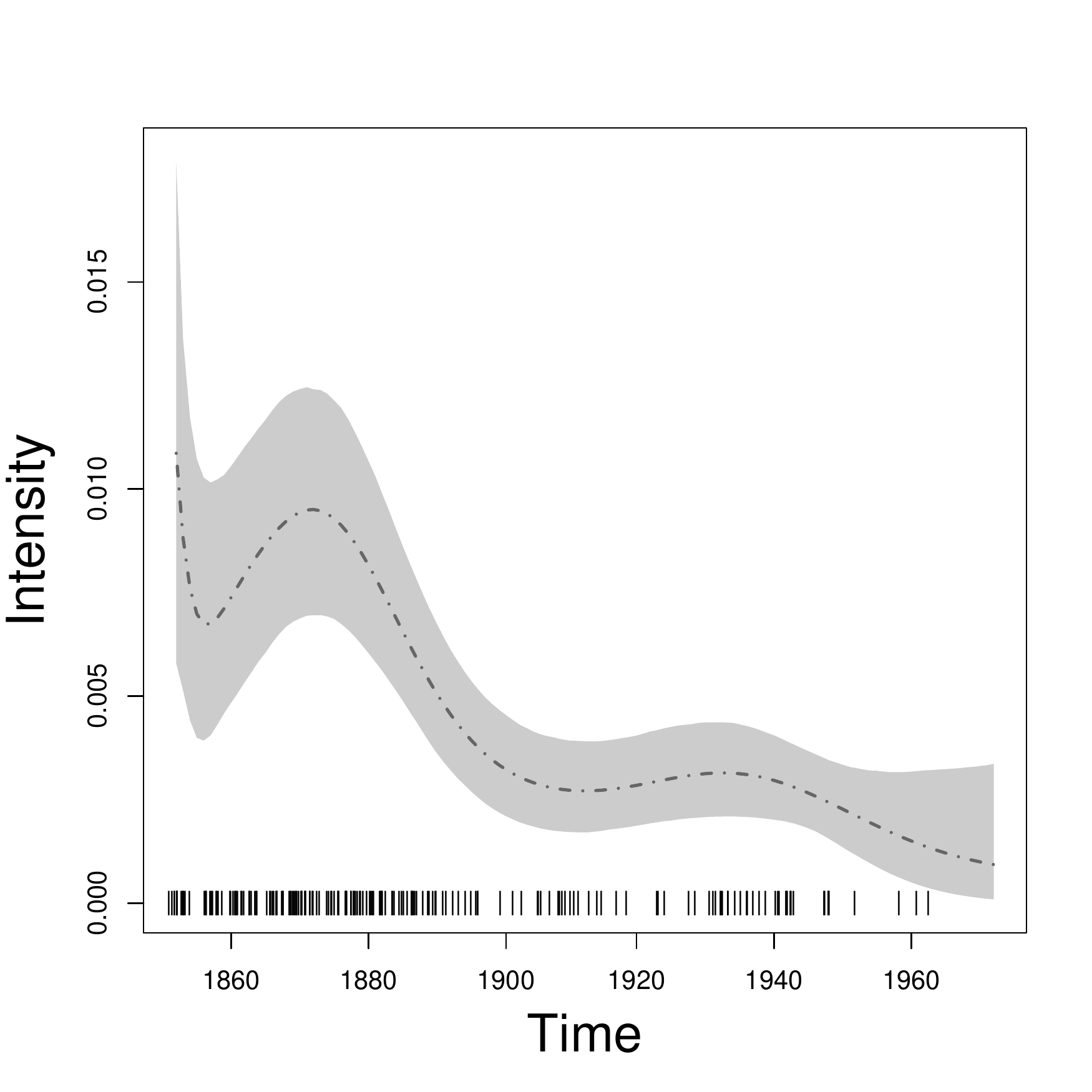}
\includegraphics[width=0.475\textwidth]{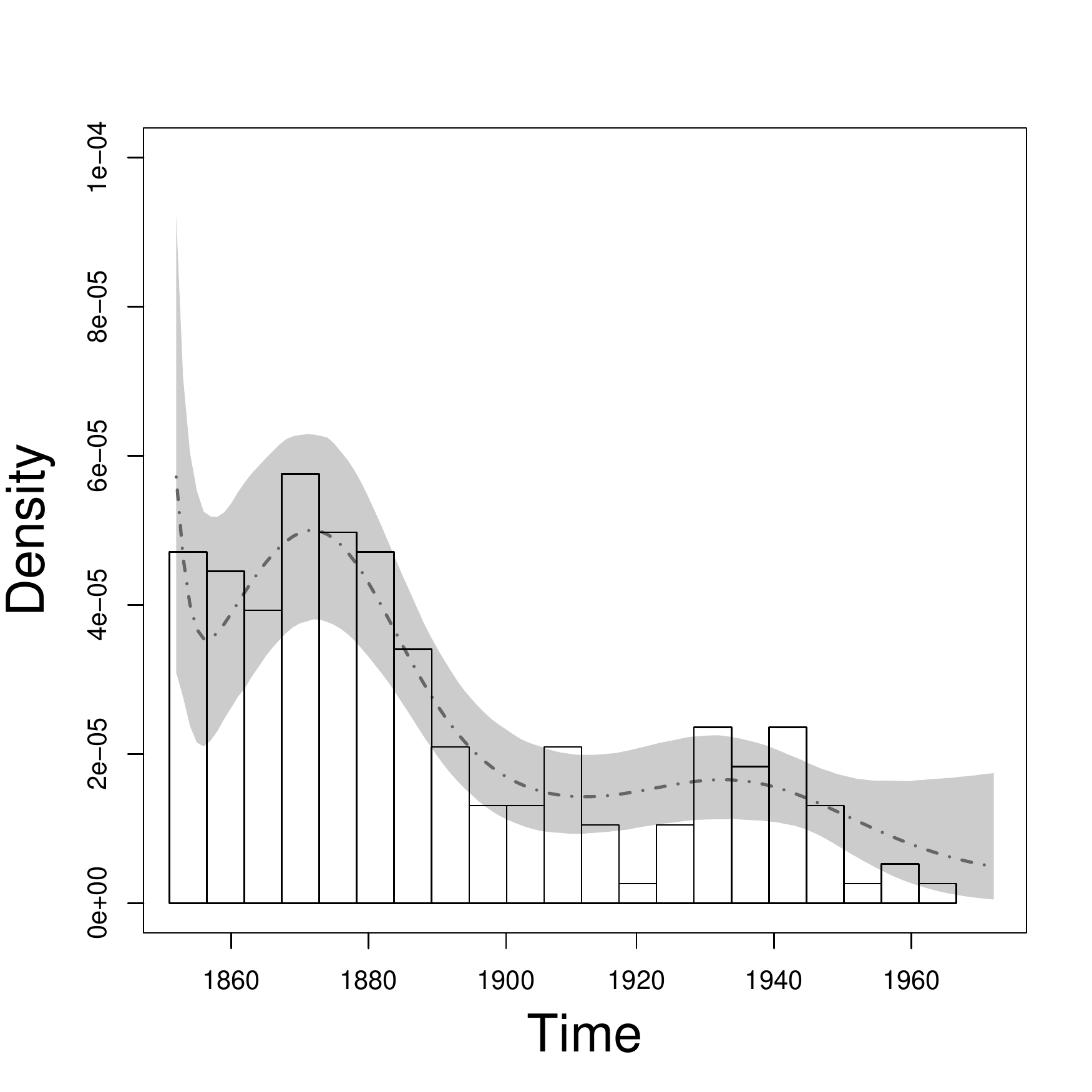}\\
\includegraphics[width=0.475\textwidth]{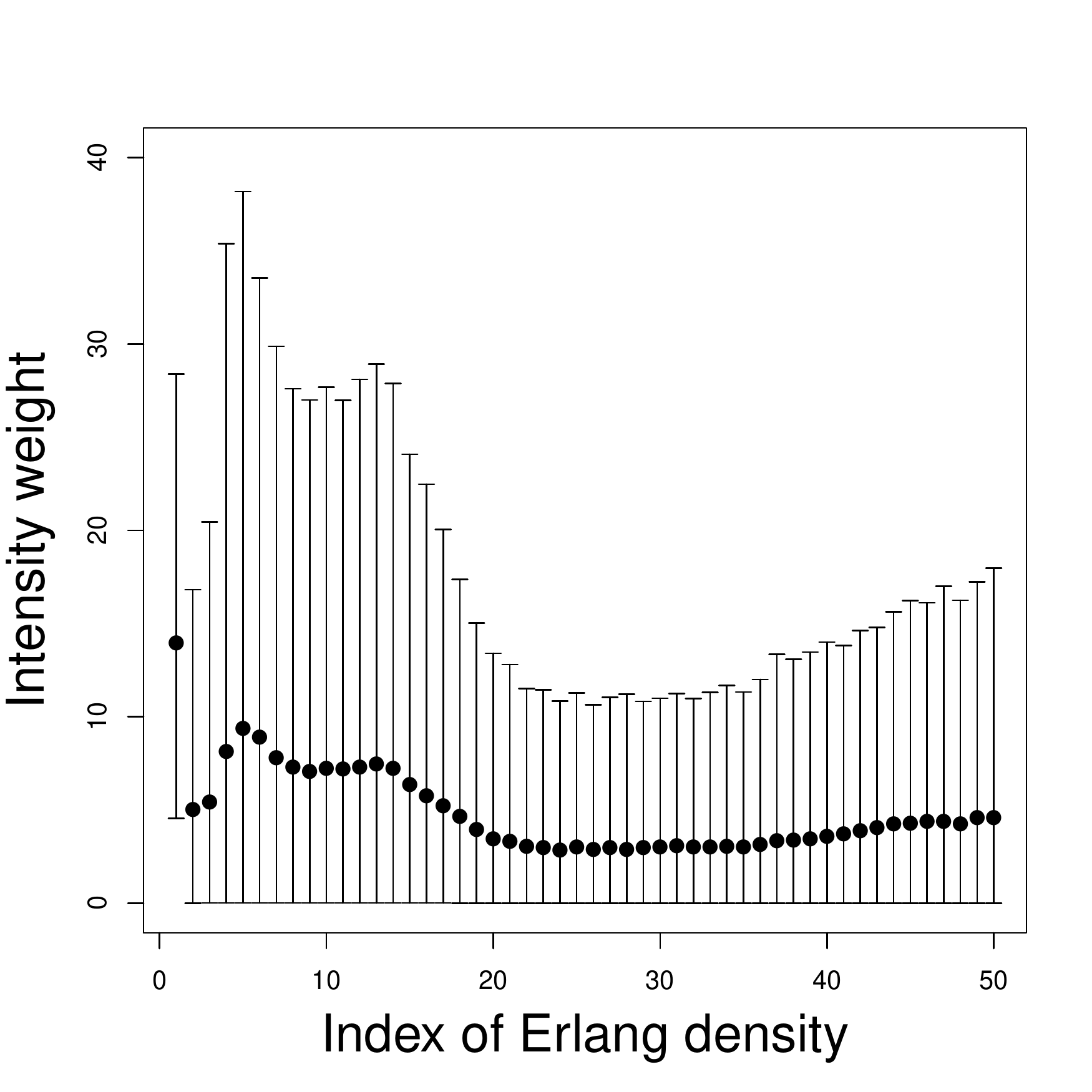}
\includegraphics[width=0.475\textwidth]{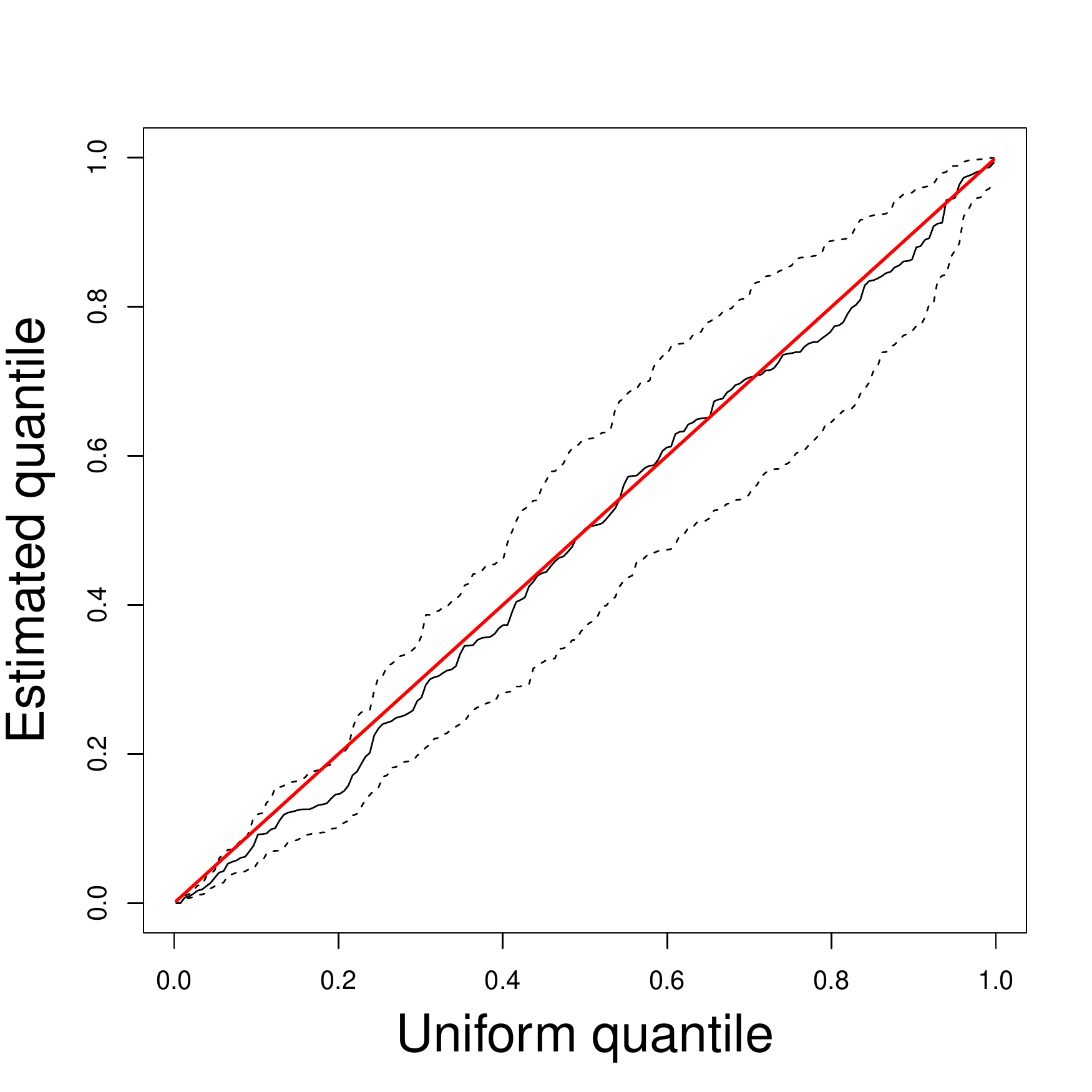}\\	
\caption{Coal-mining disasters data. The top left panel shows the posterior mean estimate 
(dashed-dotted line) and 95\% interval bands (shaded area) for the intensity function. 
The bars at the bottom indicate the observed point pattern. The top right panel plots 
the posterior mean (dashed-dotted line) and 95\% interval bands (shaded area) for the 
NHPP density, overlaid on the histogram of the accident times. The bottom left panel 
presents the posterior means (circles) and 90\% interval estimates (bars) of the mixture
weights. The bottom right panel plots the posterior mean and 95\% interval bands for 
the time-rescaling model checking Q-Q plot.}
\label{fig:real_est}
\end{figure}

We fit the Erlang mixture model with $J=50$, using a Lomax prior for $\theta$
with shape parameter $2$ and scale parameter $2,000$, such that 
$\text{Pr}(0< \theta <40,550) \approx 0.998$, and an exponential prior for $b$
with mean $213$. We also implemented the model with $J=130$,
obtaining essentially the same inference results for the NHPP functionals with 
the ones reported in Figure \ref{fig:real_est}.

The estimates for the point process intensity and density functions 
(Figure \ref{fig:real_est}, top row) suggest that the model successfully captures 
the multimodal intensity shape suggested by the data. The estimates for the 
mixture weights (Figure \ref{fig:real_est}, bottom left panel) indicate the 
Erlang basis densities that are more influential to the model fit.

%
%From the left panels of Figure \ref{fig:bim_est}, we can pair the point pattern and the major %components affecting the pattern, along with the posterior estimate of $\theta$. For example, given %$\hat{\theta} \approx 1057$, the first Erlang density component is responsible for the decreasing %pattern at an early stage, and the fourth to fifteenth components, which have %$(4228,5285,\ldots,15855)$ as their means $j\hat{\theta}$, are used for capturing the first mode at %around $10000$ days ($1880$ in year). The two panels in the right side of Figure \ref{fig:bim_est} are %employed to assess the goodness of fit of the model. Although the true intensity function is not %known, we can presume the model performance through the comparison of the density estimates with the %data distribution (see, top right panel). Besides, in the bottom right panel, the closer the posterior %mean (dot-dash line) is to the red line, the better intensity estimates the model yields. In other %words, as the model performs better in intensity estimation, the estimated quantile from ``the %time-rescaling theorem'' gets closer to the theoretical uniform quantile. In this example, the %posterior mean quantile is comparable to the theoretical quantile, and the interval bands effectively %cover the red straight line, which justifies our model performance for the data set.
%

The bottom right panel of Figure \ref{fig:real_est} reports results from graphical 
model checking, using the ``time-rescaling" theorem (e.g., \citeh{Daley_Vere-Jones:2003}).
If the point pattern $\{ 0 = t_{0} < t_{1} < ... < t_{n} < T \}$ is a realization from a 
NHPP with cumulative intensity function $\Lambda(t)=$ $\int_0^t \lambda(u) \text{d}u$,
then the transformed point pattern $\{\Lambda(t_i): i=1,...,n \}$ is a realization from
a unit rate homogeneous Poisson process. Therefore, if we further transform to $U_i =$
$1-\exp\{ -(\Lambda(t_i)-\Lambda(t_{i-1})) \}$, where $\Lambda(0) \equiv 0$, then the 
$\{ U_{i}: i=1,...,n \}$ are independent uniform$(0,1)$ random variables.  
Hence, graphical model checking can be based 
on quantile–quantile (Q-Q) plots to assess agreement of the estimated $U_{i}$ with
the uniform distribution on the unit interval. Under the Bayesian inference framework,
we can obtain a posterior sample for the $U_{i}$ for each posterior realization for 
the NHPP intensity, and we can thus plot posterior point and interval estimates for 
the Q-Q graph. These estimates suggest that the NHPP model with the Erlang mixture 
intensity provides a good fit for the coal-mining disasters data.

%
%----------------------------------------------------------------------------
%

\section{Modeling for spatial Poisson process intensities}
\label{spatial_NHPP}

In Section \ref{subsec:spatial_model}, we extend the modeling framework to spatial NHPPs
with intensities defined on $\mathbb{R}^{+}\times \mathbb{R}^{+}$. The resulting inference
method is illustrated with synthetic and real data examples in Section 
\ref{subsec:spatial_synthetic} and \ref{subsec:spatial_real}, respectively.

\subsection{The Erlang mixture model for spatial NHPPs}
\label{subsec:spatial_model}

A spatial NHPP is again characterized by its intensity function, 
$\lambda(\bm{s})$, for $\bm{s}=(s_1,s_2) \in \mathbb{R}^{+}\times \mathbb{R}^{+}$.
The NHPP intensity is a non-negative and locally integrable function such that:  
(a) for any bounded $B \subset \mathbb{R}^{+} \times \mathbb{R}^{+}$, the number of 
points in $B$, $N(B)$, follows a Poisson distribution with mean 
$\int_{B} \lambda(\bm{u}) \, \text{d}\bm{u}$; and (b) given $N(B)=n$, the random  
locations $\bm{s}_{i}=$ $(s_{i1},s_{i2})$, for $i=1,...,n$, that form the spatial 
point pattern in $B$ are i.i.d. with density 
$\lambda(\bm{s})/\{ \int_{B} \lambda(\bm{u}) \, \text{d}\bm{u} \}$.
Therefore, the structure of the likelihood for the intensity function is similar 
to the temporal NHPP case. In particular, for spatial point pattern, 
$\{\bm{s}_1,\ldots,\bm{s}_n\}$, observed in bounded region 
$D \subset \mathbb{R}^{+}\times \mathbb{R}^{+}$, the likelihood is proportional to 
$\exp\{ - \int_{D} \lambda(\bm{u}) \, \text{d}\bm{u} \} \,
\prod_{i=1}^{n} \lambda(\bm{s}_i)$.
As is typically the case in standard applications involving spatial NHPPs, we 
consider a regular, rectangular domain for the observation region $D$, which 
can therefore be taken without loss of generality to be the unit square.

Extending the Erlang mixture model in (\ref{mixture_model}), we build the basis 
representation for the spatial NHPP intensity from products of Erlang densities,  
$\{ \text{ga}(s_1 \mid j_1,\theta^{-1}_1) \, \text{ga}(s_2 \mid j_2,\theta^{-1}_2):
j_{1},j_{2} = 1,...,J \}$. Mixing is again with respect to the shape parameters 
$(j_1,j_2)$, and the basis densities share a pair of scale parameters $(\theta_1,\theta_2)$. 
Therefore, the model can be expressed as
%\begin{equation}
\[
\lambda(s_1,s_2) = \sum^{J}_{j_{1}=1} \sum^{J}_{j_{2}=1} \omega_{j_{1} j_{2}} \, 
\text{ga}(s_1 \mid j_1,\theta^{-1}_1) \, \text{ga}(s_2 \mid j_2,\theta^{-1}_2), 
\,\,\,\,\,\,\,  (s_1, s_2) \in \mathbb{R}^{+} \times \mathbb{R}^{+} .
\]
%\label{spatial_equation}
%\end{equation}
Again, a key model feature is the prior for the mixture weights. Here, the basis
density indexed by $(j_1,j_2)$ is associated with rectangle $A_{j_1 j_2} =$  
$[(j_{1}-1)\theta_1,j_{1} \theta_1) \times [(j_{2}-1) \theta_2, j_{2} \theta_2)$.
The corresponding weight is defined through a random measure $H$ supported on 
$\mathbb{R}^{+} \times \mathbb{R}^{+}$, such that $\omega_{j_{1} j_{2}}=$ 
$H(A_{j_{1} j_{2}})$. This construction extends the one for the weights of the 
temporal NHPP model. We again place a gamma process prior, $\mathcal{G}(H_0,c_0)$,
on $H$, where $c_0$ is the precision parameter and $H_{0}$ is the centering 
measure on $\mathbb{R}^{+} \times \mathbb{R}^{+}$. As a natural extension of the 
exponential cumulative hazard used in Section \ref{subsec:model} for the gamma 
process prior mean, we specify $H_{0}$ to be proportional to area. In particular, 
$H_0(A_{j_{1} j_{2}})=$ $|A_{j_1 j_2}|/b =$ $\theta_1 \theta_2 / b$, where $b>0$. 
Using the independent increments property of the gamma process, and under the 
specific choice of $H_{0}$, the prior for the mixture weights is given by 
\[
\omega_{j_{1} j_{2}} \mid \theta_{1},\theta_{2},c_{0},b \, \stackrel{i.i.d.}{\sim} 
\, \text{ga}(\omega_{j_{1} j_{2}} \mid c_{0} \, \theta_1 \, \theta_2 \, b^{-1},c_{0}),
\,\,\,\,\,\,\, j_{1},j_{2} = 1,...,J ,
\]
which, as before, is a practically important feature of the model construction
as it pertains to MCMC posterior simulation.

To complete the full Bayesian model, we place priors on the common scale parameters
for the basis densities, $(\theta_{1},\theta_{2})$, and on the gamma process prior
hyperparameters $c_0$ and $b$. The role played by these model parameters is directly 
analogous to the one of the corresponding parameters for the temporal NHPP model,
as detailed in Section \ref{subsec:model}. Therefore, we apply similar arguments 
to the ones in Section \ref{subsec:priors} to specify the model hyperpriors.
More specifically, we work with (independent) Lomax prior distributions for scale 
parameters $\theta_1$ and $\theta_2$, where the shape parameter of the Lomax prior 
is set equal to $2$ and the scale parameter is specified such that $\text{Pr}(0<\theta_1<1)\text{Pr}(0<\theta_2<1) \approx 0.999$. Recall that the 
observation region is taken to be the unit square; in general, for a square observation 
region, this approach implies the same Lomax prior for $\theta_1$ and $\theta_2$.
The gamma process precision parameter $c_{0}$ is assigned an exponential prior
with mean $10$. The result of Section \ref{subsec:model} for the prior mean of 
the NHPP intensity can be extended to show that 
$\text{E}(\lambda(s_1,s_2) \mid b,\theta_1,\theta_2)$ converges to $b^{-1}$, 
as $J \rightarrow \infty$, for any $(s_1,s_2) \in \mathbb{R}^{+} \times \mathbb{R}^{+}$,
and for any $(\theta_1,\theta_2)$ (and $c_{0}$). The prior mean for the spatial 
NHPP intensity is practically constant at $b^{-1}$ within its effective support 
given roughly by $(0,J \theta_1) \times (0,J \theta_2)$. Hence, taking the size
of the observed spatial point pattern as a proxy for the total intensity, $b$
is assigned an exponential prior distribution with mean $1/n$.
Finally, the choice of the value for $J$ can be guided from the approximate 
effective support for the intensity, which is controlled by $J$ along with 
$\theta_1$ and $\theta_2$. Analogously to the approach discussed in 
Section \ref{subsec:priors}, the value of $J$ (or perhaps a lower bound for $J$)
can be specified through the integer part of $1/\theta^\ast$, where $\theta^\ast$
is the median of the common Lomax prior for $\theta_1$ and $\theta_2$.

The posterior simulation method for the spatial NHPP model is developed through 
a straightforward extension of the approach detailed in Section \ref{subsec:MCMC}.
We work again with the augmented model that involves latent variables 
$\{\bm{\gamma}_i: i=1,\ldots,n\}$, where $\bm{\gamma}_i=$ $(\gamma_{i1},\gamma_{i2})$
identifies the basis density to which observed point location $(s_{i1},s_{i2})$ is 
assigned. The spatial NHPP model retains the practically relevant feature of efficient 
updates for the mixture weights, which, given the other model parameters and the data, 
have independent gamma posterior full conditional distributions. 
Details of the MCMC posterior simulation algorithm are provided 
in the Supplementary Material.

\subsection{Synthetic data example}
\label{subsec:spatial_synthetic}

\begin{figure}[t!]
\centering
\includegraphics[width=0.31\textwidth]{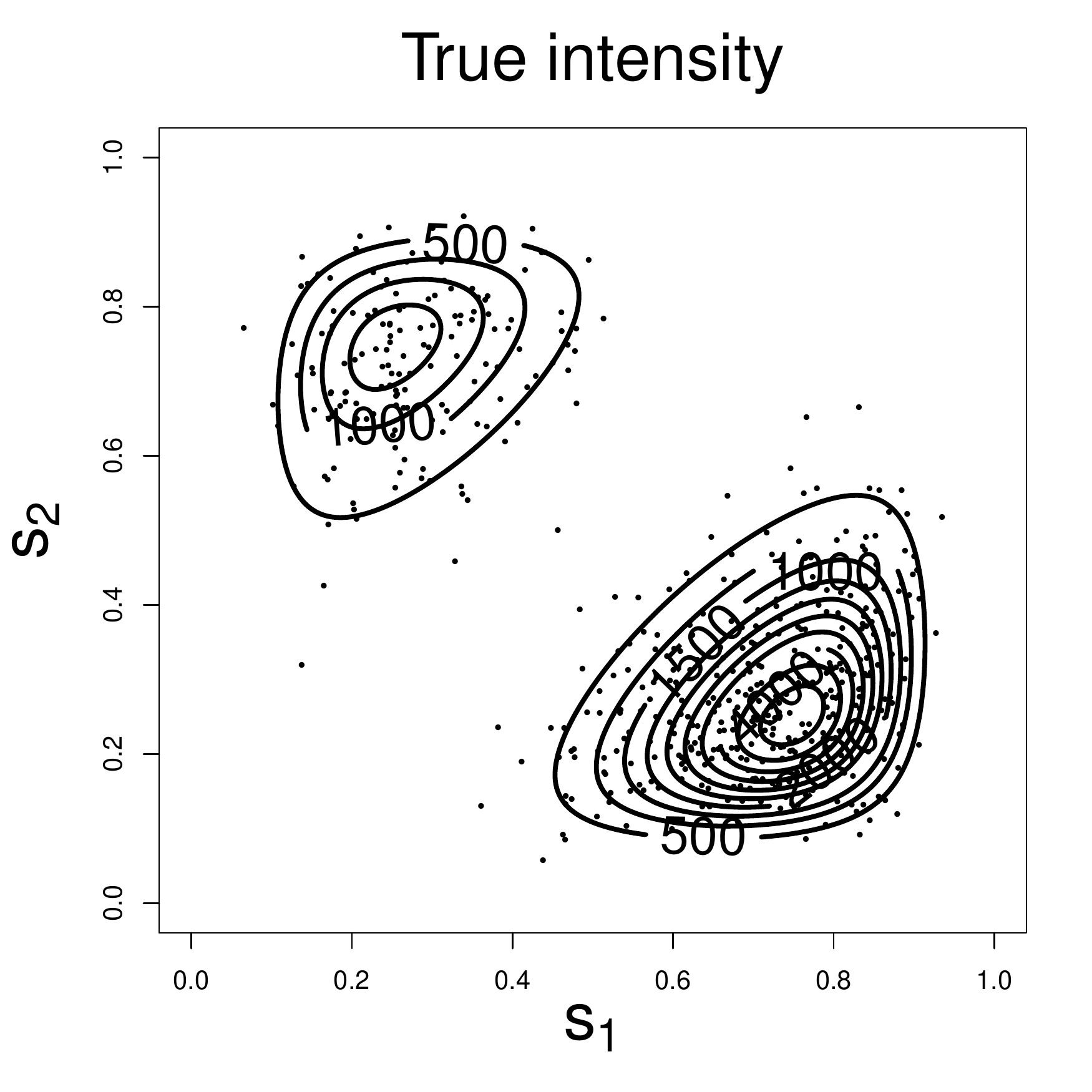}
\includegraphics[width=0.31\textwidth]{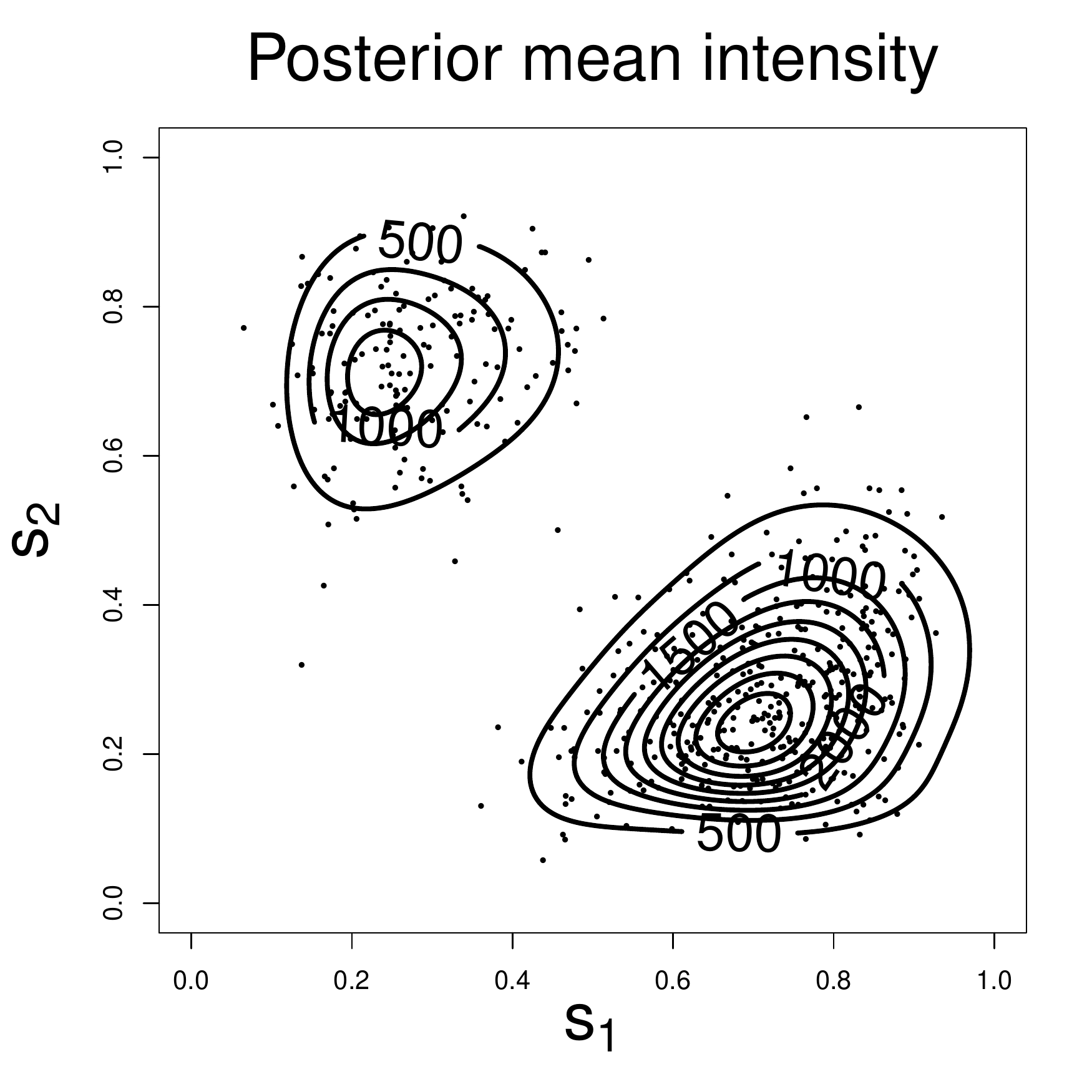}
\includegraphics[width=0.31\textwidth]{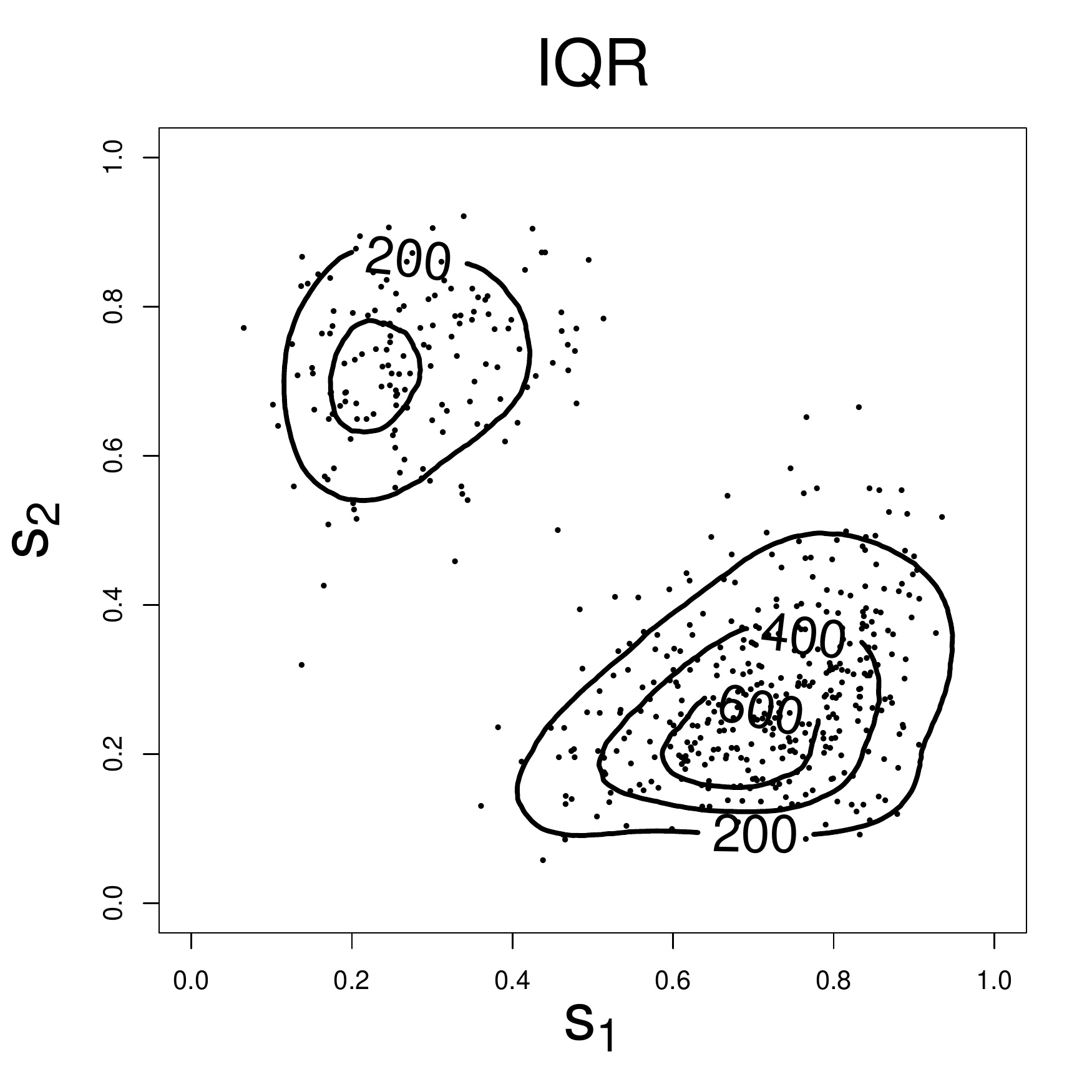}\\
\includegraphics[width=0.31\textwidth]{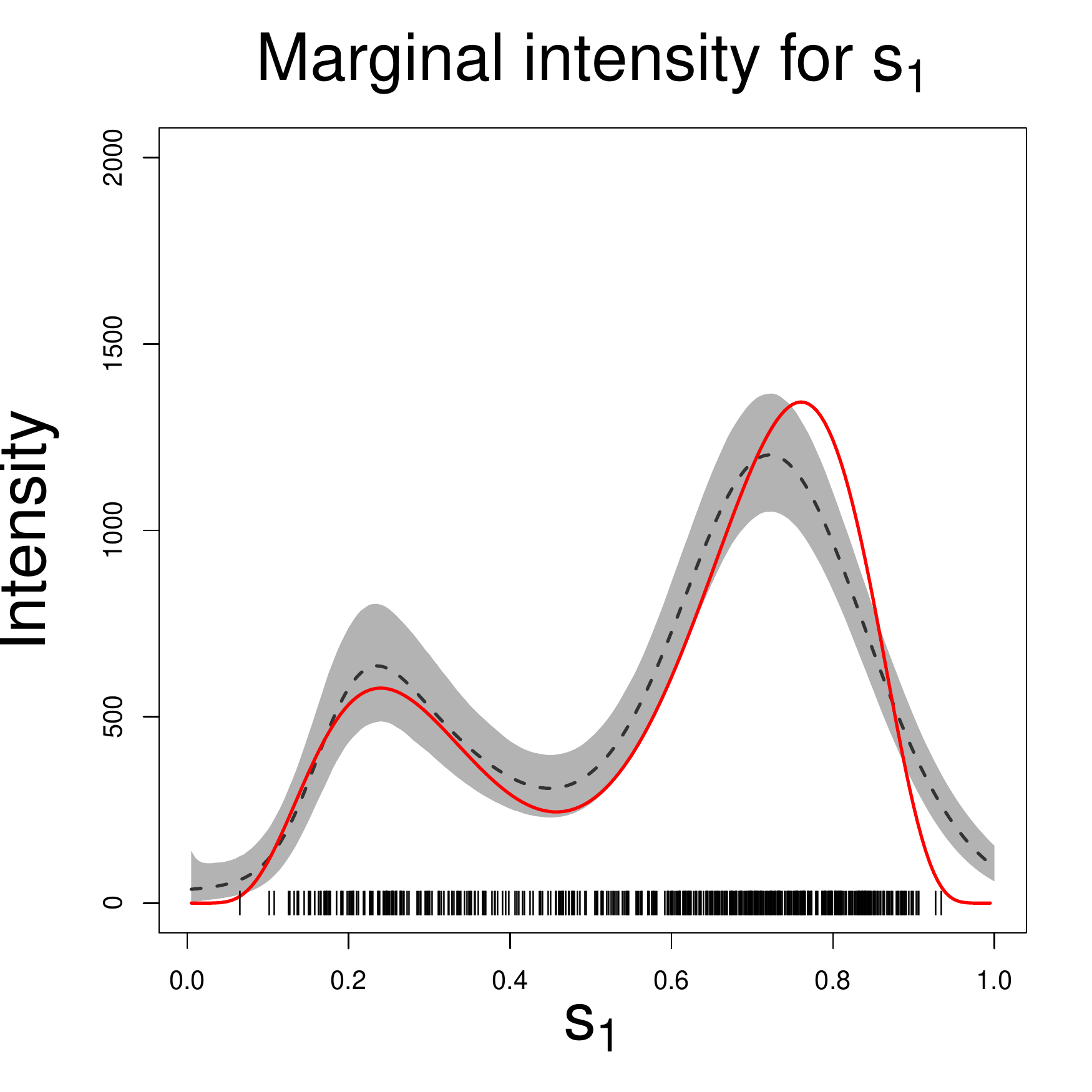}
\includegraphics[width=0.31\textwidth]{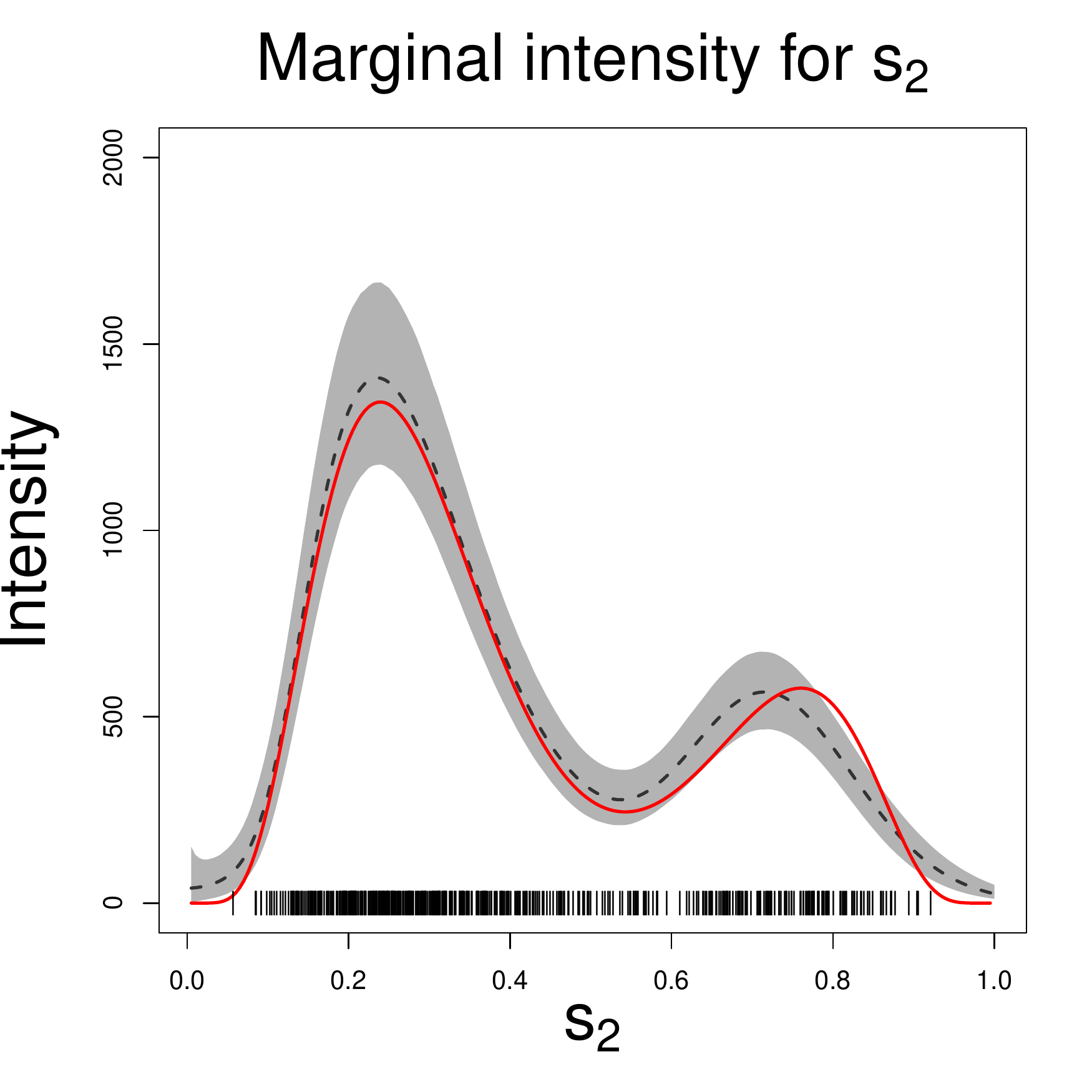}
\includegraphics[width=0.31\textwidth]{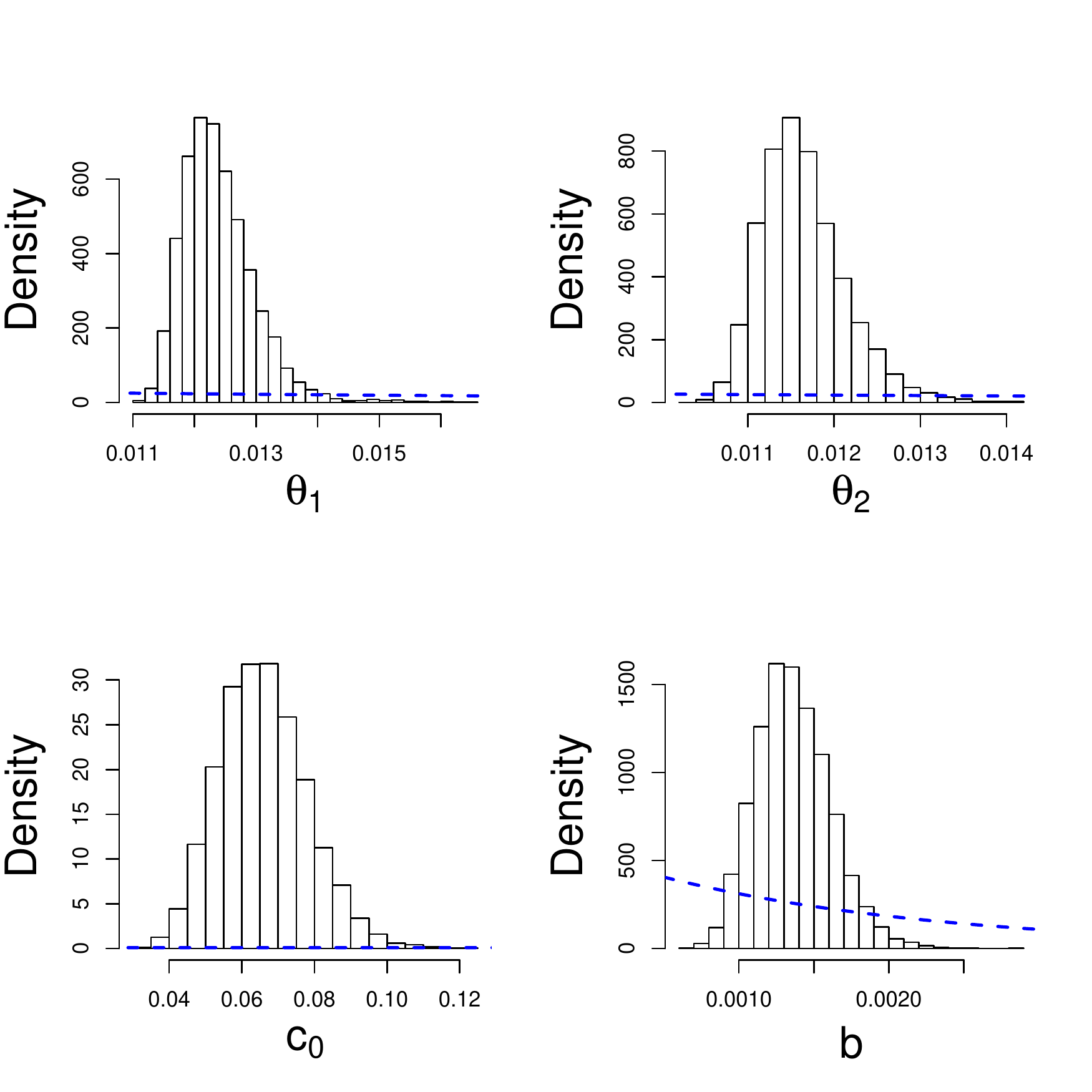}\\
\caption{Synthetic data example from spatial NHPP. The top row panels show
contour plots of the true intensity, and of the posterior mean and 
interquartile range estimates. The points in each panel indicate the observed 
point pattern. The first two panels at the bottom row show the marginal intensity 
estimates -- posterior mean (dashed line) and 95\% uncertainty bands (shaded area) --
along with the true function (red solid line) and corresponding point pattern 
(bars at the bottom of each panel). The bottom right panel displays histograms 
of posterior samples for the model hyperparameters along with the corresponding 
prior densities (dashed lines).}
	\label{fig:spatial_synthetic}
\end{figure}

Here, we illustrate the spatial NHPP model using synthetic data
based on a bimodal intensity function built from a 
two-component mixture of bivariate logit-normal densities. 
Denote by $\text{BLN}(\bm{\mu},\Sigma)$ the bivariate logit-normal 
density arising from the logistic transformation of a bivariate normal 
with mean vector $\bm{\mu}$ and covariance matrix $\Sigma$.
A spatial point pattern of size $528$ was generated over the unit 
square from a NHPP with intensity $\lambda(s_1,s_2)=$ 
$150 \, \text{BLN}((s_1,s_2) \mid \bm{\mu}_1,\Sigma) \, + \, 350 \, \text{BLN}((s_1,s_2) \mid \bm{\mu}_2,\Sigma)$, where $\bm{\mu}_1=$ $(-1,1)$, $\bm{\mu}_2=$ $(1,-1)$, and $\Sigma=$ $(\sigma_{11},\sigma_{12},\sigma_{21},\sigma_{22})=$ $(0.3,0.1,0.1,0.3)$.
The intensity function and the generated spatial point pattern are shown 
in the top left panel of Figure \ref{fig:spatial_synthetic}.

The Erlang mixture model was applied setting $J=70$ and using the hyperpriors
for $\theta_1$, $\theta_2$, $c_0$ and $b$ discussed in Section \ref{subsec:spatial_model}.
%a $\text{Lo}(2,0.035)$ prior for $\theta_1$ and $\theta_2$ (such that $\text{Pr}(0<\theta_1<1, %0<\theta_2<1) \approx 0.998$), $J=1/\theta^\ast \approx 70$ (with the median $\theta^\ast$ of the %prior for $\theta_l$, $l=1,2$), an $\text{Exp}(0.1)$ prior for $c_0$ (with mean 10), and an %$\text{Exp}(528)$ prior for $b$ (with mean 528), which follows the prior specification given in %Section \ref{subsec:spatial_model}
Figure \ref{fig:spatial_synthetic} reports inference results. The posterior mean 
intensity estimate successfully captures the shape of the underlying intensity function.
The structure of the Erlang mixture model enables ready inference for the marginal NHPP
intensities associated with the two-dimensional NHPP. Although such inference is 
generally not of direct interest for spatial NHPPs, in the context of a synthetic data 
example it provides an additional means to check the model fit. The marginal 
intensity estimates effectively retrieve the bimodality of the true marginal intensity 
functions; the slight discrepancy at the second mode can be explained by inspection of 
the generated data for which the second mode clusters are located slightly to the left 
of the theoretical mode. Finally, we note the substantial prior-to-posterior learning 
for all model hyperparameters.
%
%The inference results are based on 10,000 posterior samples, obtained after (conservative) 
%values for burn-in of 500,000 iterations and thinning of 150 iterations.
%

\subsection{Real data illustration}
\label{subsec:spatial_real}

%\begin{figure}[t!]
%	\centering
%	\includegraphics[width=0.5\textwidth]{./figs/contour_maples_100j.pdf}\includegraphics[width=0.47\textwidth]{./figs/heatmap_maples_100j.pdf}\\
%	\includegraphics[width=0.5\textwidth]{./figs/perspective_maples_100j.pdf}\includegraphics[width=0.5\textwidth]{./figs/iqr_maples_100j.pdf}\\	  
%	\caption{The top left depicts the contour plot of the posterior mean intensity with observations (dots). The Heat map (top right) and perspective plot (bottom left) of the posterior mean intensity will help you understand the intensity estimates along with the contour plot. The posterior variability of the intensity estimates can be found in the interquartile range in the top right panel.}
%	\label{fig:spatial_real}
%\end{figure}

\begin{figure}[t!]
\centering
\includegraphics[width=0.49\textwidth]{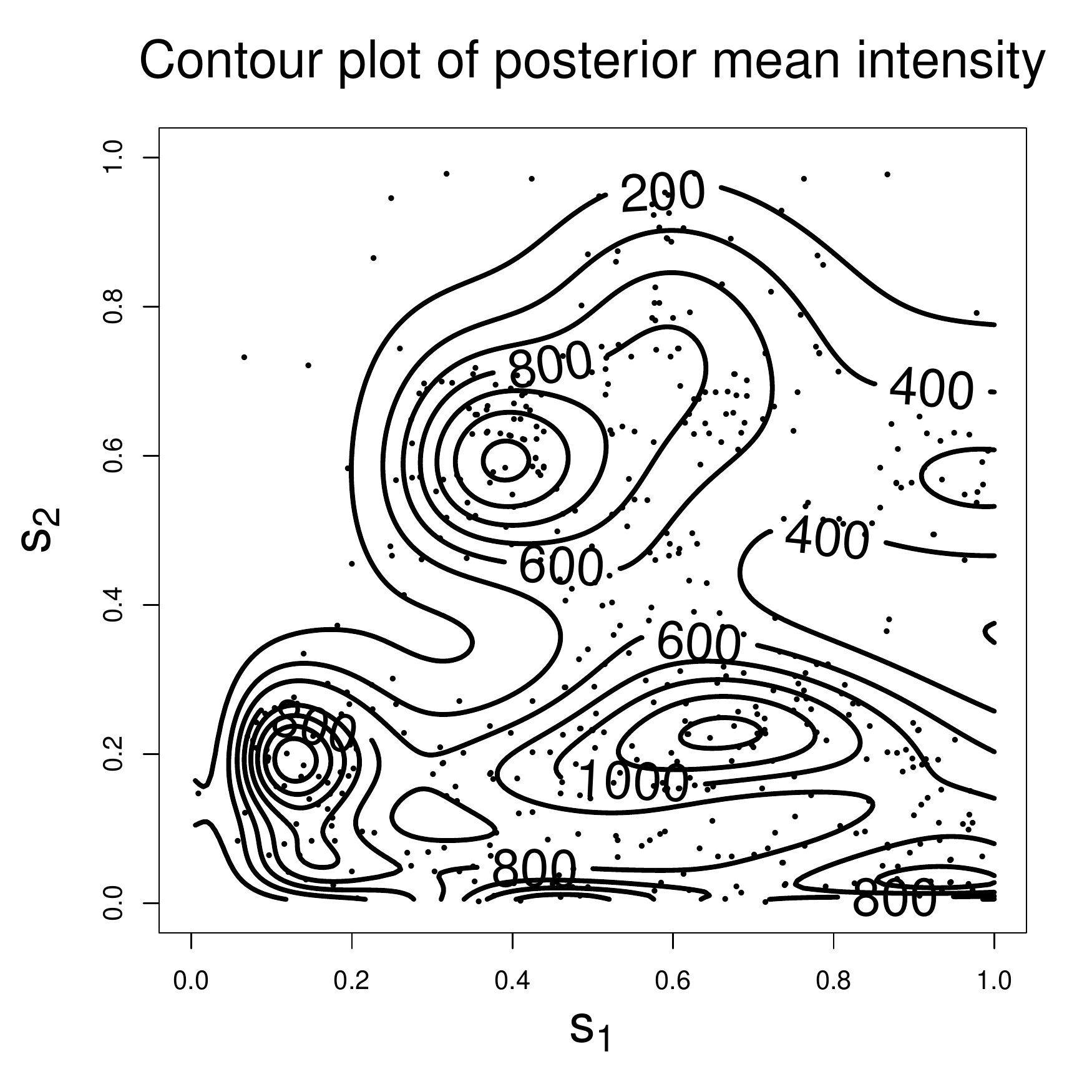}
\includegraphics[width=0.49\textwidth]{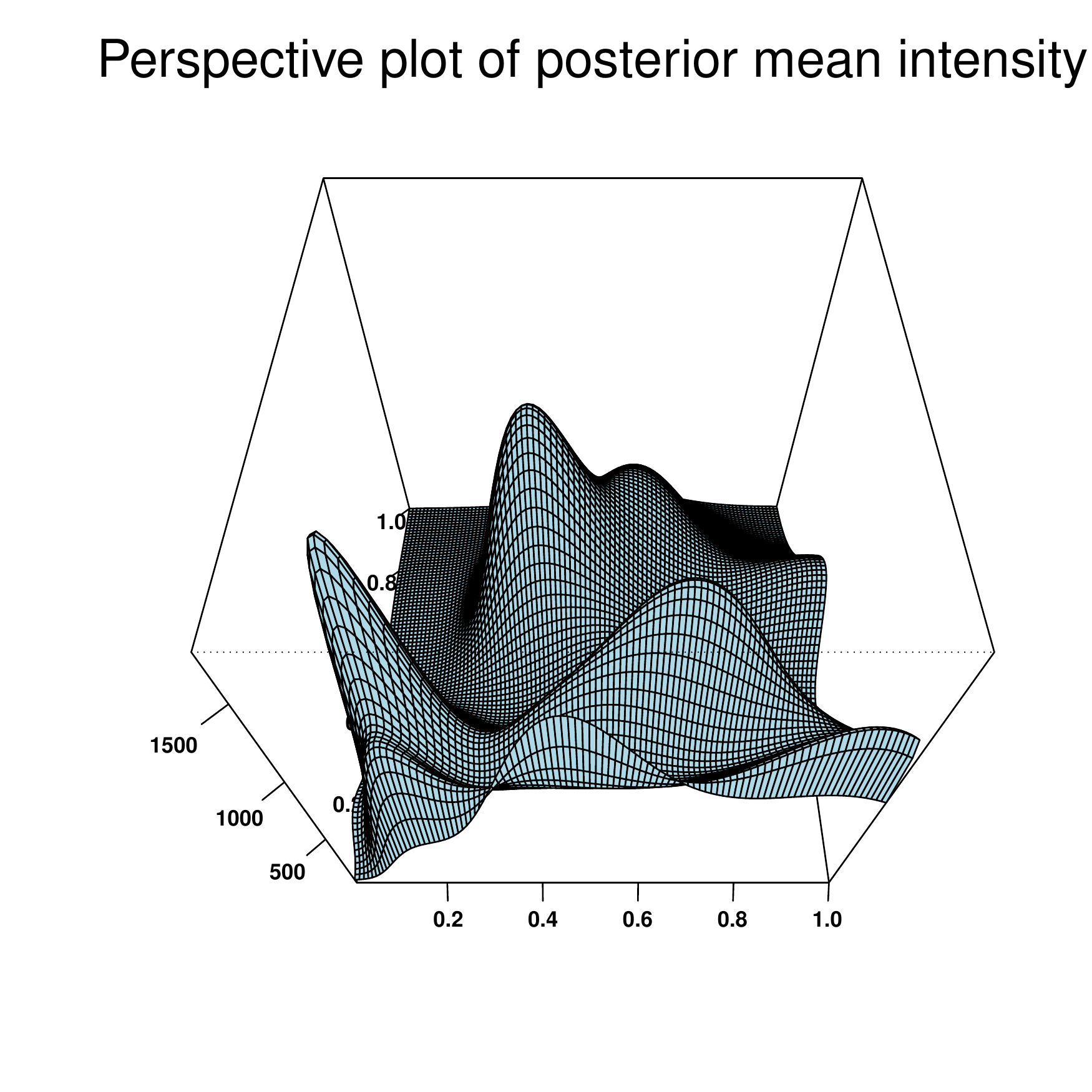}\\
\includegraphics[width=0.49\textwidth]{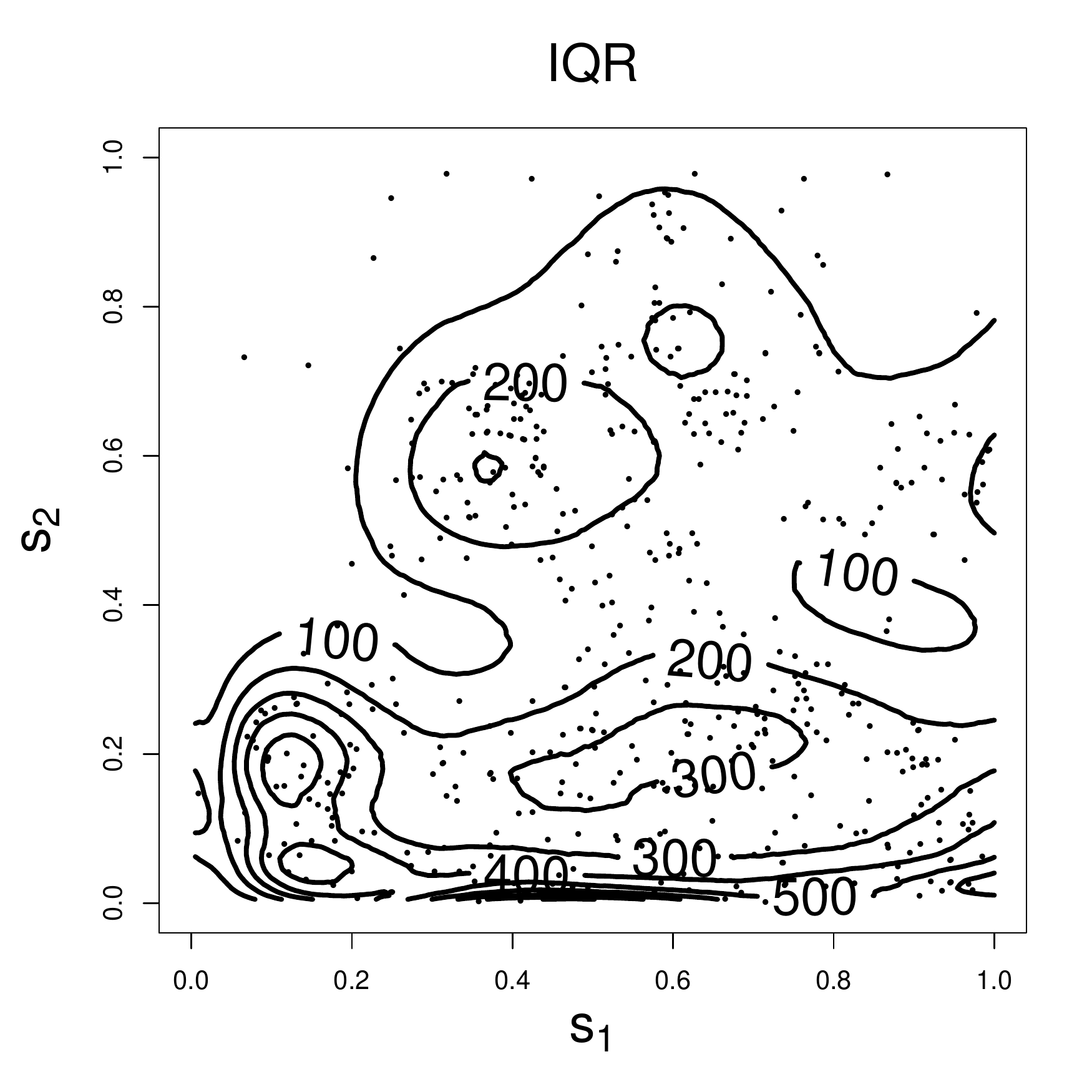}
\includegraphics[width=0.49\textwidth]{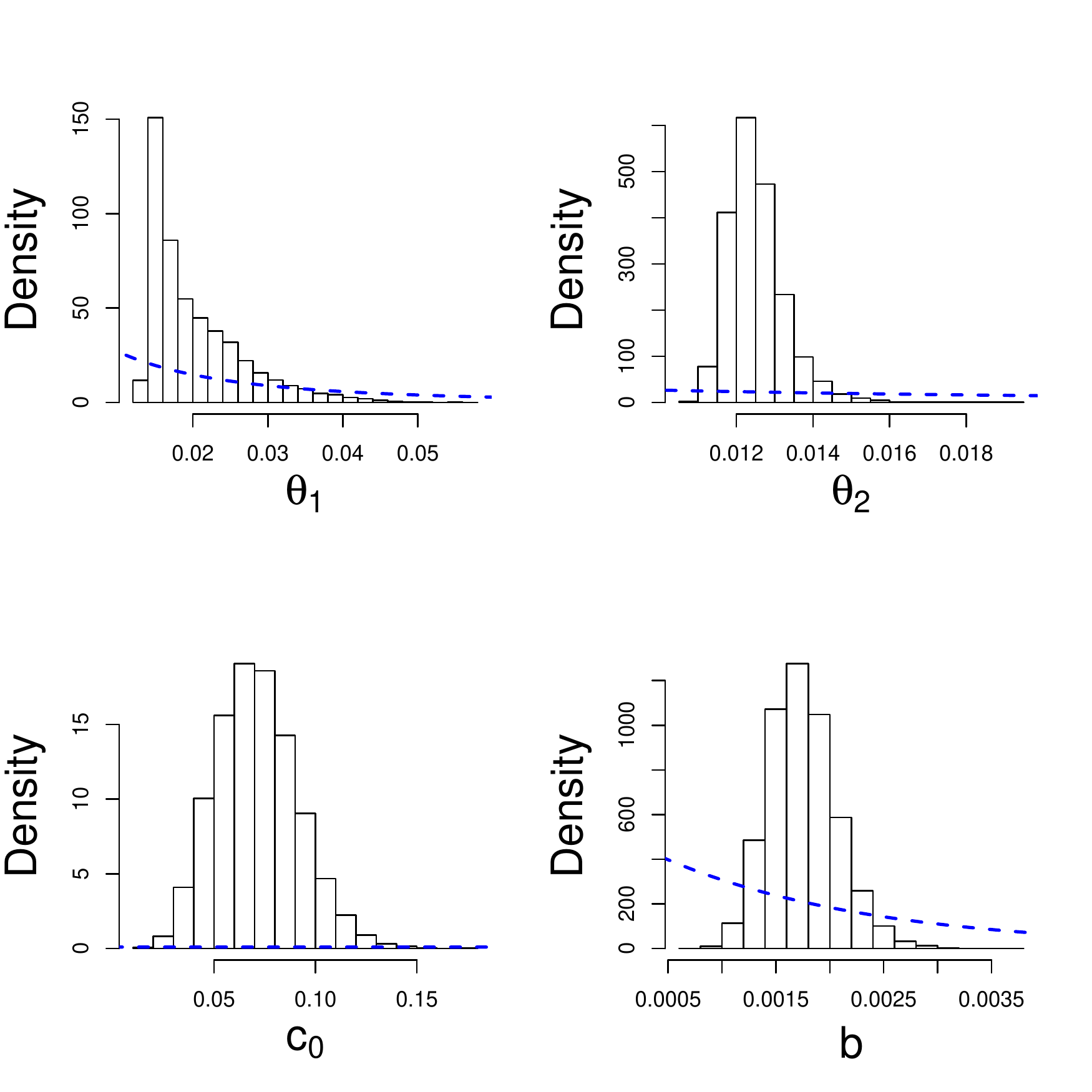}\\
\caption{Maple trees data. The top row panels show the posterior mean estimate 
for the intensity function in the form of contour and perspective plots.
The bottom left panel displays the corresponding posterior interquartile 
range contour plot. The bottom right panel plots histograms of posterior 
samples for the model hyperparameters along with the corresponding 
prior densities (dashed lines). The points in the left column plots 
indicate the locations of the 514 maple trees.}
	\label{fig:spatial_real}
\end{figure}

Our final data example involves a spatial point pattern that has been previously used to 
illustrate NHPP intensity estimation methods \cite[e.g.,][]{Diggle:2014,KS2007}.
The data set involves the locations of 514 maple trees in a 19.6 acre square plot 
in Lansing Woods, Clinton County, Michigan, USA; the maple trees point pattern is 
included in the left column panels of Figure \ref{fig:spatial_real}.

To apply the spatial Erlang mixture model, we specified the hyperpriors for $\theta_1$, 
$\theta_2$, $c_0$ and $b$ following the approach discussed in Section 
\ref{subsec:spatial_model}, and set $J=70$. As with the synthetic data example,
the posterior distributions for model hyperparameters are substantially concentrated 
relative to their priors; see the bottom right panel of Figure \ref{fig:spatial_real}.
The estimates for the spatial intensity of maple tree locations reported in 
Figure \ref{fig:spatial_real} demonstrate the model's capacity to uncover 
non-standard, multimodal intensity surfaces.
%
%Following the strategy given in Section \ref{subsec:spatial_model}, we specify the parameters 
%of the Erlang mixture model such that a $\text{Lo}(2,0.035)$ prior for $\theta_1$ and $\theta_2$ %(such that $\text{Pr}(0<\theta_1<1, 0<\theta_2<1) \approx 0.998$), $J=1/\theta^\ast \approx 70$ 
%(with the median $\theta^\ast$ of the prior for $\theta_l$, $l=1,2$), an $\text{Exp}(0.1)$ prior 
%for $c_0$ (with mean 10), and an $\text{Exp}(514)$ prior for $b$ (with mean 514).
%
%Figure \ref{fig:spatial_real} shows inference results; the posterior mean intensity 
%estimates in the two panels on top have the global pattern that is comparable with 
%the results presented by \cite{Diggle:2014} in the second panel of Figure 5.3 , p. 89. 
%It has smoother and less modes than in \cite{Diggle:2014}, and this feature of our 
%intensity estimates can be seen more clearly by comparison to \cite{KS2007}, which 
%exploited a different model for the same data set. 
%Our posterior inference is based on 10,000 posterior samples, obtained after (conservative) values %for burn-in of 100,000 iterations and thinning of 200 iterations.
%

%
%-----------------------------------------------------------------------
%

\section{Discussion}
\label{sec:disc}

We have proposed a Bayesian nonparametric modeling approach for
Poisson processes over time or space. The approach is based on a 
mixture representation of the point process intensity through Erlang 
basis densities, which are fully specified save for a scale parameter shared
by all of them. The weights assigned to the Erlang densities are
defined through increments of a random measure (a random cumulative 
intensity function in the temporal NHPP case) which is modeled with 
a gamma process prior. A key feature of the methodology
%, and the main motivation for its development, 
is that it offers a good balance between model flexibility and computational 
efficiency in implementation of posterior inference. Such inference has been 
illustrated with synthetic and real data for both temporal and spatial Poisson 
process intensities.

To discuss our contribution in the context of Bayesian nonparametric
modeling methods for NHPPs (briefly reviewed in the Introduction), note that the 
main approaches can be grouped into two broad categories: placing the prior model
on the NHPP intensity function; or, assigning separate priors to the total intensity
and the NHPP density (both defined over the observation window).

In terms of applications, 
especially for spatial point patterns, the most commonly explored class of models
falling in the former category involves Gaussian process (GP) priors for logarithmic (or logit) 
transformations of the NHPP intensity \cite[e.g.,][]{Moller_et_al.:1998,AMM2009}.
The NHPP likelihood normalizing term renders full posterior inference under GP-based models 
particularly challenging. This challenge has been bypassed using approximations of the 
stochastic integral that defines the likelihood normalizing term (\citeh{Brix_Diggle:2001}, \citeh{Brix_Moller:2001}), data augmentation techniques \citep{AMM2009}, and different types 
of approximations of the NHPP likelihood along with integrated nested Laplace approximation 
for approximate Bayesian inference (\citeh{ISR_2012}, \citeh{SILSR_2016}). In contrast,
the Erlang mixture model can be readily implemented with 
MCMC algorithms that do not involve approximations to the NHPP likelihood or complex
computational techniques. The Supplementary Material includes comparison of the 
proposed model with two GP-based models: the sigmoidal Gaussian Cox 
process (SGCP) model \citep{AMM2009} for temporal NHPPs; and the log-Gaussian Cox 
process (LGCP) model for spatial NHPPs, as implemented in the {\tt R} package 
{\tt lgcp} \citep{TDRD2013}. The results, based on the synthetic data considered
in Sections \ref{subsec:bim} and \ref{subsec:spatial_synthetic}, suggest that the 
Erlang mixture model is substantially more computationally efficient than the 
SGCP model, as well as less sensitive to model/prior specification than LGCP models 
for which the choice of the GP covariance function can have a large effect on the 
intensity surface estimates.

Since it involves a mixture formulation for the NHPP intensity, the 
proposed modeling approach is closer in spirit to methods based on Dirichlet process 
mixture priors for the NHPP density \cite[e.g.,][]{KS2007,Taddy_Kottas:2012}. 
Both types of approaches build posterior simulation from standard MCMC techniques for 
mixture models, using latent variables that configure the observed points to the mixture 
components. Models that build from density estimation with Dirichlet process mixtures 
benefit from the wide availability of related posterior simulation methods (e.g., the number 
of mixture components in the NHPP density representation does not need to be specified), 
and from the various extensions of the Dirichlet process for dependent distributions that 
can be explored to develop flexible models for hierarchically related point processes
\cite[e.g.,][]{Taddy2010,KBMPO2012,XKS2015,RWK2017}. 
However, by construction, this approach is restricted to modeling the NHPP intensity only 
on the observation window, in fact, with a separate prior for the NHPP density and for the 
total intensity over the observation window. The Erlang mixture model overcomes this 
limitation. For instance, in the temporal case, the prior model supports the intensity on 
$\mathbb{R}^{+}$, and the priors for the total intensity and the NHPP density over 
$(0,T)$ (given in Equation (\ref{formula_NHPP_density})) 
are compatible with the prior for the NHPP intensity.
%
%..... DP mixture modeling approach easier to extend to fully np models
%for marked NHPPs (though marks and covariates can be incorporated to the Erlang
%mixture model in a semiparametric fashion) .........
%

The proposed model admits a 
parsimonious representation for the NHPP intensity with the Erlang basis densities 
defined through a single parameter, the common scale parameter $\theta$. Such 
intensity representations offer a nonparametric Bayesian modeling perspective for 
point processes that may be attractive in other contexts and for different types
of applications. For instance, \cite{ZK_2021} study representations for the 
intensity through weighted combinations of structured beta densities (with different 
priors for the mixture weights), which are particularly well suited to flexible and 
efficient inference for spatial NHPP intensities over irregular domains.

Finally, we note that the Erlang mixture prior model is useful
as a building block towards Bayesian nonparametric inference for point processes that 
can be represented as hierarchically structured, clustered NHPPs. Current research 
is exploring fully nonparametric modeling for a key example, the Hawkes process 
(\citeh{Hawkes:1971}), using the Erlang mixture prior for the Hawkes process 
immigrant (background) intensity function.

\section*{Supplementary Information}
The Supplementary Material includes results from prior 
sensitivity analysis, information on computing times and effective sample
sizes, the technical details of the MCMC algorithm for the spatial NHPP model, 
and results from comparison of the Erlang mixture model with two Gaussian 
process based models for temporal or spatial NHPPs.

\section*{Acknowledgements}
The authors wish to thank an associate editor and a referee 
for several useful comments that resulted in an improved presentation of the 
material in the paper. This research was supported in part by the National 
Science Foundation under award SES 1950902.

\bibliographystyle{apalike}
\bibliography{references}

\end{document}

\title{\huge \textbf{Supplementary Material}}
%\textit{Statistics and Computing}}

% \author{Hyotae Kim and Athanasios Kottas
% \thanks{
% Hyotae Kim (hkim153@ucsc.edu) is Ph.D. student and Athanasios Kottas 
% (thanos@soe.ucsc.edu) is Professor, Department of Statistics, 
% University of California, Santa Cruz.}
% }

\date{}
\maketitle
\section{Prior sensitivity analysis and computational details}

\subsection{Prior sensitivity analysis}

We present results from sensitivity analysis to the prior choices for 
$\theta$, $c_0$, and $b$. We focus on the synthetic data set of Section 3.3 of 
the paper, mainly because it involves the smallest sample size ($n = 112$) among 
all data examples discussed in the paper, but also due to the non-standard, bimodal shape 
of the underlying non-homogeneous Poisson process (NHPP) intensity function.
We have conducted prior sensitivity analysis for all other data examples observing 
levels of robustness to the prior choice that are either higher or the same with 
the ones reported here.

\begin{figure}[t!]
\centering
\includegraphics[width=0.24\textwidth]{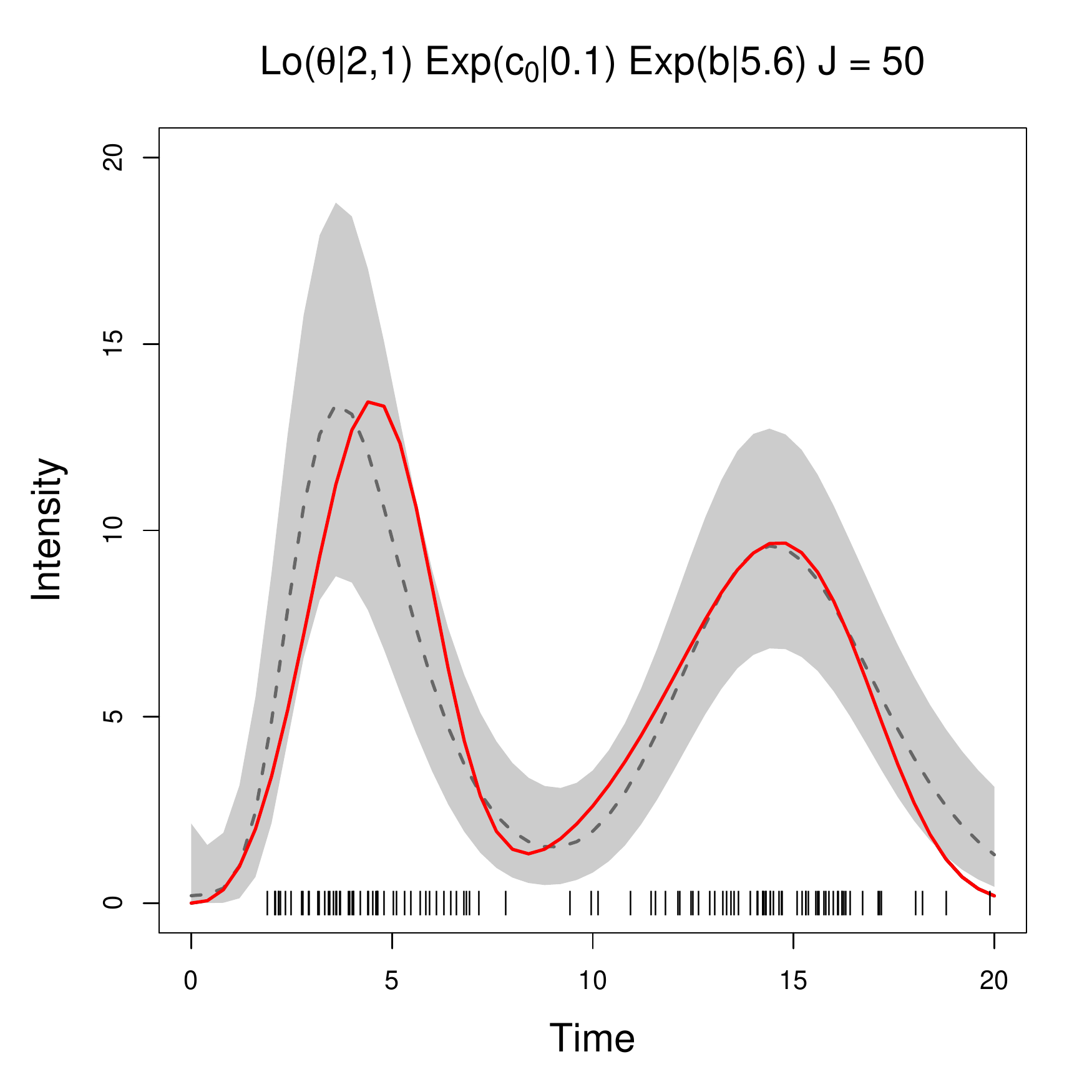}
\includegraphics[width=0.24\textwidth]{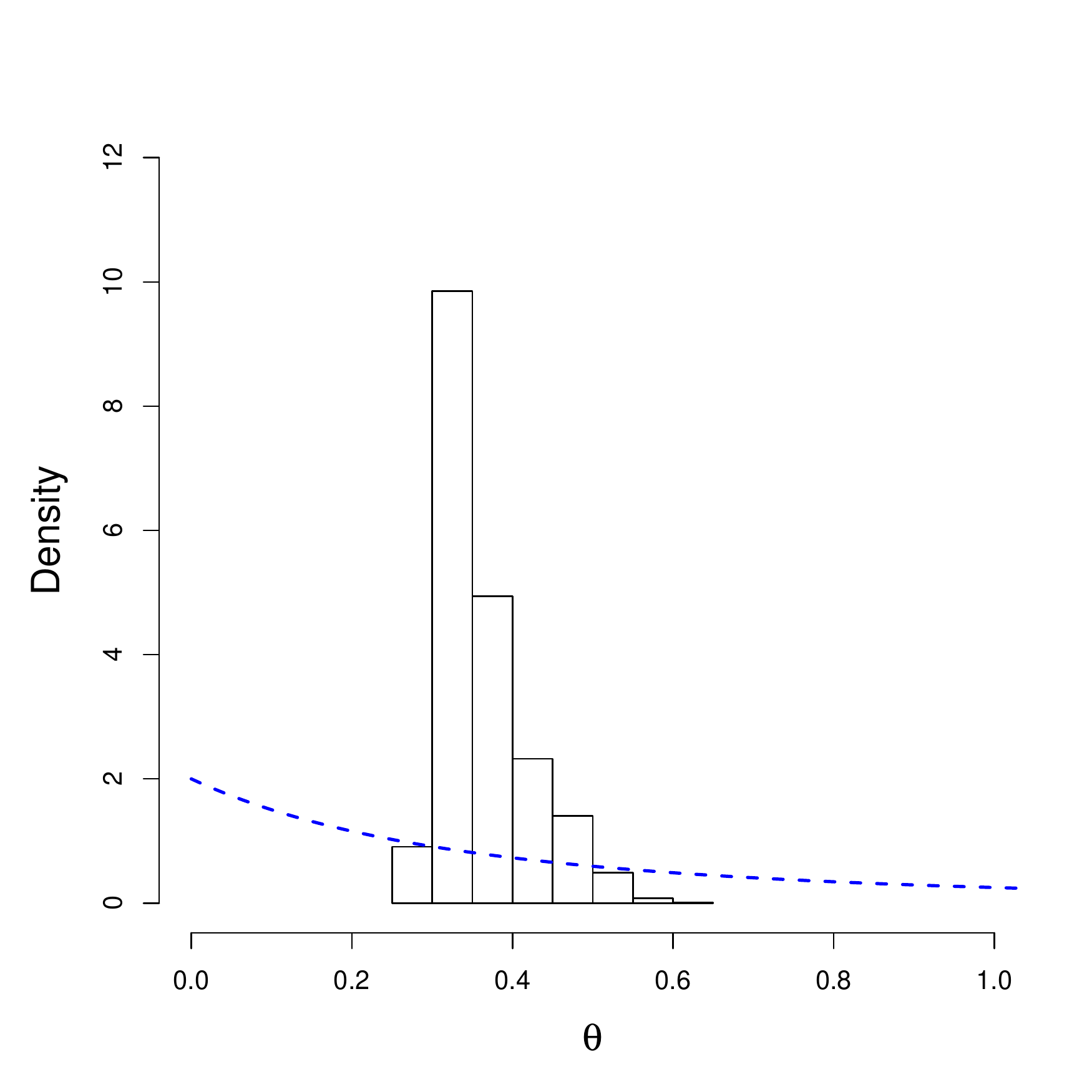}
\includegraphics[width=0.24\textwidth]{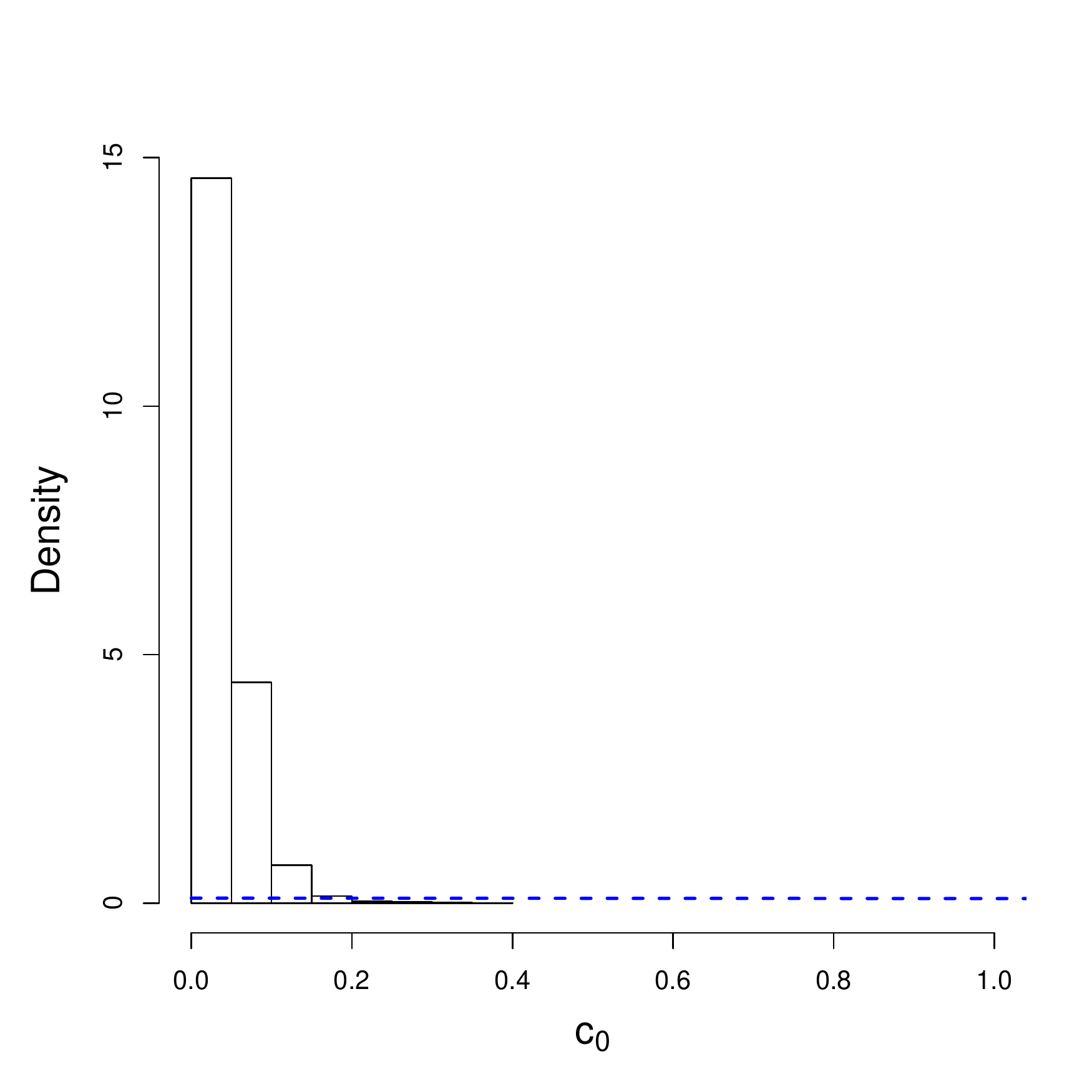}
\includegraphics[width=0.24\textwidth]{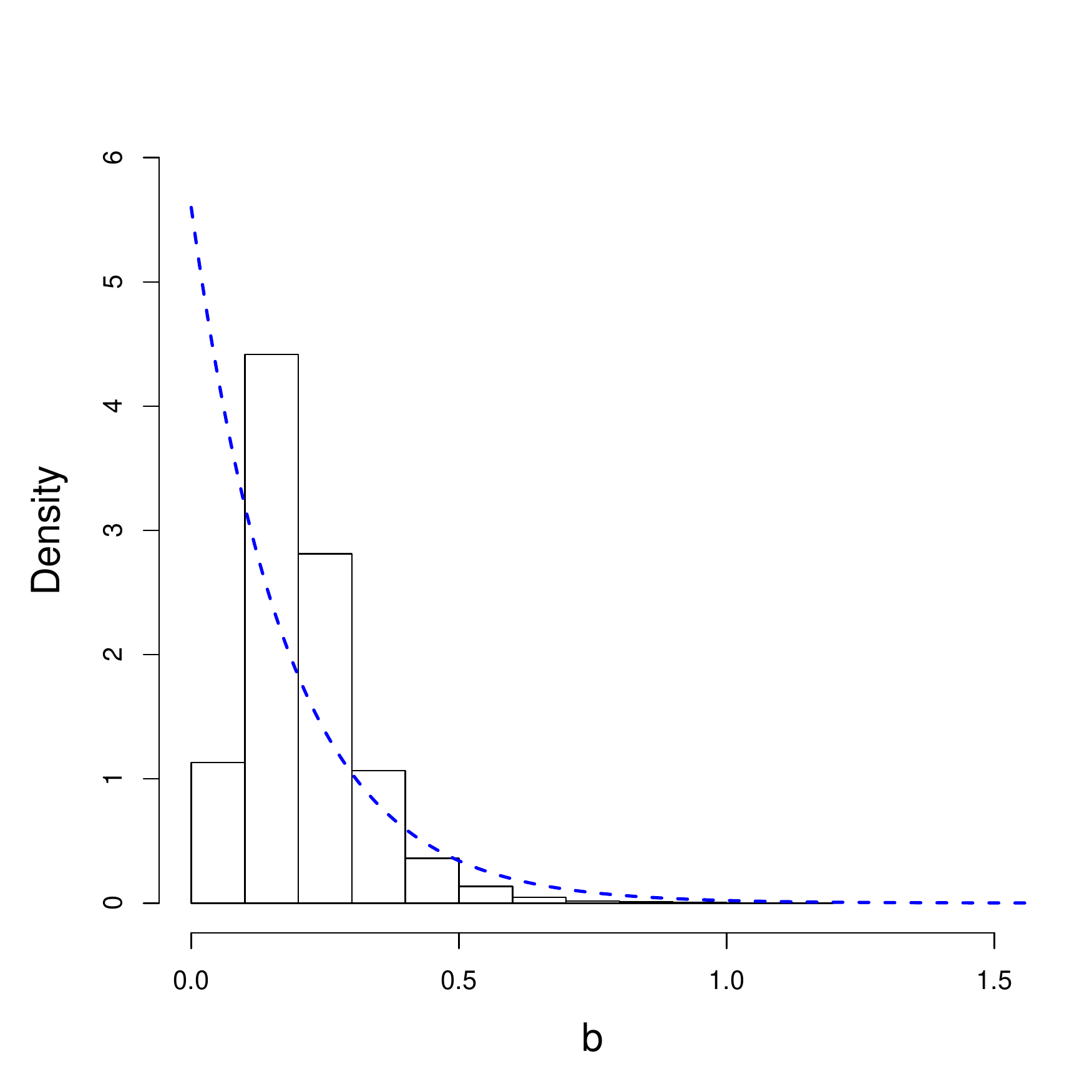}\\
\includegraphics[width=0.24\textwidth]{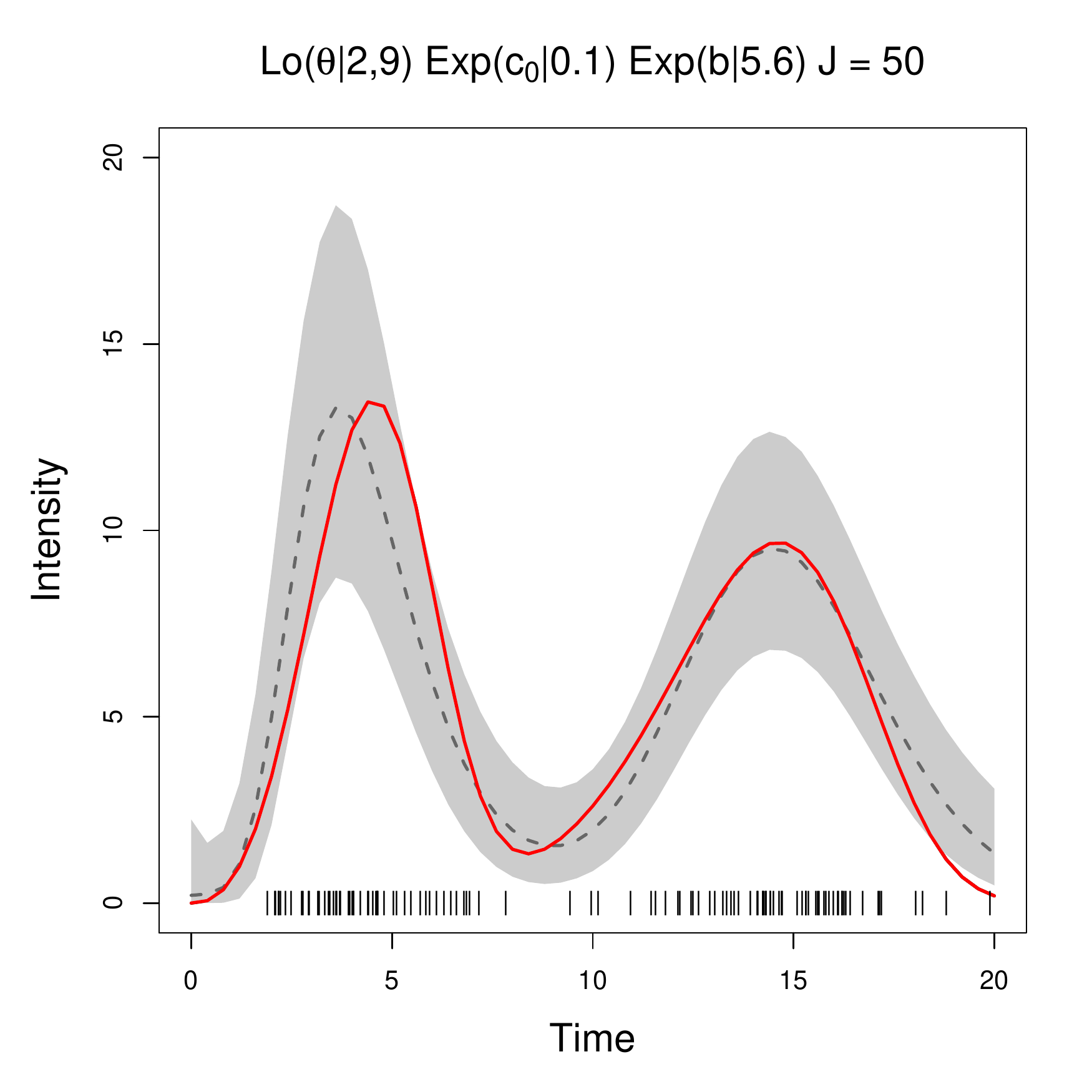}
\includegraphics[width=0.24\textwidth]{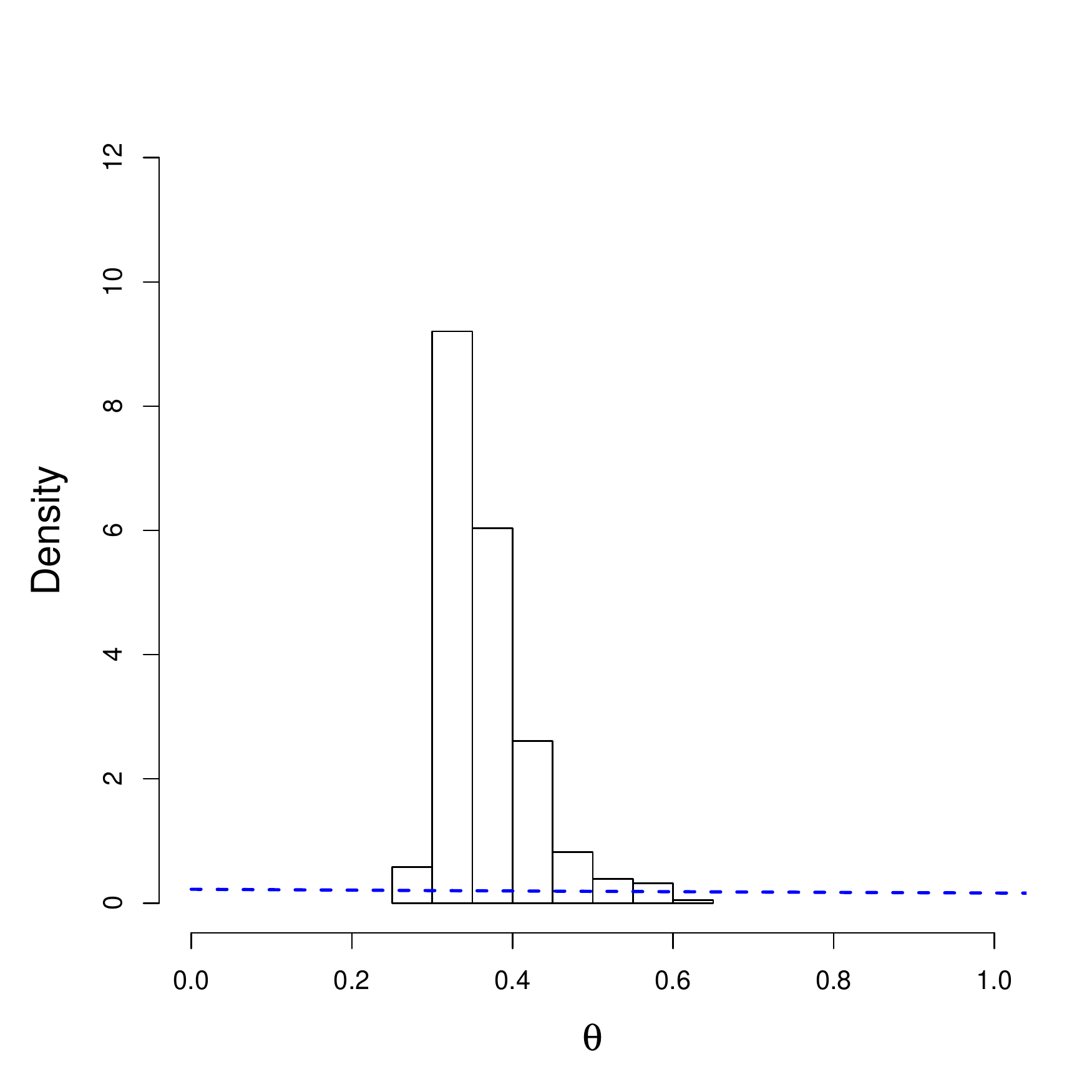}
\includegraphics[width=0.24\textwidth]{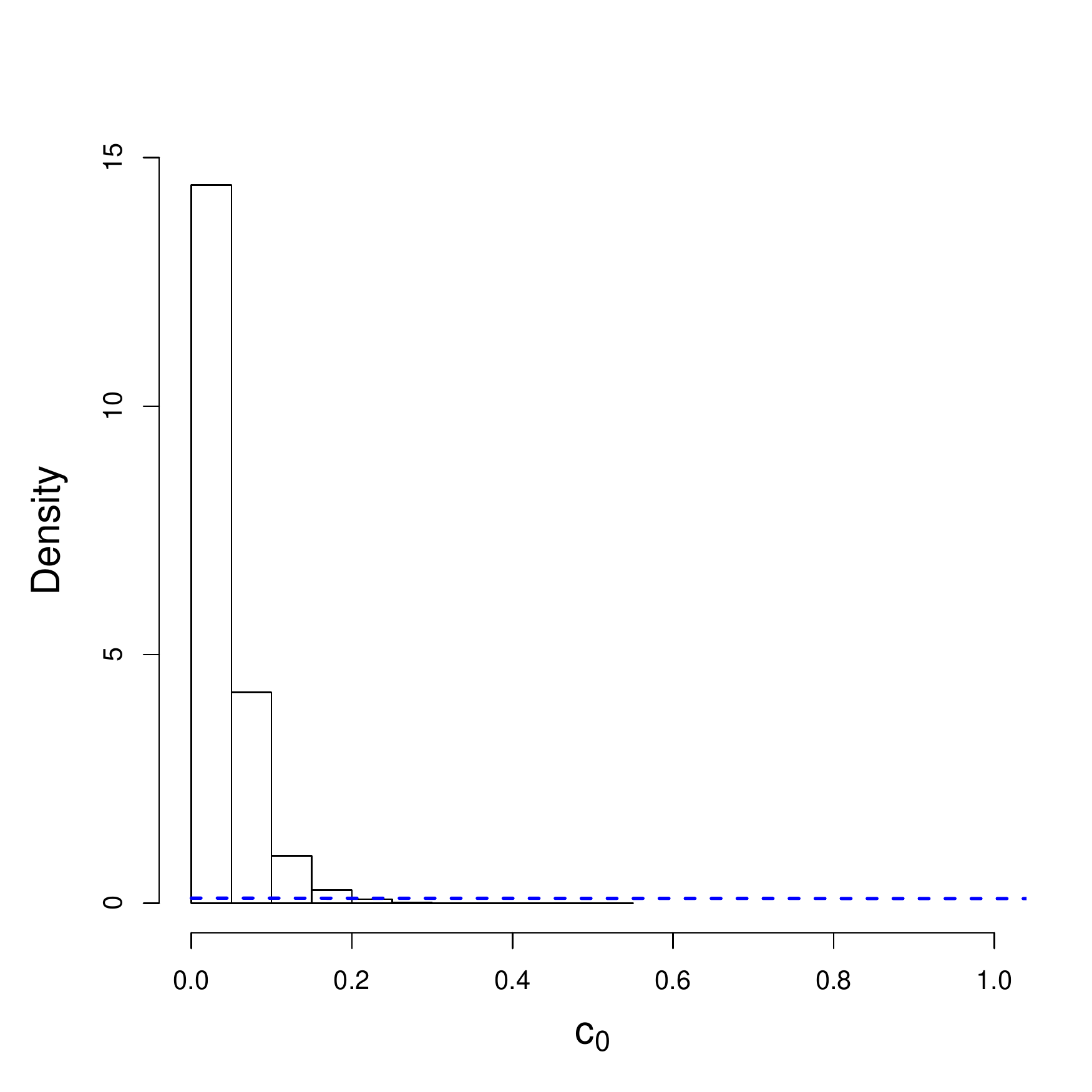}
\includegraphics[width=0.24\textwidth]{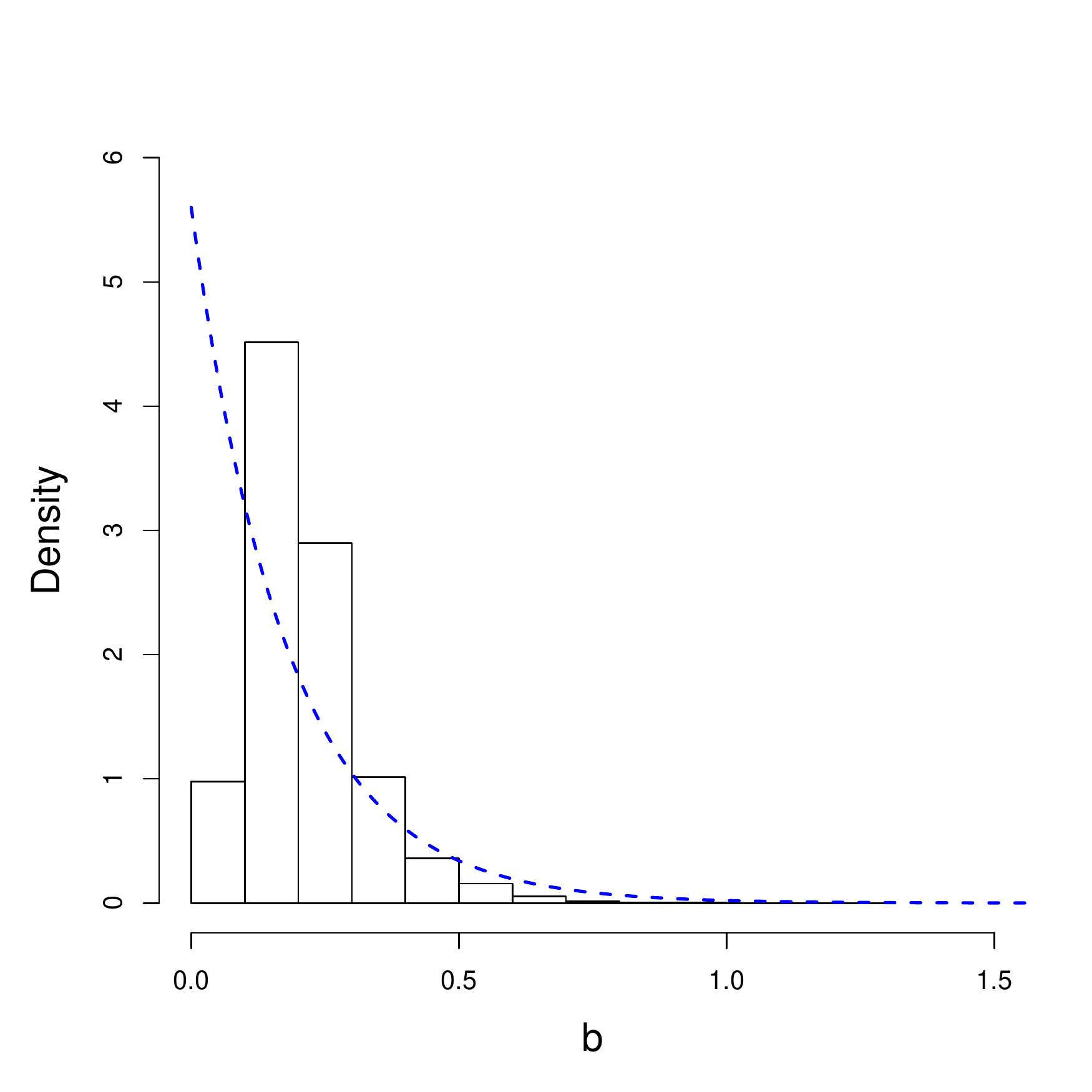}\\
\includegraphics[width=0.24\textwidth]{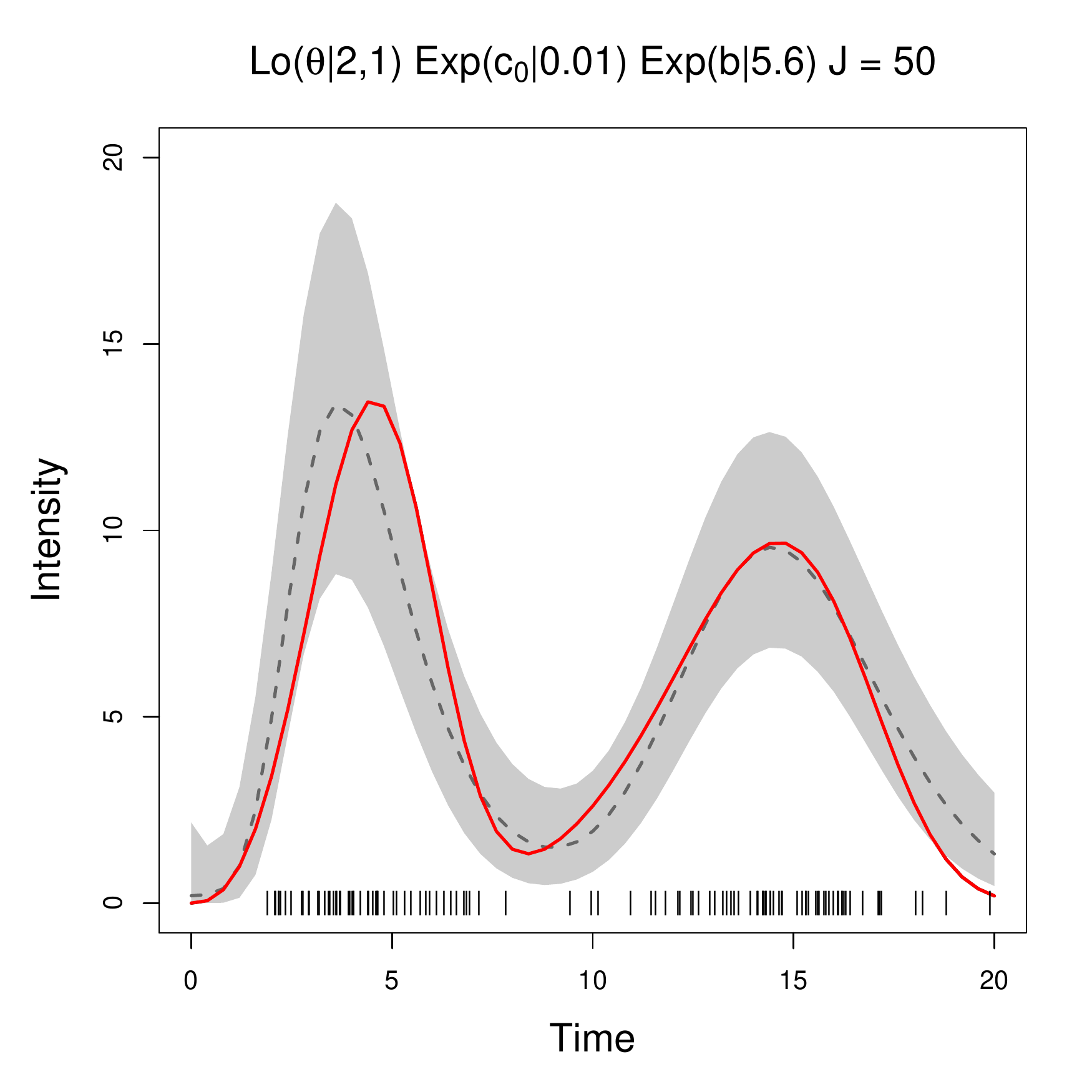}
\includegraphics[width=0.24\textwidth]{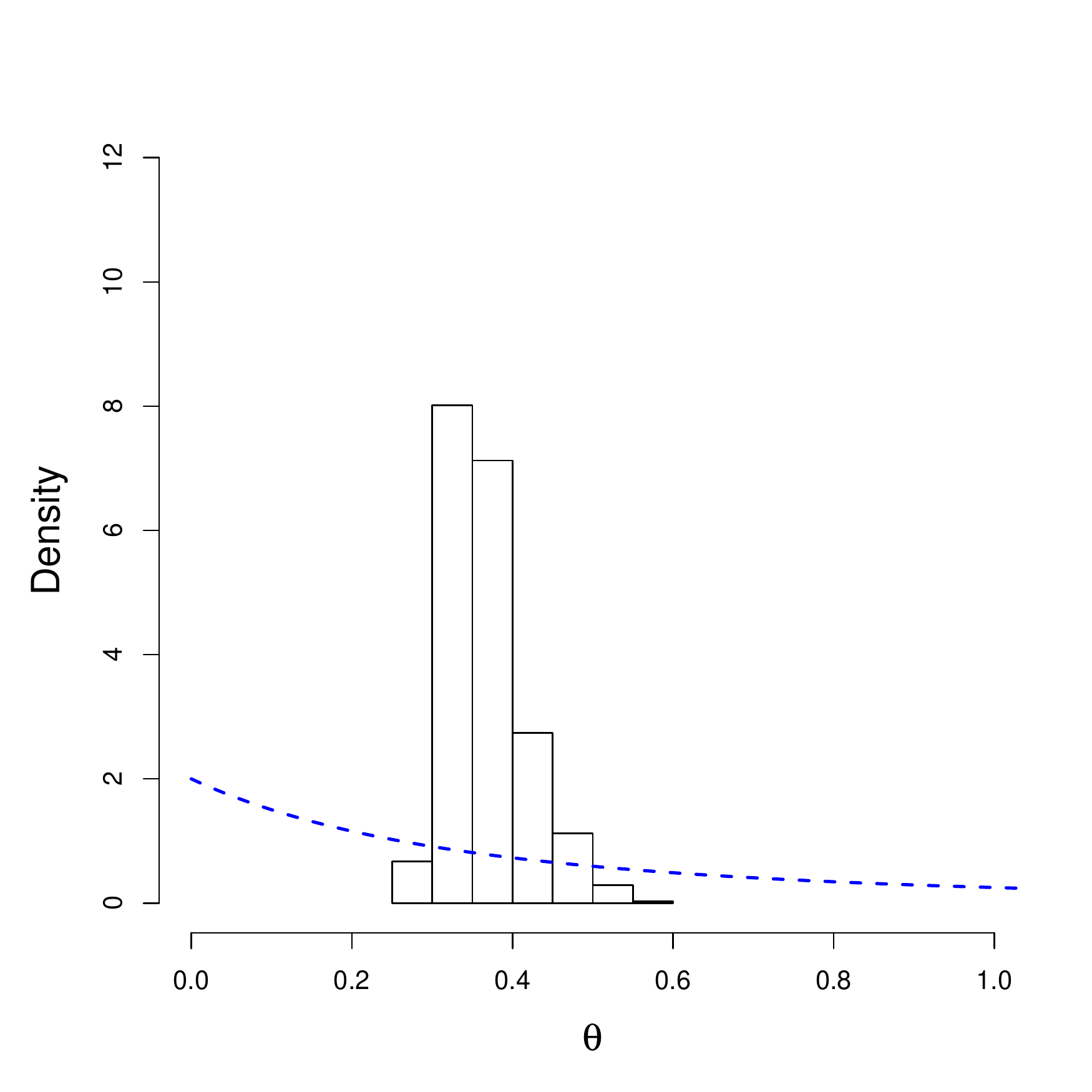}
\includegraphics[width=0.24\textwidth]{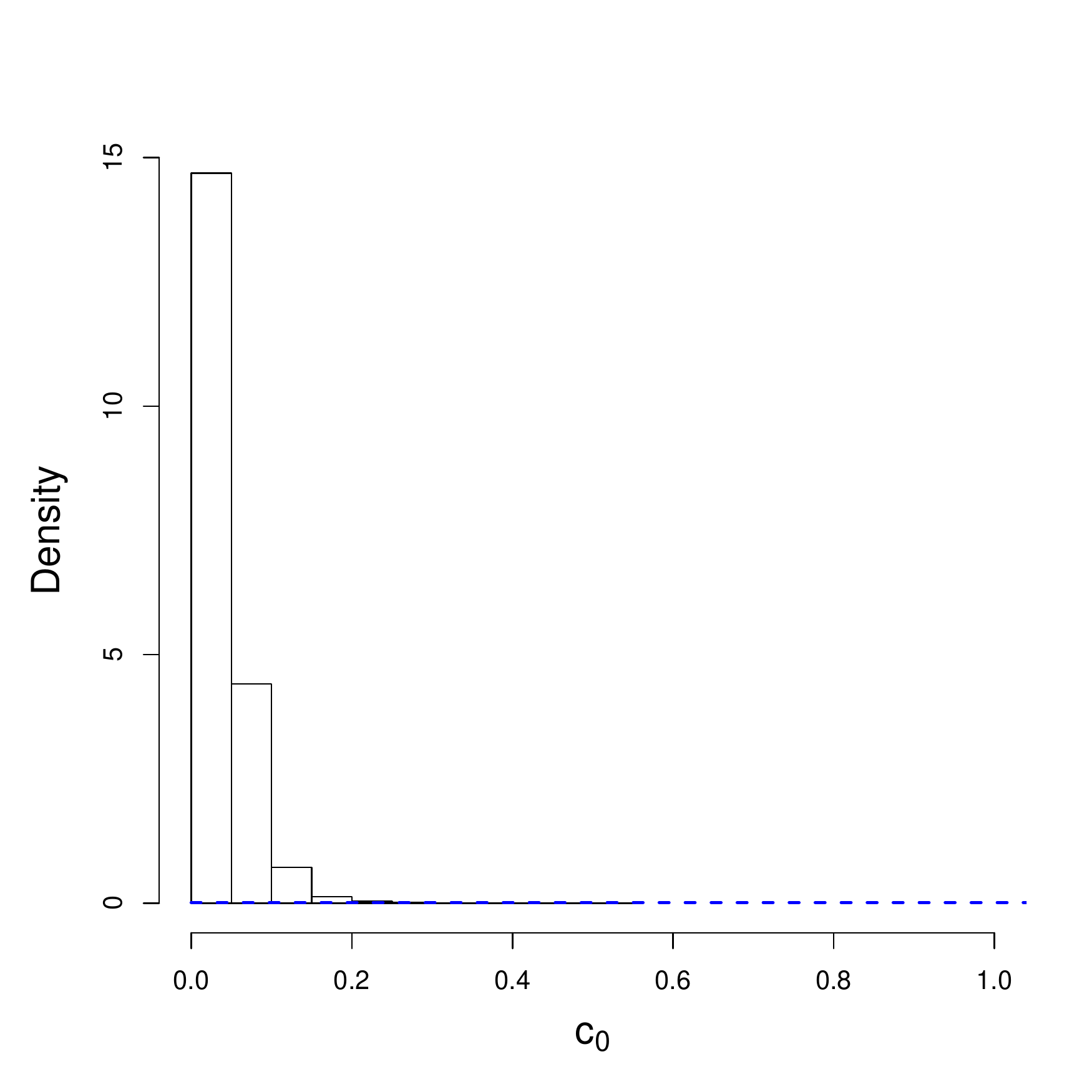}
\includegraphics[width=0.24\textwidth]{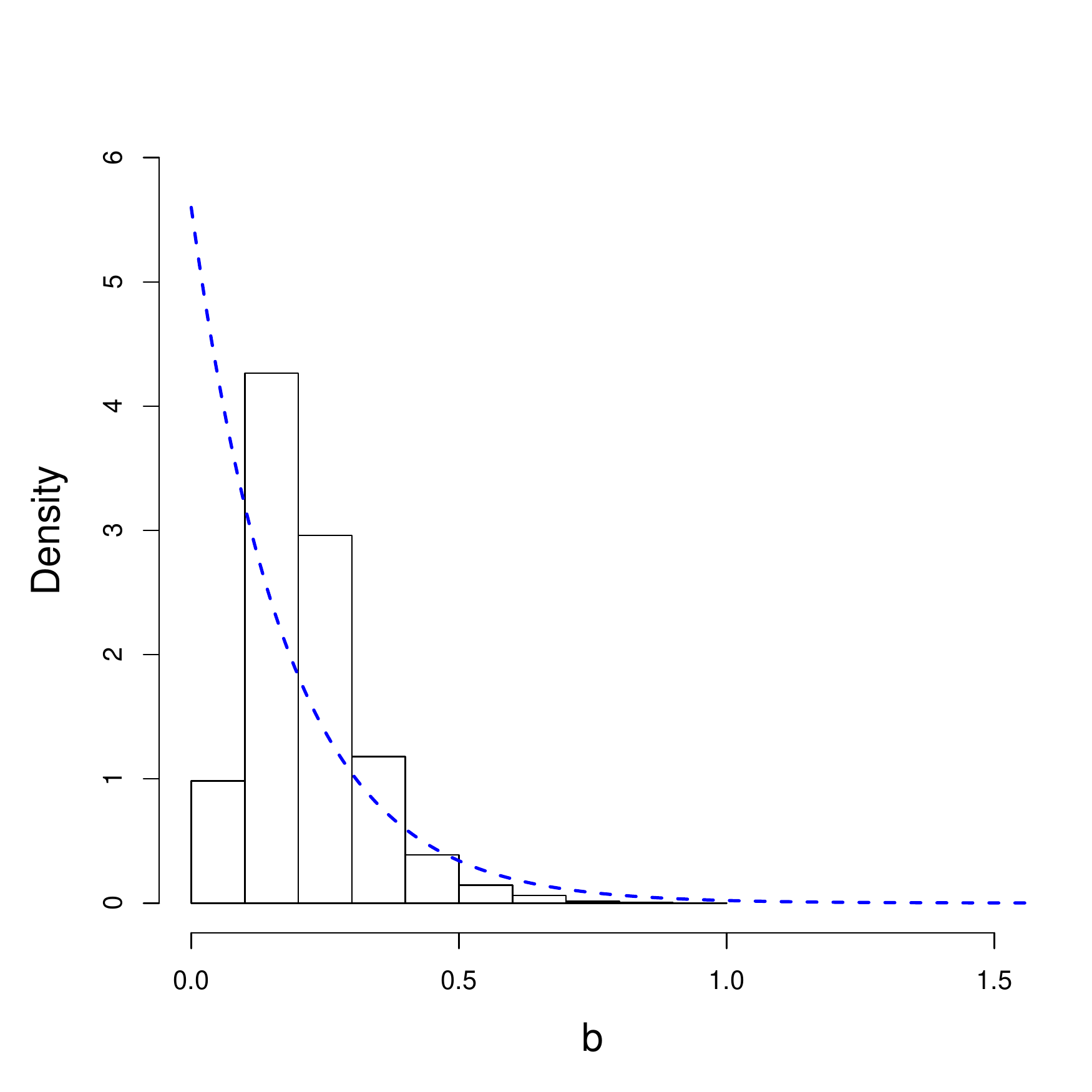}\\
\includegraphics[width=0.24\textwidth]{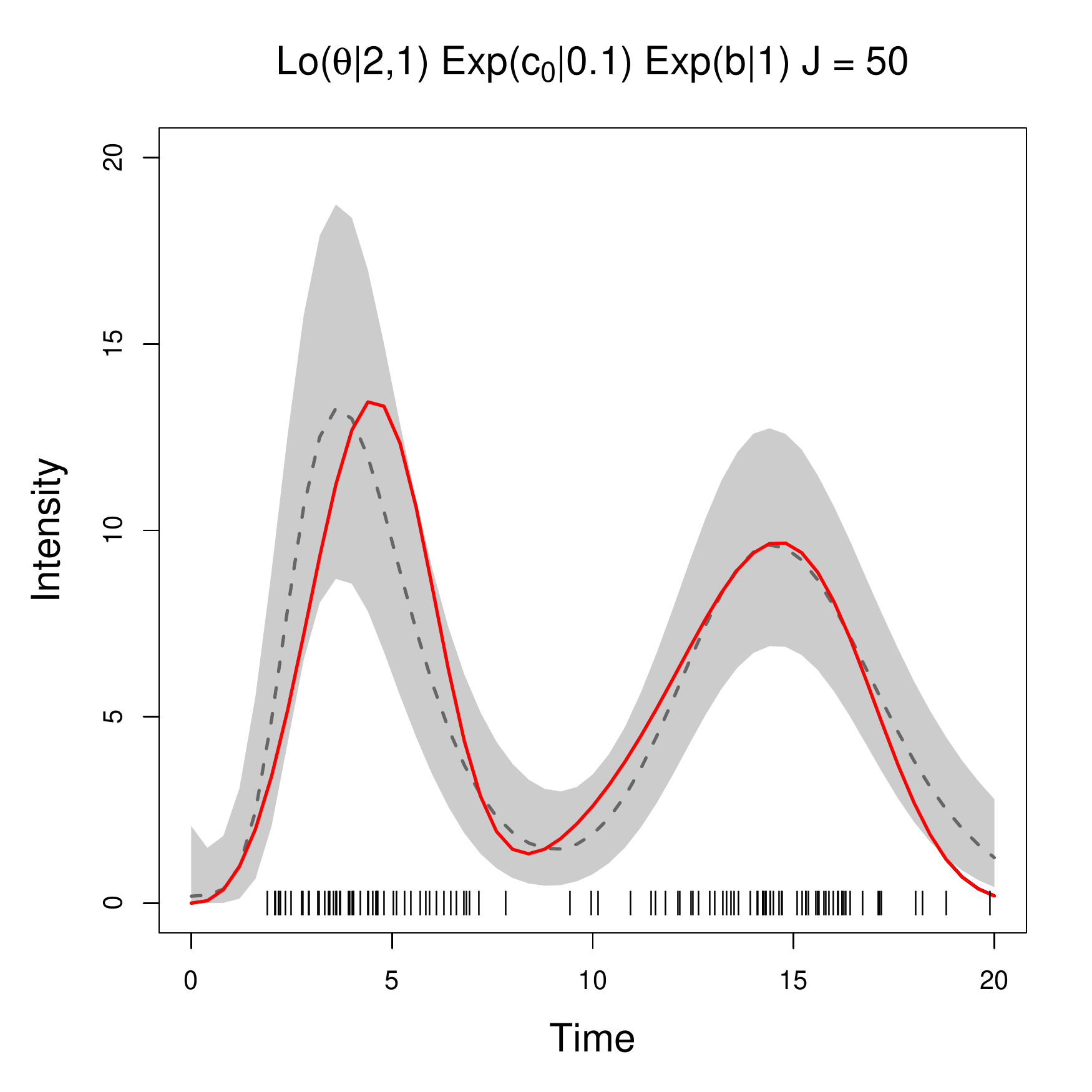}
\includegraphics[width=0.24\textwidth]{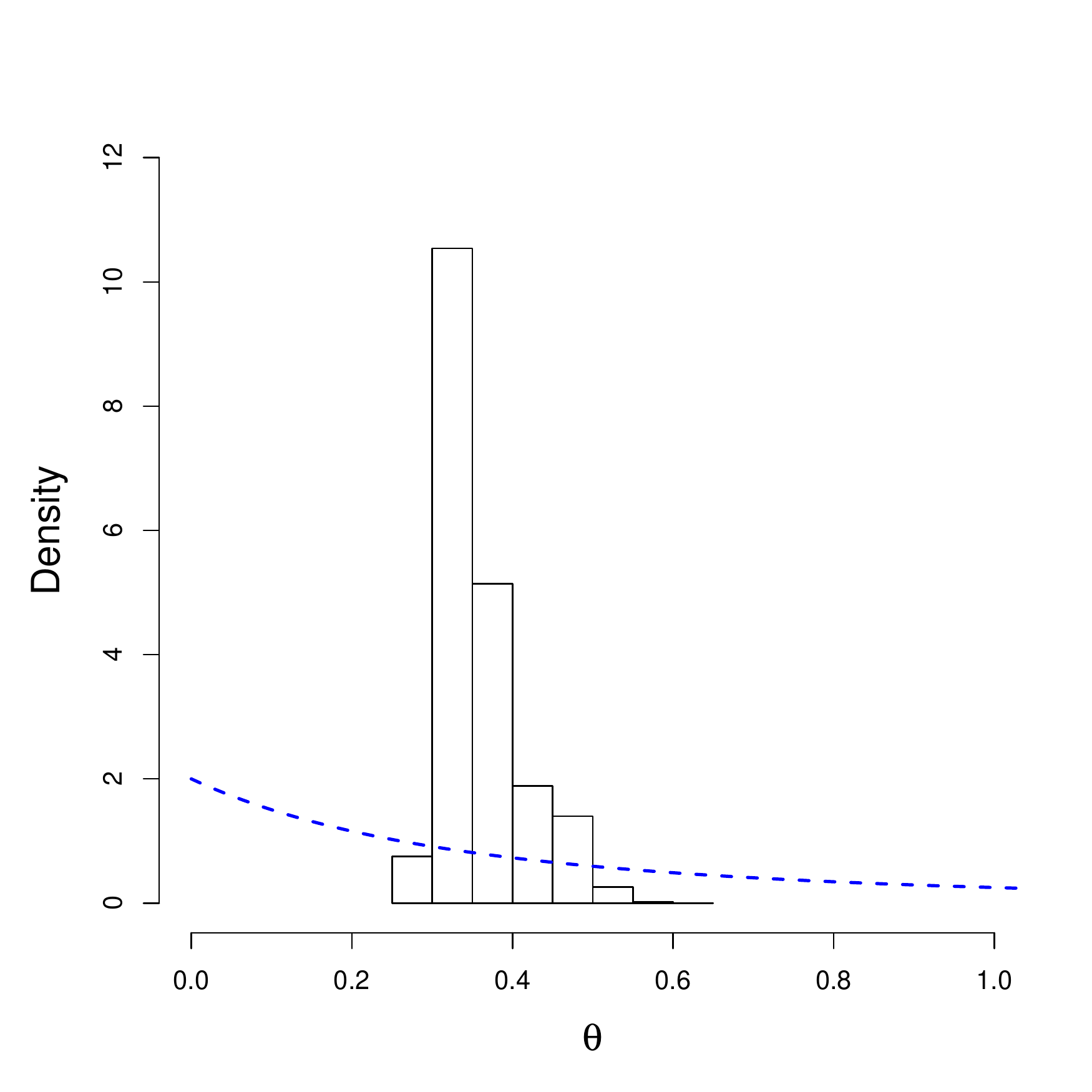}
\includegraphics[width=0.24\textwidth]{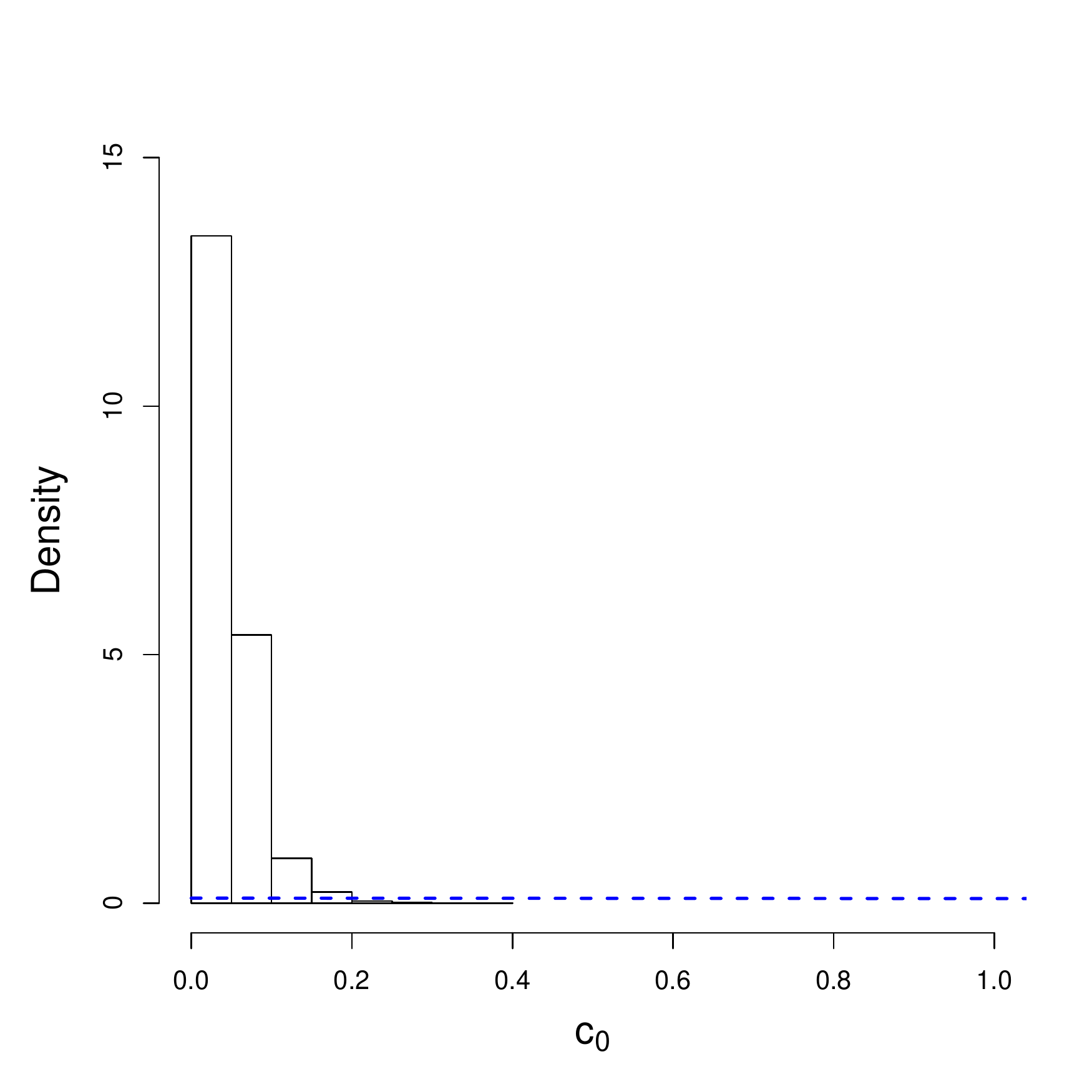}
\includegraphics[width=0.24\textwidth]{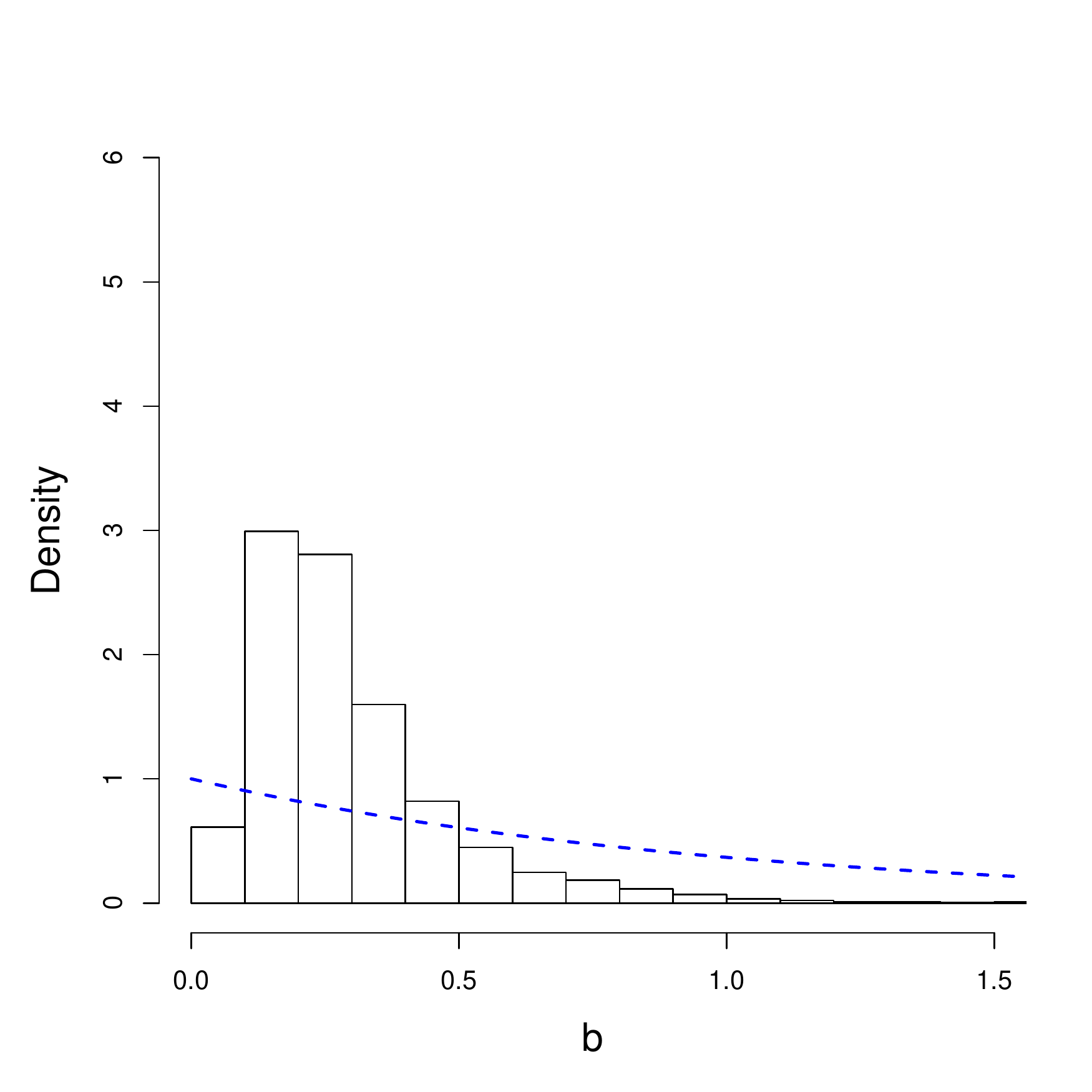}\\
\caption{Synthetic data from temporal NHPP with bimodal intensity (Section 3.3 
of the paper). Erlang mixture model inference results under four different prior 
choices for $\theta$, $c_0$, and $b$. The left column shows the posterior mean 
estimate (dashed line) and posterior 95\% interval bands (shaded area) for the 
intensity function, with the true intensity denoted by the red solid line. 
The other three columns plot histograms of the posterior samples for $\theta$, 
$c_0$, and $b$, along with the corresponding prior densities (blue dashed lines).}
	\label{fig:sensitivity}
\end{figure}

Results under different values for the number of Erlang basis densities ($J=50$ 
and $J=100$) are discussed in Section 3.3 of the paper. Here, we take $J=50$ and study the 
effect of the priors for $\theta$, $c_0$, and $b$. Inference results for the 
intensity function and for the model parameters are reported in Figure \ref{fig:sensitivity}, 
under four different prior choices. The top row corresponds to the prior 
specification approach described in Section 2.2 of the paper, and thus to the 
results reported in Section 3.3 (under $J=50$). For each of the other three 
cases, we change one of the priors for $\theta$, $c_0$, and $b$ relative to
the ``default" prior specification. In particular, results in the second row 
are based on a more dispersed Lomax prior for $\theta$, with scale parameter 
$d_{\theta} = 9$ (instead of $d_{\theta} = 1$), such that 
$\text{Pr}(0 < \theta < T) \approx 0.9$ (instead of 
$\text{Pr}(0 < \theta < T) \approx 0.999$), where $T=20$.
Analogously, the third and fourth rows correspond to more dispersed exponential 
priors for $c_0$ and $b$, respectively. Note that the posterior distribution 
for $\theta$ and $c_0$ is largely unaffected by the change in the dispersion 
of the prior distribution, whereas there is some effect on the tail of the 
posterior density for $b$. Importantly, for a point pattern with a relatively 
small size, posterior estimates for the intensity function are essentially the 
same under the different prior choices.

%
%--------------------------------------------------------------------
%

\subsection{Computational details for the MCMC algorithm}

To report on computing time for implementing the Erlang mixture model, we 
consider the three synthetic data examples of Section 3 of the paper, which 
allows us to study the effect of the point pattern size ($n$) and the number 
of Erlang basis densities ($J$).
Tables \ref{tab:comp_time_ess_bimodal} and \ref{tab:comp_time_ess_others}
include computing times (in minutes) for 70,000 MCMC posterior samples. 
%$60,000$ posterior samples obtained after discarding 
%$10,000$ burn-in samples from a total of $70,000$ MCMC iterations. 
We also provide estimates for the effective sample size (ESS), that is, the 
MCMC sample size adjusted for autocorrelation, thus estimating how many 
uncorrelated samples the posterior samples are equivalent to. The ESS was 
computed using function \textbf{effectiveSize} of the \textbf{R coda} package.
Results for the ESS are based on 60,000 posterior samples obtained after 
discarding the first 10,000 samples. The ``mean ESS" reported in the tables 
is the average of 51 effective sample sizes for $\lambda(t_k)$, for 
$k = 1,\ldots,51$, where $t_k$ are equally-spaced points on a grid over 
$(0,T)$. All MCMC posterior simulations were performed on a laptop with 
an Intel i5-8250U 1.6GHz (8 CPUs) processor.

Increasing $J$ increases the dimension of the parameter space. 
Table \ref{tab:comp_time_ess_bimodal} indicates the corresponding rate of increase 
in computing time, and decrease in mean ESS. Under $J=50$ for the three 
synthetic data examples of Section 3 of the paper, Table \ref{tab:comp_time_ess_others}
shows how computing time increases with the point pattern size. 
%(i.e., also with the size of the observation time window, $(0,T)$). 
Note that there is no evident relationship between mean ESS and the point pattern size.

\begin{table}[t!]
	\begin{tabular*}{\hsize}{@{\extracolsep{\fill}}cccc}
		\hline
		\multicolumn{1}{c}{} & \multicolumn{1}{c}{$J = 50$} & \multicolumn{1}{c}{$J = 75$} & \multicolumn{1}{c}{$J = 100$}\\		
		\hline
		\hline\\
		[-8pt]
		Computing Time  	&	19.6	&	25.8     &	33.9\\
		[5pt]
		Mean ESS	            &	3391	&	3065	&	2984\\
		[5pt]									
		\hline\\
	\end{tabular*}
\caption{Synthetic data from temporal NHPP with bimodal intensity (Section 3.3 
of the paper). Computing time (in minutes) for 70,000 MCMC iterations, and average 
of effective sample sizes for the intensity evaluated at 51 grid points in $(0,T)$,
under three different values of $J$.}
\label{tab:comp_time_ess_bimodal}
\end{table}

\begin{table}[t!]
	\begin{tabular*}{\hsize}{@{\extracolsep{\fill}}cccc}
		\hline
		\multicolumn{1}{c}{} & \multicolumn{1}{c}{Bimodal $(n = 112)$} &
		\multicolumn{1}{c}{Decreasing $(n = 491)$} & 
		\multicolumn{1}{c}{Increasing $(n = 565)$}\\		
		\hline
		\hline\\
		[-8pt]
		Computing Time    	& 19.6 &	28.5	&	29.5\\
		[5pt]
		Mean ESS	            & 3391 &	5792	&	2122\\
		[5pt]
		\hline\\
	\end{tabular*}
\caption{Synthetic data from temporal NHPP with bimodal/decreasing/increasing 
intensity (Section 3 of the paper). Computing time (in minutes) for 70,000 MCMC iterations, and average of effective sample sizes for the intensity evaluated 
at 51 grid points in $(0,T)$, under $J = 50$ for all three data examples.}
\label{tab:comp_time_ess_others}
\end{table}

\begin{figure}[t!]
\centering
\includegraphics[width=0.45\textwidth]{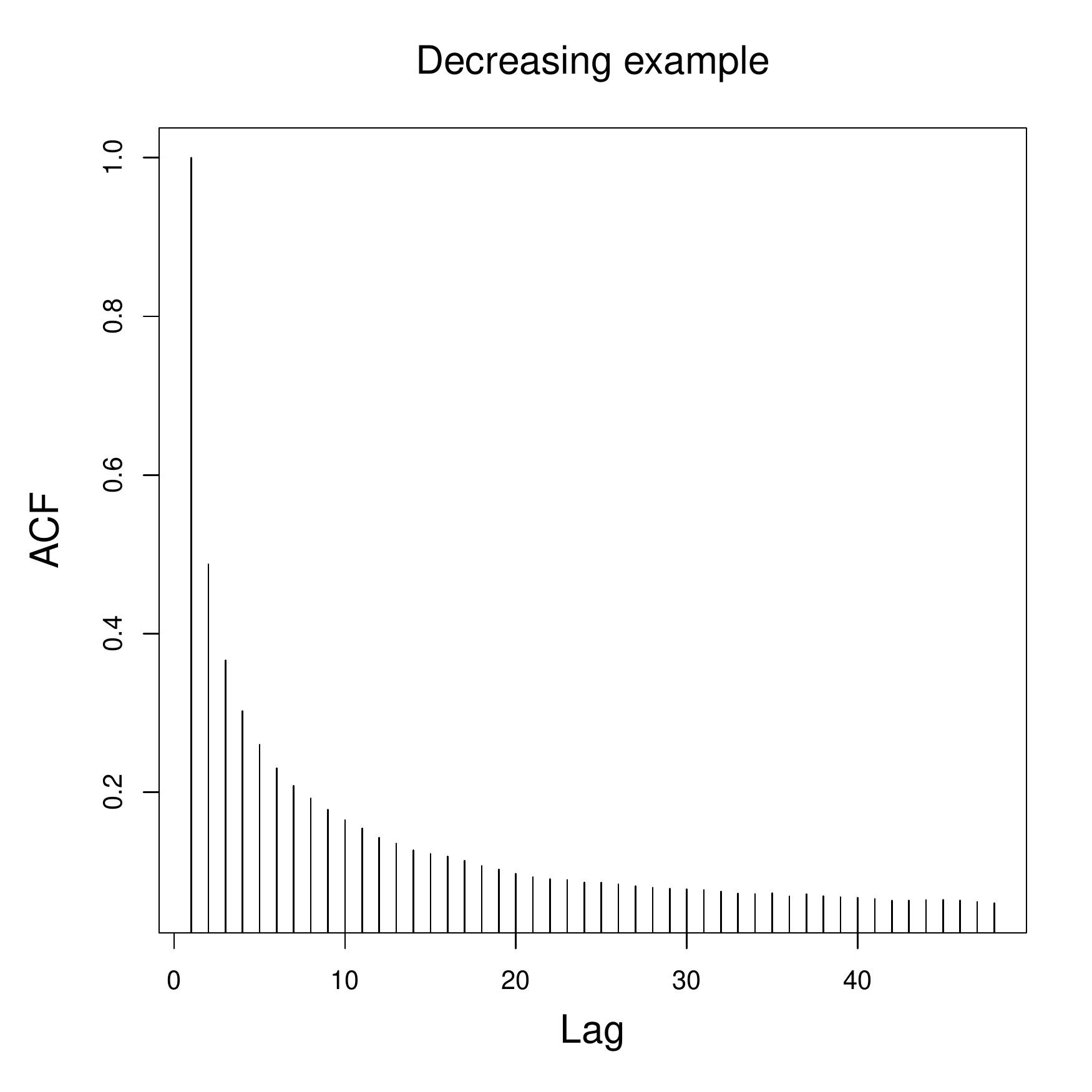}
\includegraphics[width=0.45\textwidth]{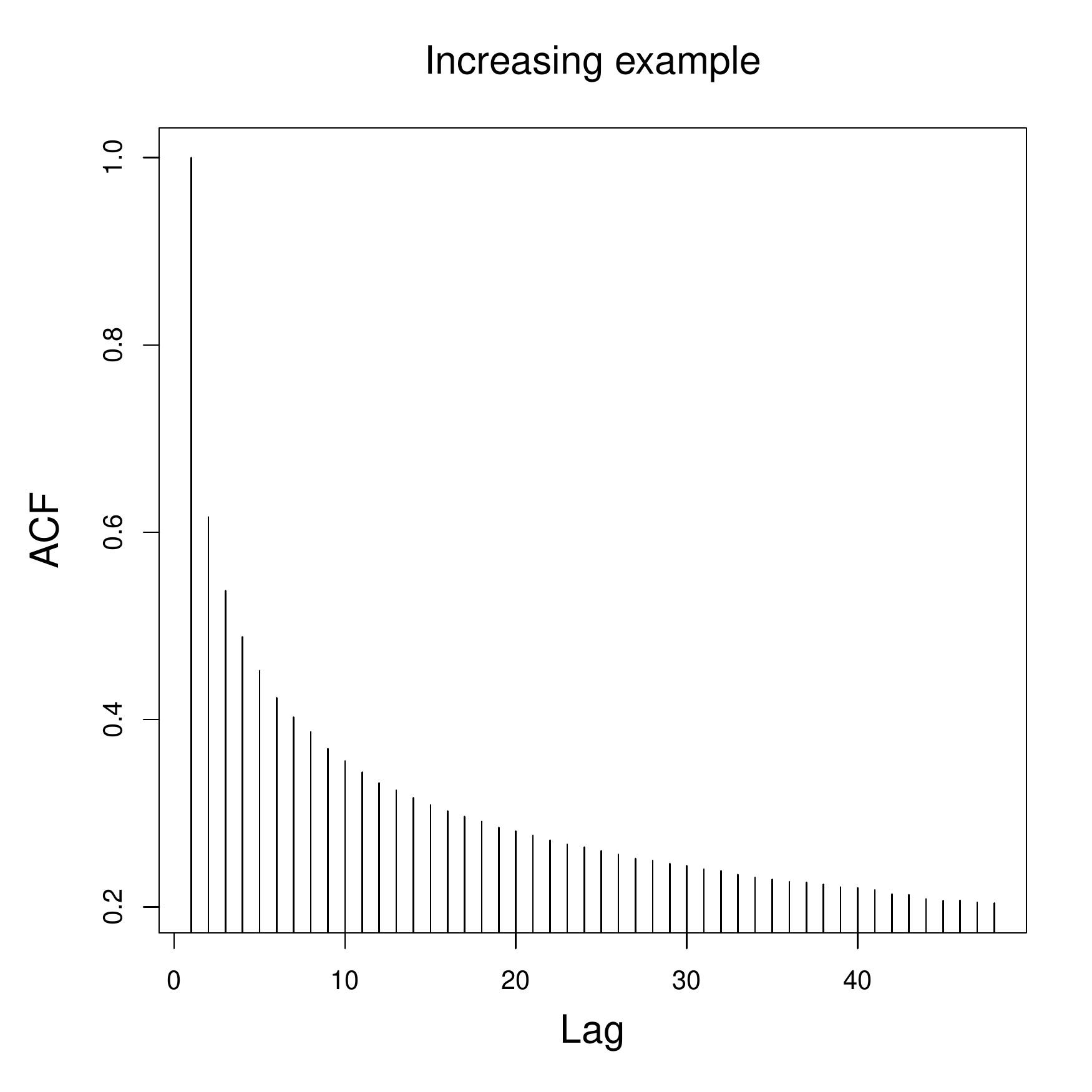}\\
\caption{Synthetic data from temporal NHPP with decreasing/increasing intensity 
(Sections 3.1 and 3.2 of the paper). Average of autocorrelation functions for the 
intensity evaluated at 51 grid points in $(0,T)$.}
	\label{fig:acf}
\end{figure}

Figure \ref{fig:acf} provides an illustration of autocorrelation in the MCMC posterior 
samples. Plotted for the synthetic data of Section 3.1 and 3.2 of the paper are averages 
of autocorrelation functions (at 48 lags) for the intensity function evaluated at the 
51 equally-spaced grid points in $(0,T)$. The difference in the autocorrelations 
in the two panels of Figure \ref{fig:acf} is compatible with the corresponding 
mean ESS given in Table \ref{tab:comp_time_ess_others}.

\begin{figure}[t!]
\centering
\includegraphics[width=12.5cm,height=9.5cm]{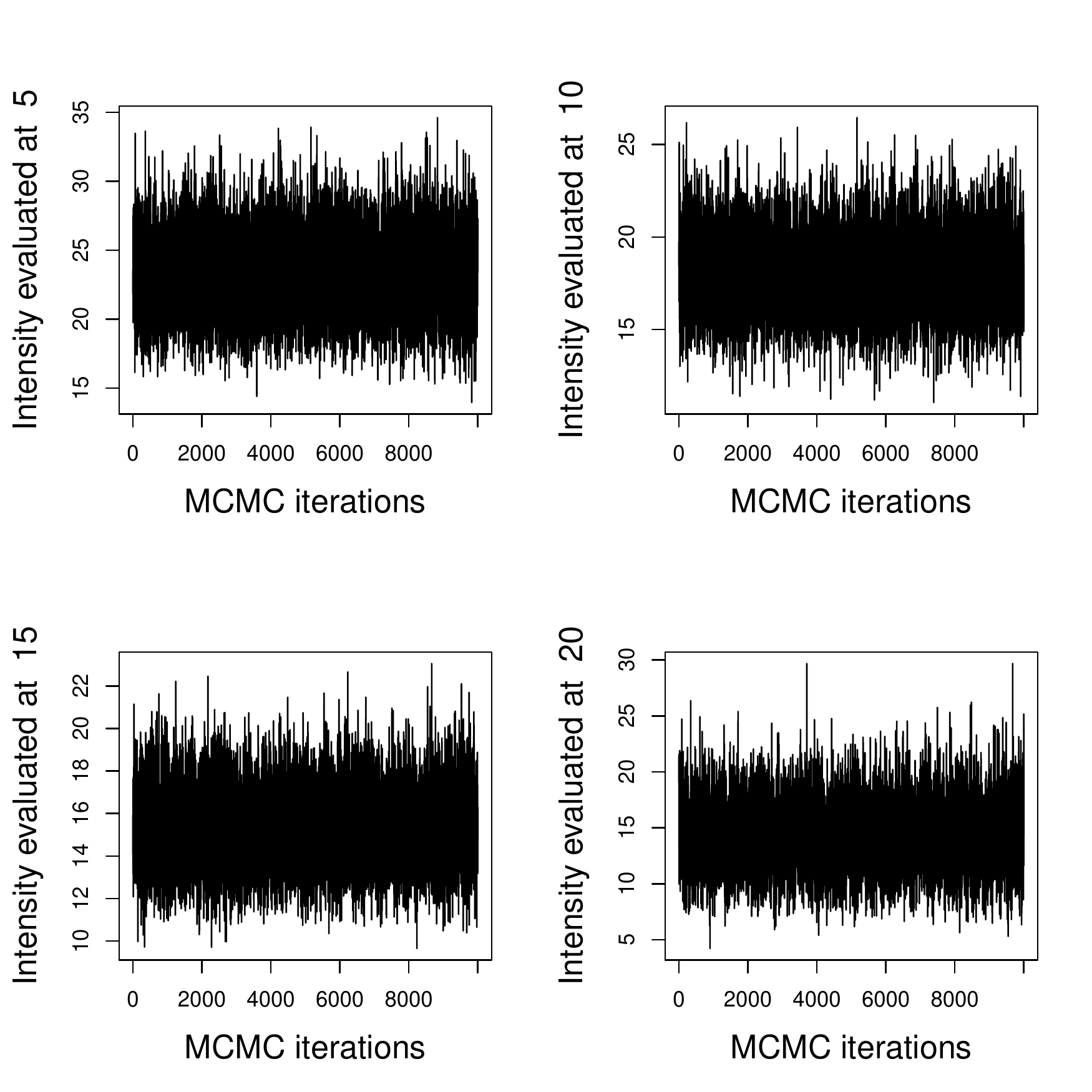}
\caption{Synthetic data from temporal NHPP with decreasing intensity (Section 
3.1 of the paper). Trace plots of posterior samples for the intensity function 
evaluated at time points $t=5$, $10$, $15$, $20$.}
	  \label{fig:dec_tra}
\end{figure}

Convergence and mixing of the MCMC algorithm can also be assessed graphically through 
trace plots of the intensity function evaluated at specific time points within the 
observation window. An example is given in Figure \ref{fig:dec_tra} for the synthetic 
data of Section 3.1 of the paper. The plots in Figure \ref{fig:dec_tra} are representative 
of intensity trace plots obtained for all other data examples. 

\section{Comparison with Gaussian process based models}

Here, we compare our model with Bayesian nonparametric models based 
on Gaussian process (GP) priors for logarithmic or logit transformations of the 
NHPP intensity \cite[e.g.,][]{Moller_et_al.:1998,Brix_Diggle:2001,AMM2009}.
Such models are popular in both the statistics and the machine learning literature. 
For the temporal case, Section \ref{temporal} considers comparison with the 
sigmoidal Gaussian Cox process (SGCP) model (\citeh{AMM2009}). Since we were
not able to find publicly available software, results are based on our implementation 
of the SGCP model, which allows for a detailed comparison involving full inference 
results and computational efficiency. Software in the form of \textbf{R} packages is 
available for spatial and spatio-temporal log-Gaussian Cox process (LGCP) models 
(\citeh{BT2005}; \citeh{TDRD2013}), although its output is limited in terms of 
inferences that are of interest in our setting. Therefore, for the spatial NHPP case, 
Section \ref{spatial} focuses on graphical comparison of point estimates for the intensity 
surface in the context of the synthetic data considered in Section 4.2 of the paper.
  
%
%Many of the Bayesian nonparametric modeling 
%approaches discussed in Section 1, including our model, are of a mixture form with taking 
%stochastic process priors for the mixture weights. On the other hand, the GP-based models 
%directly map the Gaussian process onto transformed intensity functions. We expect this 
%structural dissimilarity between our model and the GP-based competitors can bring about 
%a discernible difference in posterior inference.
%

\subsection{Temporal Poisson process models}
\label{temporal}

Under the SGCP model, the temporal NHPP intensity is represented as $\lambda(t) =$ 
$\lambda^\ast \, \sigma(g(t))$, where $\lambda^\ast$ is an upper bound on the intensity 
function, $\sigma(z) =$ $(1+e^{-z})^{-1}$ is the logistic function, and $g$ is a real-valued 
random function assigned a GP prior. \cite{AMM2009} develop MCMC posterior simulation using 
latent variables (associated with ``thinned'' events) to handle the intractable NHPP 
likelihood normalizing term.

%
%Posterior inference with the standard MCMC methods is not feasible because of an intractable 
%integral in the normalizing constant term in the likelihood under the model. \cite{AMM2009} 
%addresses the issue via introducing latent variables. The augmented likelihood does not require 
%the integral of the stochastic process in the normalizing constant, facilitating the posterior 
%inference through the standard MCMC methods. 
%

\begin{figure}[t!]
\centering
\includegraphics[width=0.325\textwidth]{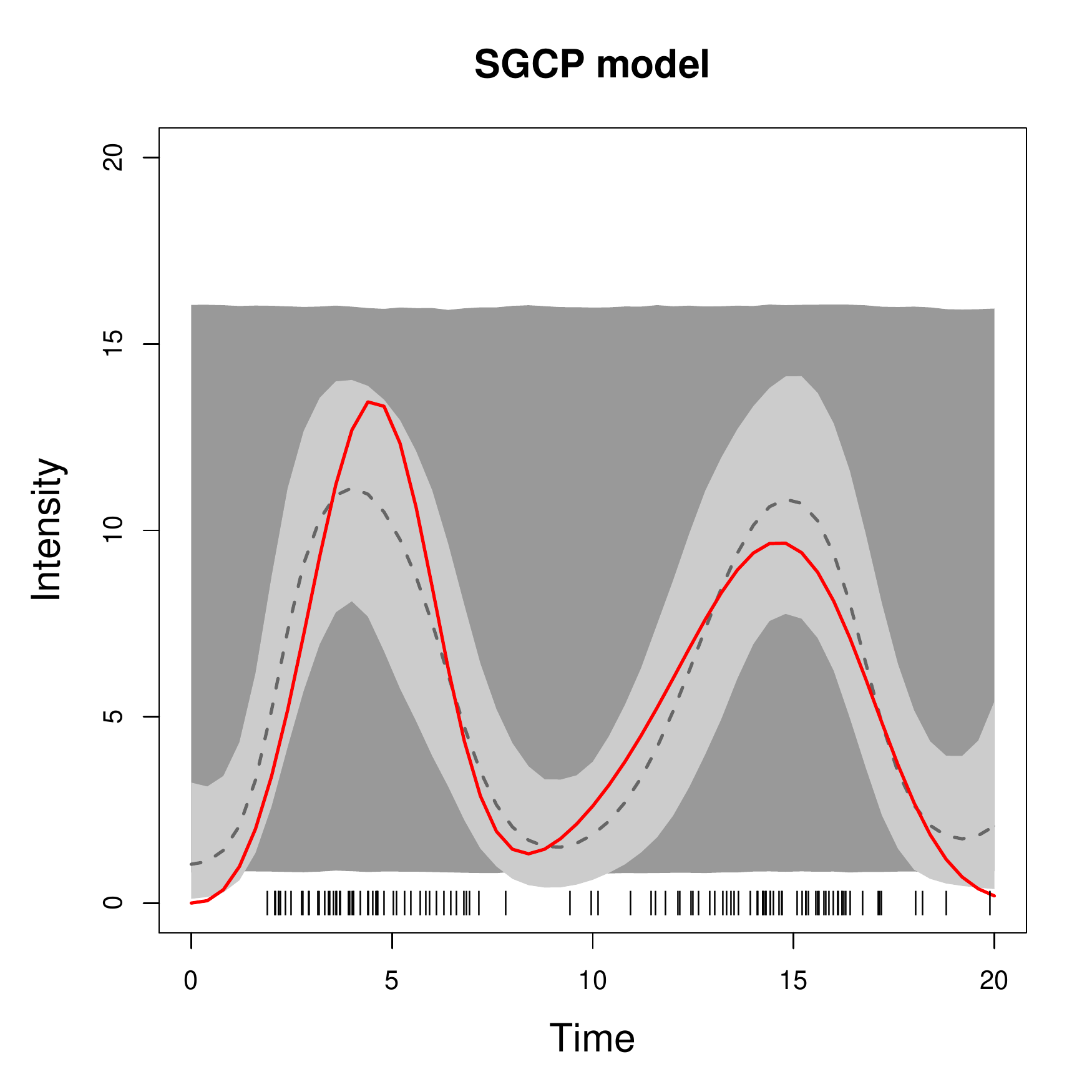}
\includegraphics[width=0.325\textwidth]{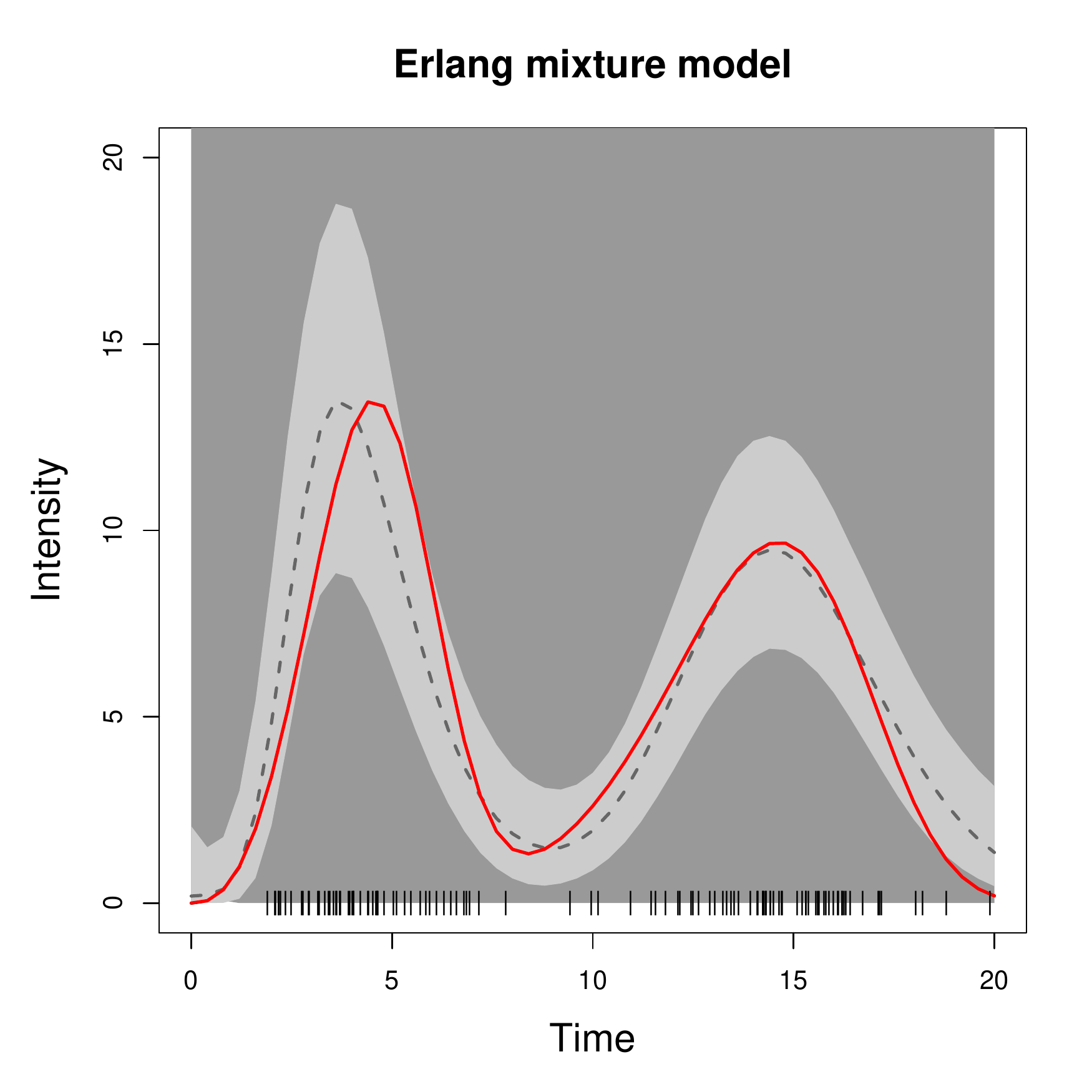}
\includegraphics[width=0.325\textwidth]{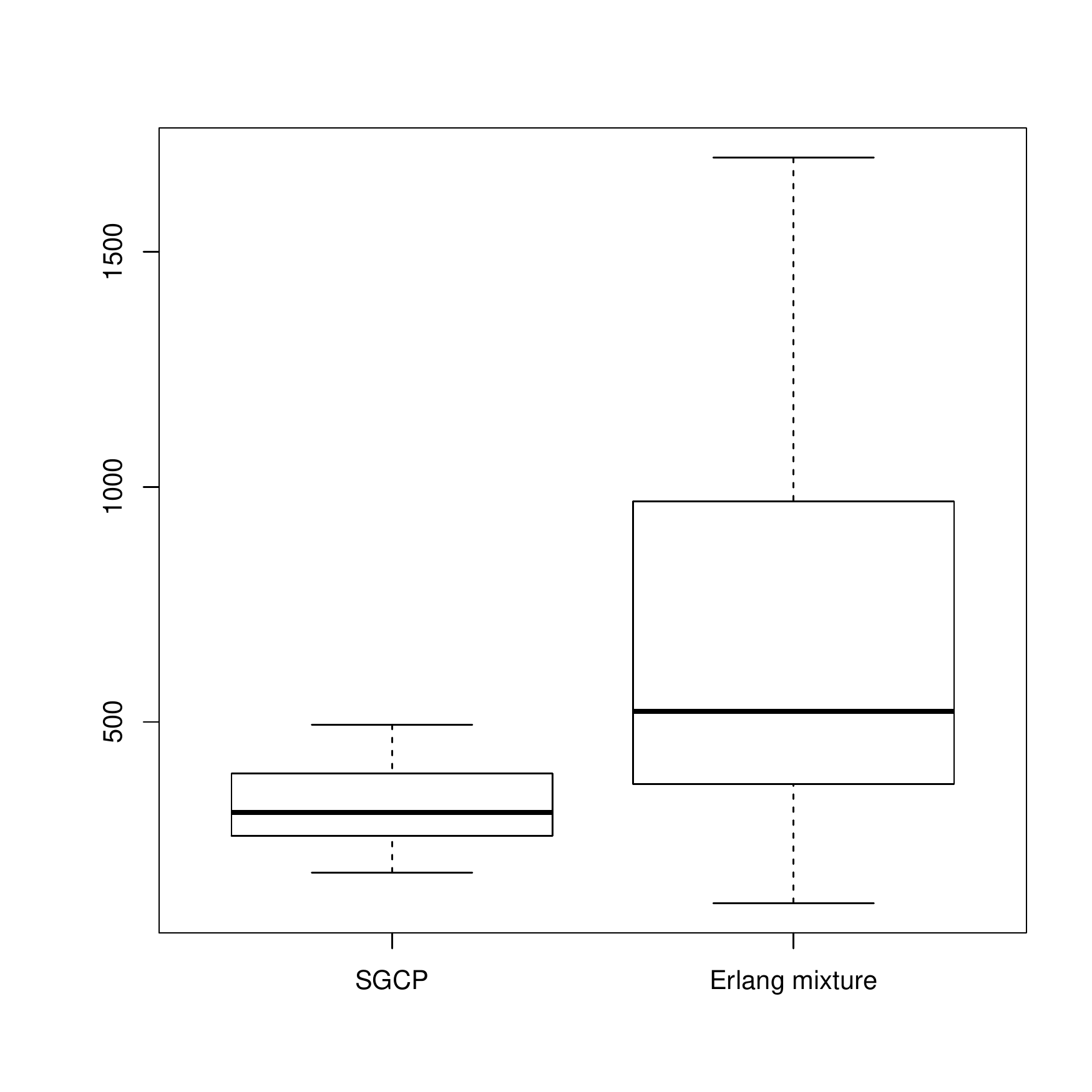}\\
\caption{Synthetic data from temporal NHPP with bimodal intensity (Section 3.3 of the paper). 
The left and middle panels provide estimates for the underlying intensity (red solid line) 
under the SGCP and Erlang mixture model, respectively: prior 95\% interval bands (dark-gray 
shaded area), posterior mean (dashed line), and posterior 95\% interval bands (light-gray 
shaded area). The right panel shows for each model boxplots of 51 ESS values based on 
posterior samples for the intensity at 51 equally-spaced time points.}
	\label{fig:simulation_temporal_nhpp}
\end{figure}

To compare the Erlang mixture and SGCP models, we work with the synthetic data set of 
Section 3.3 of the paper. This is a choice arising from practical considerations; as 
discussed below, the SGCP model is particularly computationally intensive, and we thus 
consider the data example with the smallest point pattern size.

We applied the SGCP model using a GP prior for function $g$ with constant mean, $\mu$, 
and squared-exponential covariance function, $\text{cov}(g(t_1),g(t_2))=$ 
$\theta_1e^{-\theta_2(t_1-t_2)^2}$.
Therefore, the SGCP model parameters comprise $\lambda^\ast$ and $(\mu,\theta_1,\theta_2)$, all 
four of which are assigned priors. We note that a prior specification strategy is not provided in 
\cite{AMM2009}. We observed that, at least for this data set, inference for all SGCP model 
parameters is sensitive to the prior choice. Moreover, there is a conflict regarding the 
role of parameter $\lambda^\ast$: it controls the extent of prior uncertainty, with larger 
$\lambda^\ast$ values resulting in wider prior interval bands for the intensity, but at
the same time, increasing $\lambda^\ast$ increases the number of latent variables and thus 
also the MCMC algorithm computing time.

We tuned the priors for the SGCP model parameters to obtain the best possible estimates for 
the intensity function. The resulting estimates are reported in the left panel of Figure \ref{fig:simulation_temporal_nhpp}.
%the corresponding posterior (and prior) densities for the model hyperparameters can be 
%found in the left column of Figure \ref{fig:simulation_temporal_nhpp_sensitivity}. 
Even though we intentionally favored the SGCP model
through the prior selection, the Erlang mixture model posterior mean estimate captures the peaks 
of the underlying intensity more accurately than the SGCP model, indeed, under larger levels of 
prior uncertainty (middle panel of Figure \ref{fig:simulation_temporal_nhpp}).

The right panel of Figure \ref{fig:simulation_temporal_nhpp} shows boxplots of 51 ESS values 
computed from the posterior samples for $\lambda(t)$ at 51 equally-spaced time points on a 
grid over $(0,T) = (0,20)$. The ESS values are based on 15,000 posterior samples, obtained 
after discarding 5,000 burn-in samples. Evidently, the Erlang mixture model outperforms the 
SGCP model in terms of mixing of the MCMC algorithm as measured by the ESS. The benefit is 
more emphatic considering computing times: completing the 20,000 MCMC iterations took around 
310 minutes under the SGCP model, while the corresponding computing time for the Erlang mixture 
model was about 5 minutes.

\subsection{Spatial Poisson process models}
\label{spatial}

%The intensity function under the LGCP model is defined as $\lambda(s)=$
%$\tilde{\lambda}(s)\exp(g(s))$, where the real-valued function $g$ is assigned a GP prior 
%with covariance function $\text{cov}(g(s_1),g(s_2)) = \sigma^2 r(||s_2-s_1||;\phi)$ with a suitable norm $|| %\cdot ||$ on $\mathbb{R}^2$ and the isotropic correlation function $r$ with the positive scalars $\sigma$, %$\phi > 0$. The mean of $g$ is set to be $-\sigma^2/2$, which implies $\text{E}(\exp(g)) = 1$ and so %$\text{E}(\lambda(s)) = \tilde{\lambda}(s)$.

The LGCP model for spatial NHPP intensities can be expressed as $\lambda(\bm{s})=$
$\tilde{\lambda}(\bm{s})\exp(g(\bm{s}))$, where function $g$ is assigned a GP prior with 
isotropic correlation function, and variance $\sigma^{2}$. The correlation function includes 
parameter $\phi > 0$, which controls the rate at which correlation decreases with distance, 
and it may contain additional parameters (as in, e.g., the Mat\'{e}rn case). The mean of the 
GP prior is set to $-\sigma^2/2$, which implies $\text{E}(\exp(g(\bm{s}))) = 1$, and thus $\text{E}(\lambda(\bm{s}))=$ $\tilde{\lambda}(\bm{s})$.

To obtain point estimates for the spatial intensity function under the LGCP model, one can use 
two \textbf{R} packages: \textbf{spatstat} \citep{BT2005} for parameter estimation, 
and \textbf{lgcp} \citep{TDRD2013} for intensity estimation. 

The function \textbf{lgcpPredictSpatial} of the \textbf{lgcp} package performs posterior 
inference for the intensity, obtaining posterior samples for a discretized version of 
function $g$ through Metropolis-adjusted Langevin algorithms (\citeh{TD2014}). However, to 
use \textbf{lgcpPredictSpatial}, values for $\sigma^2$ and $\phi$, and function 
$\tilde{\lambda}(\bm{s})$ need to be provided. Since the \textbf{lgcp} package does not contain 
any functions for inference about the LGCP model hyperparameters, we use the \textbf{spatstat} 
package, which provides non-Bayesian estimates for $\sigma^2$, $\phi$, and $\tilde{\lambda}(\bm{s})$.
In particular, \textbf{spatstat} function \textbf{density.ppp} computes a nonparametric kernel 
intensity estimator, which can be used to estimate $\tilde{\lambda}(\bm{s})$. Moreover, function \textbf{lgcp.estpcf} yields estimates for $\sigma^2$ and $\phi$ through a non-parametric kernel 
estimator for the pair correlation function.

Evidently, the approach described above is not fully Bayesian. And, since it involves a two-stage
estimation procedure, comparison with the Erlang mixture model in terms of uncertainty estimates
is not particularly meaningful. We thus consider graphical comparison of the Erlang mixture model
posterior mean intensity estimate with the point estimates obtained from the LGCP model under 
two different GP correlation functions. The comparison is based on the synthetic data example
presented in Section 4.2 of the paper, for which we have the true underlying intensity as the 
point of reference.

\begin{figure}[t!]
\centering
\includegraphics[width=0.45\textwidth]{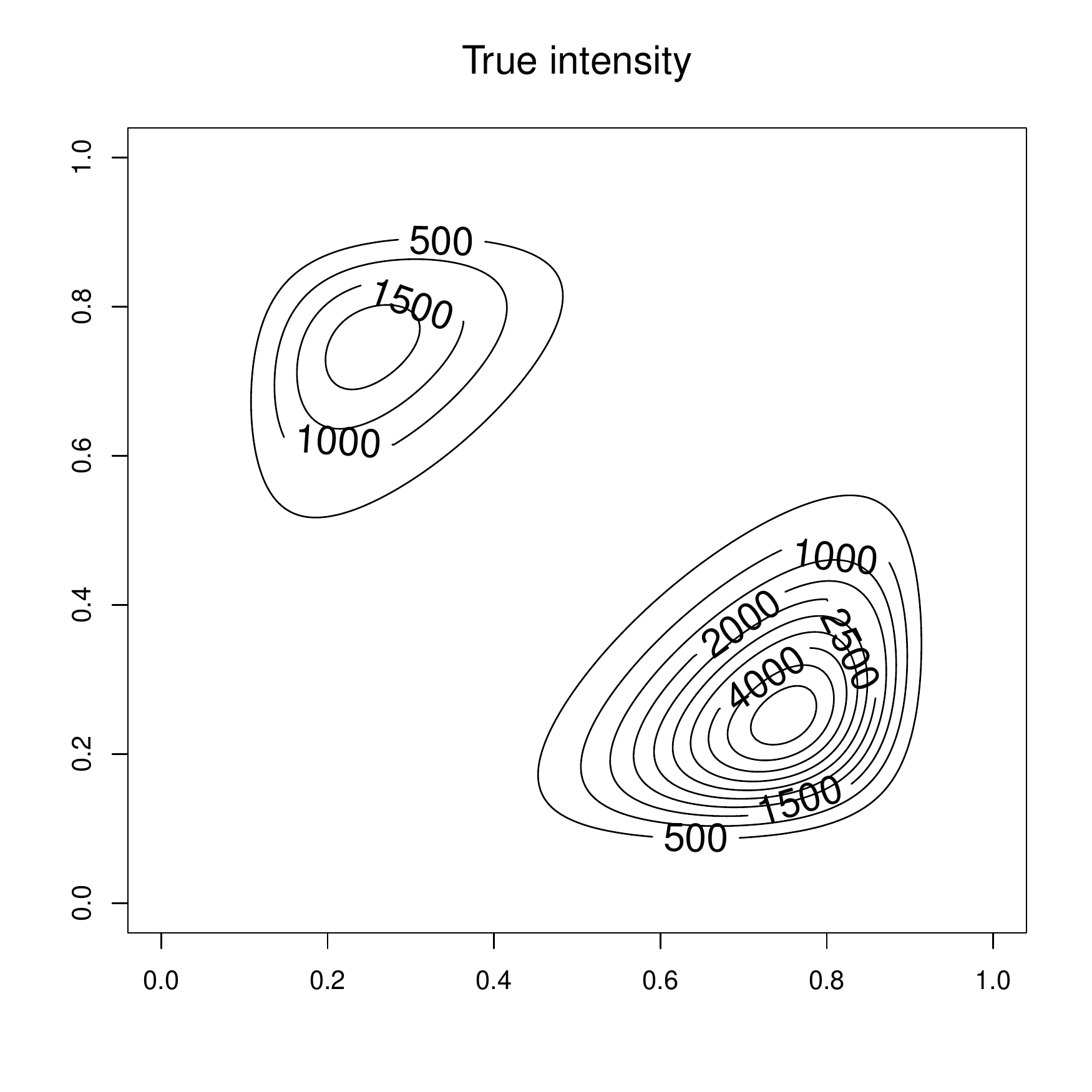}
\includegraphics[width=0.45\textwidth]{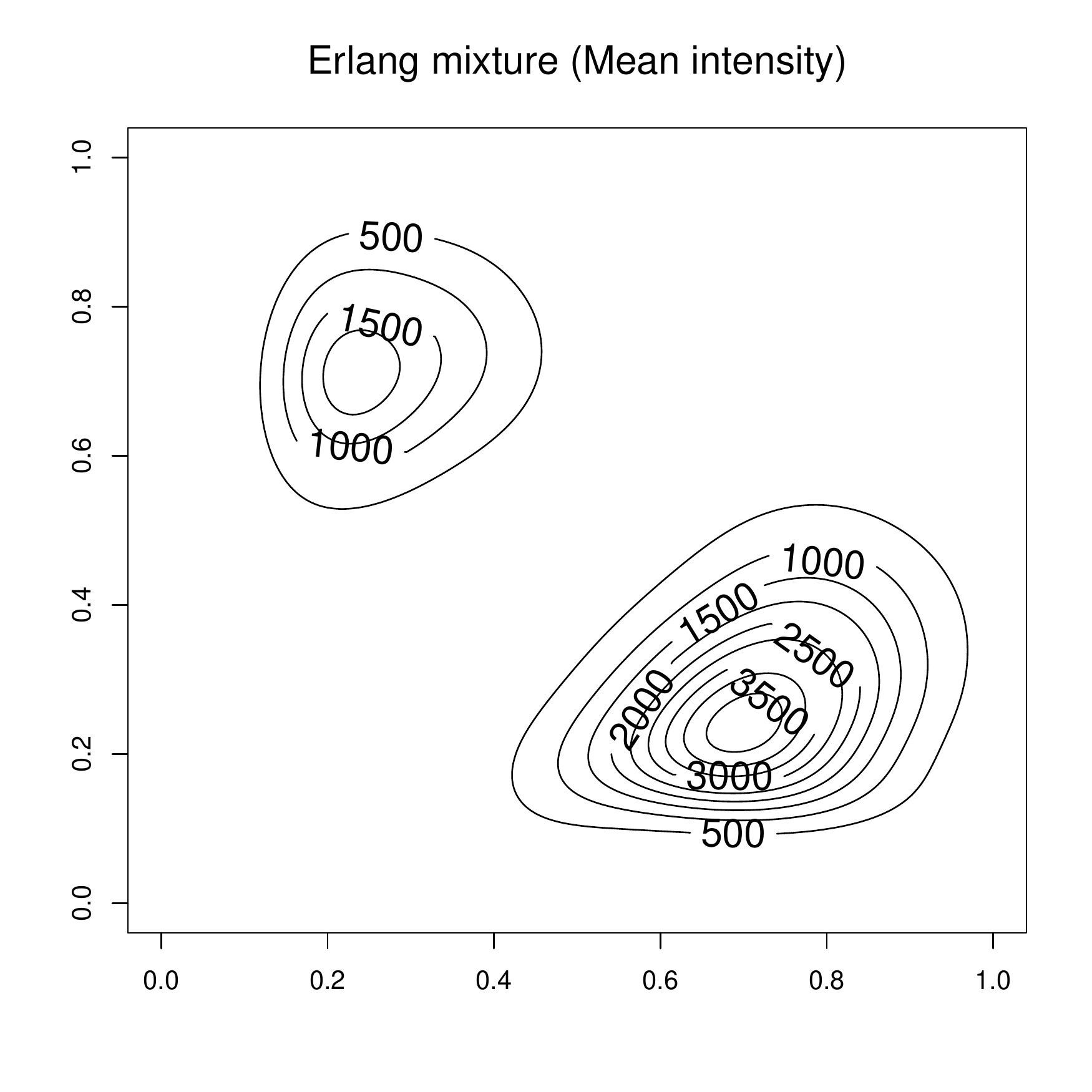} \\
\includegraphics[width=0.45\textwidth]{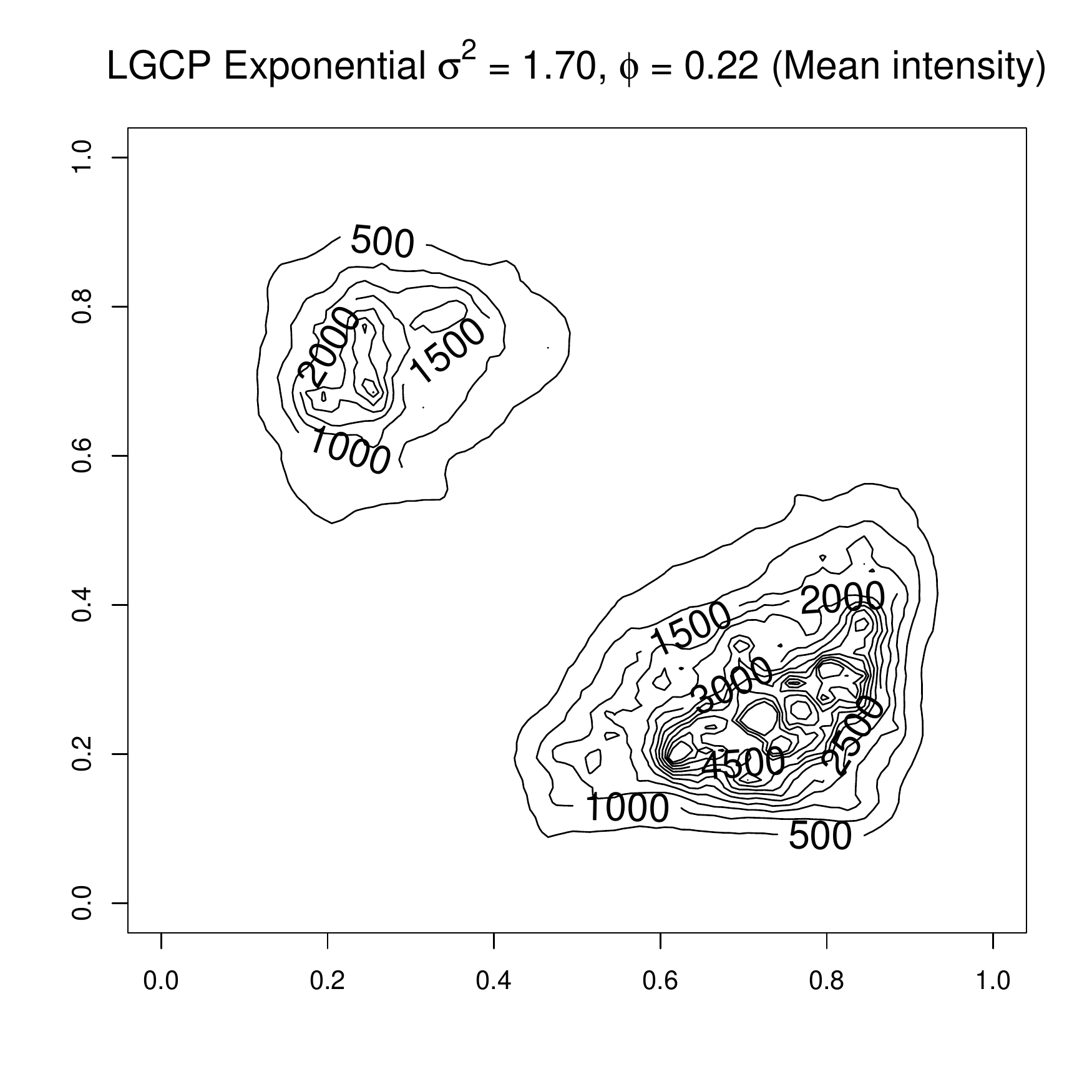}
%\qquad\qquad\includegraphics[width=0.38\textwidth]{./figs/lgnmix_lgcp_iqr_exp_092421.pdf}\\
\includegraphics[width=0.45\textwidth]{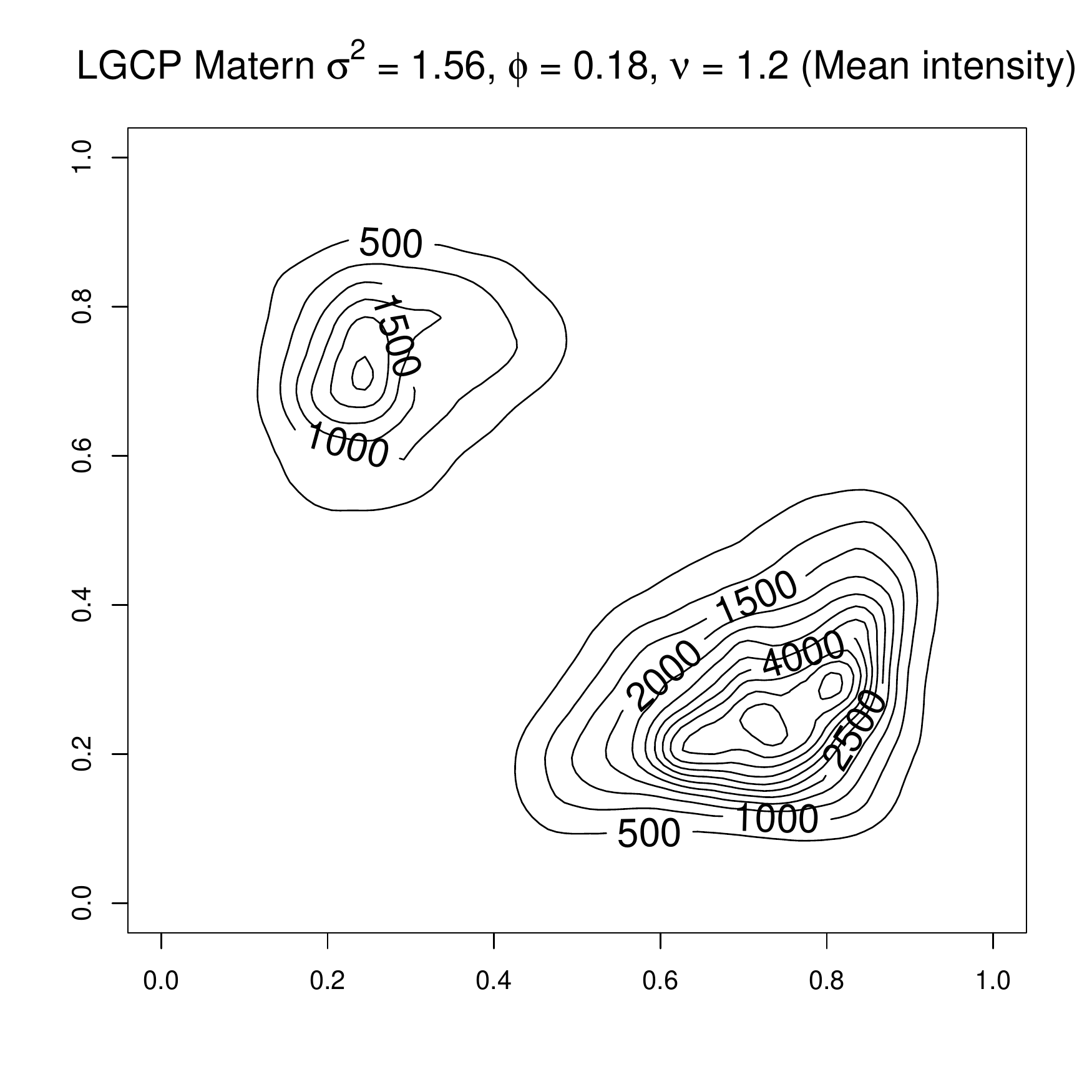} \\
%\qquad\qquad\includegraphics[width=0.38\textwidth]{./figs/lgnmix_lgcp_iqr_matern_092421.pdf}\\
\caption{Synthetic data from spatial NHPP with bimodal intensity defined through a 
two-component mixture of bivariate logit-normal densities (Section 4.2 of the paper). 
The top left panel plots the true intensity function, the top right panel the posterior
mean intensity under the Erlang mixture model, and the two bottom panels the point 
estimate for the intensity under the LGCP model with exponential and Mat\'{e}rn 
GP correlation function.}
	\label{fig:lgcp}
\end{figure}

The top row of Figure \ref{fig:lgcp} shows the true intensity, and the posterior mean intensity
under the Erlang mixture model, obtained under the prior specification discussed in Section 4.2
of the paper. Note that the Erlang mixture model estimates are fairly robust to the prior choice 
for the model hyperparameters, indeed, there is substantial learning for the hyperparameters 
under very dispersed priors (refer to Figure 7 of the paper). 

The bottom row of Figure \ref{fig:lgcp} plots the LGCP model point estimates for the intensity,
using either an exponential or a Mat\'{e}rn GP correlation function. We note that a further 
challenge with the use of LGCP models is that the \textbf{spatstat} package does not provide 
estimation for the additional parameters of correlation functions more general than the 
exponential, in particular, it does not provide an estimate for the smoothness parameter 
($\nu$) of the Mat\'{e}rn correlation function. As suggested in \cite{TDRD2013}, we selected 
the value for $\nu$ by comparing graphically the pair correlation function estimator and the 
covariance function (with $\sigma^2$ and $\phi$ estimated). 

The LGCP model estimates retrieve the bimodal global pattern of the true intensity, but 
with localized behavior (especially in the case of the exponential correlation function) 
that is not present in the underlying intensity surface. In addition to the challenge of 
obtaining full inference with appropriate uncertainty quantification, the sensitivity 
of the point estimates to the choice of the GP correlation function is evident.

\section{MCMC posterior simulation for the spatial NHPP Erlang mixture model}

Denote by $\{ \bm{s}_i: i=1,...,n \}$, where $\bm{s}_i = (s_{i1},s_{i2})$, the 
spatial point pattern observed in the unit square. Under the Erlang mixture model 
for spatial NHPPs, developed in Section 4.1 of the paper, the likelihood is 
proportional to
\begin{equation*}
    \begin{split}
        &\exp\Big(-\int_0^1\int_0^1\lambda(u_1,u_2) \, \text{d}u_1 \text{d}u_2\Big)
        \prod_{i=1}^n\lambda(s_{i1},s_{i2})\\
        &= \exp\Big(-\sum_{j_1=1}^J\sum_{j_2=1}^J \omega_{j_1j_2} \, K_{j_1,\theta_1}(1)K_{j_2,\theta_2}(1)\Big)\prod_{i=1}^n\Big\{\sum_{j_1=1}^J\sum_{j_2=1}^J \omega_{j_1j_2} \, \text{ga}(s_{i1}\mid j_1,\theta_1^{-1})\text{ga}(s_{i2}\mid j_2,\theta_2^{-1})\Big\}\\
        &= \prod_{j_1=1}^J\prod_{j_2=1}^J \exp\Big(-\omega_{j_1j_2}K_{j_1,\theta_1}(1)K_{j_2,\theta_2}(1)\Big)\prod_{i=1}^n\Big\{\Big(\sum_{r_1=1}^J\sum_{r_2=1}^J\omega_{r_1r_2}\Big)\sum_{j_1=1}^J\sum_{j_2=1}^J \Big(\frac{\omega_{j_1j_2}}{\sum_{r_1=1}^J\sum_{r_2=1}^J\omega_{r_1r_2}}\Big)\\
        & \times\text{ga}(s_{i1}\mid j_1,\theta_1^{-1})\text{ga}(s_{i2}\mid j_2,\theta_2^{-1})\Big\}.\\
    \end{split}
\end{equation*}
where $K_{j,\theta}(1)=$ $\int_{0}^{1} \text{ga}(u \mid j,\theta^{-1})\, \text{d}u$.

% As in Section 2.3 for the posterior inference method for the temporal NHPP model, 
Next, we introduce auxiliary variables $\Gamma =$ 
$\{\bm{\gamma}_i: i=1,\ldots,n\}$, where $\bm{\gamma}_i=$ $(\gamma_{i1},\gamma_{i2})$, 
to obtain the following hierarchical model representation: 
\begin{equation*}
    \begin{split}
        \{\bm{s}_1,...,\bm{s}_n\}\mid \Gamma,\bm{\omega},\bm{\theta} &\sim \prod_{j_1=1}^J\prod_{j_2=1}^J \exp\Big(-\omega_{j_1j_2}K_{j_1,\theta_1}(1)K_{j_2,\theta_2}(1)\Big)\\
        &\times \prod_{i=1}^n\Big\{\Big(\sum_{r_1=1}^J\sum_{r_2=1}^J\omega_{r_1r_2}\Big)
        \, \text{ga}(s_{i1}\mid \gamma_{i1},\theta_1^{-1})\text{ga}(s_{i2}\mid \gamma_{i2},\theta_2^{-1})\Big\}\\
        \bm{\gamma}_i\mid \bm{\omega} &\iidsim \sum_{j_1=1}^J\sum_{j_2=1}^J \Big(\frac{\omega_{j_1j_2}}{\sum_{r_1=1}^J\sum_{r_2=1}^J\omega_{r_1r_2}}\Big)
        \, \delta_{j_1}(\gamma_{i1})\delta_{j_2}(\gamma_{i2}), \quad i=1,\ldots,n\\
        \bm{\theta},c_{0},b,\bm{\omega} &\sim p(\theta_1)p(\theta_2)p(c_{0})p(b)\prod_{j_1=1}^{J}\prod_{j_2=1}^{J} \text{ga}(\omega_{j_1j_2} \mid c_0\theta_1\theta_2 b^{-1},c_0)
    \end{split}
\end{equation*}
where $\bm{\omega}=\{ \omega_{j_1j_2}: j_1,j_2=1,...,J \}$, $\bm{\theta} = (\theta_1,\theta_2)$, and $p(\theta_1)$, $p(\theta_2)$, $p(c_{0})$, and $p(b)$ denote the priors for $\theta_1$, $\theta_2$, $c_{0}$, and $b$.

As in the posterior inference method for the temporal NHPP model (Section 2.3 of the paper), 
we explore the posterior distribution using Gibbs sampling. The posterior full conditional for each $\bm{\gamma}_i$ is a discrete distribution on $\{ 1,..., J \} \times \{ 1,..., J \}$ such that 
$\text{Pr}(\gamma_{i1} = j_1, \gamma_{i2} = j_2 \mid \bm{\theta},\bm{\omega},\text{data}) \propto$ $\omega_{j_1j_2} \, \text{ga}(s_{i1} \mid j_1,\theta_1^{-1})\text{ga}(s_{i2} \mid j_2,\theta_2^{-1})$, 
for $j_1, j_2 =1,...,J$.

Let $N_{j_1j_2}=$ $| \{ \bm{s}_{i}: \gamma_{i1} = j_1, \enspace \gamma_{i2} = j_2 \} |$,
for $j_1, j_2 = 1,...,J$. With the conditionally conjugate priors for $\omega_{j_1j_2}$, 
implied by the gamma process prior, the posterior full conditional distribution for 
the mixture weights is derived as follows:
\begin{equation*}
    \begin{split}
        p(\bm{\omega} \mid \Gamma,\bm{\theta},c_{0},b,\text{data}) &\propto \Big\{\prod_{j_1=1}^J\prod_{j_2=1}^J \exp\Big(-\omega_{j_1j_2}K_{j_1,\theta_1}(1)K_{j_2,\theta_2}(1)\Big)\Big\}\Big(\sum_{r_1=1}^J\sum_{r_2=1}^J\omega_{r_1r_2}\Big)^n\\
        &\times \Big\{ \prod_{j_1=1}^{J}\prod_{j_2=1}^{J} \omega_{j_1j_2}^{N_{j_1j_2}} \big(\sum_{r_1=1}^J\sum_{r_2=1}^J\omega_{r_1r_2}\big)^{-N_{j_1j_2}}\Big\}\big\{\prod_{j_1=1}^{J}\prod_{j_2=1}^{J} \text{ga}(\omega_{j_1j_2} \mid c_0\theta_1\theta_2 b^{-1},c_0)\big\}\\
        &\propto \prod_{j_1=1}^J\prod_{j_2=1}^J \exp\Big(-\omega_{j_1j_2}K_{j_1,\theta_1}(1)K_{j_2,\theta_2}(1)\Big)\omega_{j_1j_2}^{N_{j_1j_2}}\text{ga}(\omega_{j_1j_2} \mid c_0\theta_1\theta_2 b^{-1},c_0)\\
        &= \prod_{j_1=1}^J\prod_{j_2=1}^J\text{ga}(\omega_{j_1j_2} \mid N_{j_1j_2}+c_0\theta_1\theta_2 b^{-1},K_{j_1,\theta_1}(1)K_{j_2,\theta_2}(1)+c_0)
    \end{split}
\end{equation*}
with  $\sum_{j_1=1}^{J}\sum_{j_2=1}^{J} N_{j_1j_2} = n$. As in the temporal case,
the mixture weights can be updated independently, given the other parameters and the data, 
from gamma posterior full conditional distributions. Therefore, the practical benefits of 
the Erlang mixture model structure -- convenient updates for the mixture weights and 
computational efficiency of the MCMC algorithm -- carry over to inference for spatial NHPPs.

Finally, parameters $\theta_1$ and $\theta_2$ and the hyperparameters, $c_{0}$ and $b$, 
of the gamma process prior for $H$ are updated with Metropolis-Hastings steps, using log-normal 
proposal distributions.

\bibliographystyle{apalike}
\bibliography{references}